\documentclass[a4paper,12pt]{article}
\usepackage{amsmath}
\usepackage{amssymb}
\usepackage{amsfonts}
\usepackage{amssymb,amsmath}
\usepackage{cancel}
\usepackage[dvips]{lscape,graphicx,epsfig}
\usepackage{fleqn}
\usepackage[footnotesize]{caption}
\usepackage{mathrsfs}

\def\beq{\begin{equation}}
\def\eeq{\end{equation}}
\def\bea{\begin{eqnarray}}
\def\eea{\end{eqnarray}}

\def\<{\left\langle}
\def\>{\right\rangle}

\newcommand{\bc}{\begin{center}}
\newcommand{\ec}{\end{center}}
\newcommand{\bd}{\begin{displaymath}}
\newcommand{\ed}{\end{displaymath}}
\newcommand{\be}{\begin{equation}}
\newcommand{\ee}{\end{equation}}
\newcommand{\ba}{\begin{array}}
\newcommand{\ea}{\end{array}}
\newcommand{\bt}{\begin{tabular}}
\newcommand{\et}{\end{tabular}}

\newcommand{\ds}{\displaystyle}

\begin{document}

\bibliographystyle{OurBibTeX}

\begin{titlepage}

 \vspace*{-15mm}
\begin{flushright}
SHEP-05-32\\
\end{flushright}
\vspace*{5mm}

\begin{center}
{
\sffamily
\LARGE
Theory and Phenomenology of an Exceptional Supersymmetric Standard Model}
\\[8mm]

S.~F.~King\footnote{E-mail: \texttt{sfk@hep.phys.soton.ac.uk}},
S.~Moretti\footnote{E-mail: \texttt{stefano@hep.phys.soton.ac.uk}},
R.~Nevzorov\footnote{E-mail: \texttt{nevzorov@phys.soton.ac.uk}.} 
\footnote{On leave of absence from the Theory Department, 
ITEP, Moscow, Russia.} 
\\[3mm]
{\small\it
School of Physics and Astronomy,
University of Southampton,\\
Highfield, Southampton, SO17 1BJ, U.K.
}\\[1mm]
\end{center}
\vspace*{0.75cm}

\begin{abstract}

\noindent 
We make a comprehensive study of the theory and
phenomenology of a low energy supersymmetric standard model
originating from a string-inspired $E_6$ grand unified gauge group.
The Exceptional Supersymmetric Standard Model (ESSM) considered here
is based on the low energy SM gauge group together with
an extra $Z'$ corresponding to an extra $U(1)_{N}$ gauge
symmetry under which right--handed neutrinos have zero charge.
The low energy matter content of the ESSM corresponds to three $27$
representations of the $E_6$ symmetry group, to ensure anomaly
cancellation, plus an additional pair of Higgs--like doublets as
required for high energy gauge coupling unification. 
The ESSM is therefore a low energy alternative to the MSSM or NMSSM.
The ESSM involves extra matter beyond the MSSM contained in
three $5+5^*$ representations of $SU(5)$, plus
three $SU(5)$ singlets which carry $U(1)_{N}$ charges, one of
which develops a VEV, providing the effective
$\mu$ term for the Higgs doublets, as well as the necessary exotic
fermion masses. We explore the RG flow of the
ESSM and examine theoretical restrictions on the values of new
Yukawa couplings caused by the validity of perturbation theory up
to the Grand Unification scale. We then discuss electroweak symmetry breaking
and Higgs phenomenology, and establish an upper
limit on the mass of the lightest Higgs particle which
can be significantly heavier than in the MSSM and NMSSM, in leading two--loop
approximation. We also discuss the phenomenology of the
$Z'$ and the extra matter, whose discovery will
provide a smoking gun signal of the model.
\end{abstract}

\end{titlepage}
\newpage
\setcounter{footnote}{0}

\section{Introduction}

Despite the absence of any evidence for new particles beyond those
contained in the Standard Model (SM), the cancellation of
quadratic divergences \cite{1} remains a compelling theoretical
argument in favour of softly broken supersymmetry (SUSY) 
which stabilises the Electro-Weak (EW) scale and solves the hierarchy
problem \cite{3} (for a recent review see \cite{Chung:2003fi}).
SUSY also facilitates the high energy convergence of the SM gauge
couplings \cite{2} which allows the SM gauge group to be embedded into
Grand Unified Theories (GUTs) \cite{4} based on simple gauge groups
such as $SU(5)$, $SO(10)$ or $E_6$. The rational $U(1)_Y$ charges,
which are postulated ``ad hoc" in the case of the SM, then appear
in a natural way in the context of SUSY GUT models after the
breakdown of the extended symmetry at some high energy scale
$M_X$, providing a simple explanation of electric charge
quantisation.

An additional motivation to consider models with softly broken
SUSY is associated with the possible incorporation of the
gravitational interactions. The local version of SUSY
(supergravity) leads to a partial unification of the SM gauge
interactions with gravity. However SUpergGRAvity (SUGRA) itself is a
non--renormalisable theory and has to be considered as an
effective low energy limit of some renormalisable or even finite
theory. Currently, the best candidate for such an underlying
theory is ten--dimensional heterotic superstring theory based on
$E_8\times E'_8$ \cite{5}. In the strong coupling regime of an
$E_8\times E'_8$ heterotic string theory described by eleven
dimensional SUGRA (M--theory) \cite{6}, the string scale
can be compatible with the unification scale $M_X$ \cite{7}.
Compactification of the extra dimensions results in the breakdown
of $E_8$ down to $E_6$ or one of its subgroups in the observable
sector \cite{7a}. The remaining $E'_8$ plays the role of a hidden sector
that gives rise to spontaneous breakdown of SUGRA, which
results in a set of soft SUSY breaking terms \cite{8}
characterised by the gravitino mass ($m_{3/2}$) of order the
EW scale\footnote{{\noindent A large mass hierarchy between 
$m_{3/2}$ and Planck scale can appear because of non--perturbative 
sources of SUSY breaking in the hidden sector gauge group
(for a review see \cite{9}).}}.

Although the theoretical argument for low energy SUSY is quite
compelling, it is worth emphasising that the choice of low energy
effective theory at the TeV scale consistent with high energy
conventional (or string) unification is not uniquely specified.
Although the Minimal Supersymmetric Standard Model (MSSM) is the
best studied and simplest candidate for such a low energy
effective theory, the MSSM suffers from the $\mu$ problem: the
superpotential of the MSSM contains one bilinear term
$\mu \hat{H}_d\hat{H}_u$
which can be present before SUSY is
broken. As a result one would naturally expect it to be of the
order of the Planck scale $M_{\rm{Pl}}$. If $\mu\simeq M_{\rm{Pl}}$ 
then the Higgs
scalars get a huge positive contribution $\sim\mu^2$ to their squared
masses and EW Symmetry Breaking (EWSB) does not occur. On the
other hand the parameter $\mu$ cannot be simply omitted. If
$\mu=0$ at some scale $Q$ the mixing between Higgs doublets is not
generated at any scale below $Q$ due to the non--renormalisation
theorems \cite{10}. In this case the minimum of the Higgs boson
potential is attained for $<H_d>=0$. Because of this down--type quarks and
charged leptons remain massless. In order to get the correct
pattern of EWSB, $\mu$ is required to be
of the order of the SUSY breaking (or EW) scale.

The Next--to--Minimal Supersymmetric Standard Model (NMSSM) is an
attempt to solve the $\mu$ problem of the MSSM in the most direct
way possible, by generating $\mu$ dynamically as the low energy
Vacuum Expectation Value
(VEV) of a singlet field. The superpotential of the NMSSM
is given by \cite{11}--\cite{111}
\be
W_{NMSSM}= \lambda\hat{S}(\hat{H}_d\hat{H}_u)
+ \frac{1}{3}\kappa\hat{S}^3+W_{MSSM}(\mu=0)\,. \label{2}
\ee
The cubic term of the new singlet superfield $\hat{S}$ in the superpotential
breaks an additional $U(1)$ global symmetry that would appear
and is a common way to avoid the axion that would result.
However the NMSSM itself is not without problems. The NMSSM superpotential
is still invariant under the transformations of a discrete $Z_3$
symmetry. This $Z_3$ symmetry should lead to the formation of
domain walls in the early universe between regions which were
causally disconnected during the period of EWSB
 \cite{12}. Such domain structure of vacuum create
unacceptably large anisotropies in the cosmic microwave background
radiation \cite{13}. In an attempt to break the $Z_3$ symmetry
operators suppressed by powers of the Planck scale could be
introduced. But it has been shown that these operators give rise
to quadratically divergent tadpole contributions, which
destabilise the mass hierarchy once again \cite{14}.

One solution to these difficulties is to
consider the simplest gauge extensions of the SM gauge group that involve an
additional non--anomalous $U(1)'$ gauge symmetry. Models with
an additional $U(1)'$ factor can arise naturally out of
string-inspired constructions \cite{141,142}. In particular one
or two extra $U(1)'$ factors may emerge in the breaking of a string-inspired
$E_6$ gauge group and the phenomenology of such models has
been extensively studied in the literature
\cite{141}. Such theories may lead to a $U(1)'$
extension of the NMSSM in which a SM singlet field $S$ couples to
the Higgs doublets and yields an effective $\mu$ parameter $\sim
\lambda <S>$, while the $\hat{S}^3$ term is forbidden by the $U(1)'$ gauge
symmetry. In such models the Peccei--Quinn (PQ) symmetry becomes
embedded in the new $U(1)'$ gauge symmetry. Clearly there are no
domain wall problems in such a model since there is no discrete
$Z_3$ symmetry. The field $S$ is charged under the $U(1)'$ so that
its expectation value also gives mass to the new $Z'$ gauge boson
breaking the $U(1)'$, in other words the would-be PQ axion is
eaten by the $Z'$. The extended gauge symmetry forbids an
elementary $\mu$ term as well as terms like $\hat S^n$ in the
superpotential. The role of the $S^3$ term in generating quartic
terms in the scalar potential, which stabilise the physical vacuum, is
played by $D$--terms.

In this paper we explore a specific $E_6$ inspired supersymmetric
realisation of the above $U(1)'$ type model which is capable of
resolving the $\mu$ problem as in the NMSSM, but without facing any
of its drawbacks. The total matter content of our model
corresponds to three families of $27_i$ representations, 
plus two Higgs--like doublets,
consistent with SUSY unification. A particular feature of our model is that
we assume that the $E_6$
gauge group is broken at high energies down to the SM gauge group
plus a particular extra $U(1)_{N}$ gauge symmetry in which right--handed
neutrinos have zero charge and so are singlets and
do not participate in the gauge interactions. Since
right--handed neutrinos have zero charges they can
acquire very heavy Majorana masses
and are thus suitable to take part in the standard seesaw mechanism
which yields small neutrino masses. Having heavy right--handed neutrinos
also avoids any stringent constraints on the mass of the
$Z'$ boson coming from the nucleosynthesis and astrophysical data
which would be present if the right--handed neutrinos were light.

The above superstring inspired $E_6$ SUSY model, henceforth 
referred to as the Exceptional Supersymmetric Standard Model
(ESSM), provides a
theoretically attractive solution to the $\mu$ problem of the MSSM
since the bilinear Higgs $\mu$ terms are forbidden by the $U(1)_N$ gauge symmetry.
This model contains three pairs of candidate Higgs doublets
${H}_{1i}$, ${H}_{2i}$, plus three singlets $S_i$, which
carry $U(1)_N$ charge. These states all originate
from three $27_i$ representations and couple together according to
a $27_i27_j27_k$ coupling resulting in NMSSM type
superpotential couplings of the form:
\be
\ba{c}
W_{H}=\sum_{ijk}\lambda_{ijk}\hat{S}_i\hat{H}_{1j}\hat{H}_{2k}\,.
\ea
\label{1}
\ee
The breaking of the EW and $U(1)_N$ gauge symmetry down to 
$U(1)_{em}$ takes place when some of these Higgs fields acquire VEVs.
It is possible to work in a basis where only one family of Higgs
fields and singlets acquire non--zero VEVs, which we can define 
to be the third family, and we can then define
$S\equiv S_3$, $H_d\equiv H_{1,3}$, $H_u\equiv H_{2,3}$.
We shall then refer to the remaining first two families 
$S_i$, $H_{1,i}$, $H_{2,i}$, with $i=1,2$ as non--Higgs doublets and singlets. 
The relation of the third family Higgs so defined 
to the third family quarks and leptons is more model dependent,
but in the context of Radiative EWSB (REWSB) 
it is natural to associate the third family Higgs to the third
family quarks and leptons, since it is the large Yukawa coupling
of the third family which drives the Higgs VEVs.
This also avoids
the appearance of Flavour Changing Neutral Currents  (FCNCs).
Then, restricting ourselves to Higgs fields which develop VEVs,
the superpotential (\ref{1}) reduces to
\beq
W_{H} \rightarrow \lambda\hat{S}(\hat{H}_d\hat{H}_u)
\eeq
which is just the NMSSM coupling in Eq.~(\ref{2}).
After the spontaneous symmetry
breakdown at the EW scale the scalar component of the 
superfield $\hat{S}$
acquires non--zero VEV  ($\langle S \rangle
=s/\sqrt{2}$) and an effective $\mu$-term ($\mu=\lambda
s/\sqrt{2}$) of the required size is automatically generated. The
Higgs sector of the considered model contains only one additional singlet field and
one extra parameter compared to the MSSM. Therefore it can be
regarded as the simplest extension of the Higgs sector of the MSSM.

It is interesting to compare the ESSM to other related models with an extra
$U(1)'$. In general
anomaly cancellation requires either the presence of exotic chiral
supermultiplets \cite{24}--\cite{25} or family--non--universal
$U(1)'$ couplings \cite{26}. Any family dependence of the
$U(1)'$ charges would result in 
FCNCs mediated by the $Z'$ which can be suppressed for the first
two generation and manifest themselves in rare $B$ decays and
$B-\overline{B}$ mixing \cite{27}.
In the ESSM, because the $U(1)_{N}$ charge assignment is
flavour--independent, the considered model does not suffer from the
FCNC problem. In the ESSM anomalies are cancelled in a flavour--independent
way since the model contains an extra $U(1)_N$ arising from
$E_6$ together with the matter content of (three) complete $27$ representations
of $E_6$ down to the TeV scale, apart from the three right--handed
neutrinos which are singlets under the low energy gauge group.
The existence of exotic supermultiplets in the ESSM
is consistent with gauge coupling unification since the
extra matter is in complete $27$ representations.
However exotic quarks and non--Higgses naturally appear in the $E_6$
inspired model, with the quantum numbers of three families
of $5+5^\ast$ $SU(5)$ representations,
which phenomenologically correspond to three families
of extra down--type quark singlets and three
families of matter with the quantum numbers of Higgs doublets,
where each multiplet is accompanied by its conjugate representation.
The large third family
coupling of the extra coloured chiral superfields ($D, \bar{D}$)
to the singlet $S$ of the form $\kappa S(D\overline{D})$ may help to
induce radiative breakdown of the $SU(2)\times U(1)_Y\times U(1)'$ symmetry
\cite{24}, \cite{28}--\cite{30}.

Before describing the phenomenological work performed 
in this paper it is worth
briefly reviewing the phenomenological
studies performed so far on related models in the literature.
Recently the implications of SUSY models with an additional
$U(1)'$ gauge symmetry have been studied for CP--violation
\cite{31}, neutrino physics \cite{311}--\cite{312}, dark matter
\cite{32}, leptogenesis \cite{321}, EW baryogenesis
\cite{33}--\cite{331}, muon anomalous magnetic moment \cite{34},
electric dipole moment of electron \cite{341}, lepton flavour
violating processes like $\mu\to e\gamma$ \cite{342} (forbidding
R--parity violating terms \cite{25}, \cite{35}). An
important property of $U(1)'$ models is that the tree--level mass
of the lightest Higgs particle can be larger than $M_Z$ even for
moderate values of $\tan\beta\simeq 1-2$
\cite{30}--\cite{31}, \cite{36}, hence the existing LEP bounds can be
satisfied with almost no need for large radiative corrections.
Models with a $U(1)_{N}$ gauge symmetry in which right--handed
neutrinos have zero charge have been studied in \cite{312} in the
context of non--standard neutrino models with extra singlets, in
\cite{401} from the point of view of $Z-Z'$ mixing and a discussion
of the neutralino sector, in \cite{29} where the RG
 was studied, in \cite{30} where a one--loop Higgs mass upper bound was
presented.

The phenomenological analysis presented here goes
well beyond what has appeared
so far in the literature.
Our analysis begins with a detailed
RG analysis of the dimensionless couplings in the ESSM,
and an examination of the fixed points of the model.
This analysis is completely new as it has not appeared before.
We later use the results of this analysis in determining
an upper bound on the lightest CP--even (or scalar) Higgs boson mass,
using the effective potential and including two--loop corrections,
which had also not been considered previously,
and we make a detailed comparison with similar bounds obtained in the MSSM and NMSSM.
We then make a comprehensive phenomenological study of the full Higgs spectrum
and couplings, which includes the low energy allowed regions
in which EWSB is successful, and comment on the
crucial phenomenological aspects of the Higgs sector of the ESSM.
The chargino and neutralino spectrum expected in the ESSM
is also studied in some depth. We then discuss the phenomenology of the
extra particles predicted by the ESSM, including the $Z'$ and some exotic 
fermions, and provide a numerical estimate and a full discussion
of their production cross sections
and signatures at the upcoming CERN Large Hadron Collider (LHC) and 
a future International Linear Collider (ILC).

The paper is organised as follows. In section 2 we specify our
model. In section 3 we examine the RG flow of
gauge and Yukawa couplings assuming that gauge coupling
unification takes place at high energies. The EWSB
and Higgs phenomenology are studied in sections 4 and 5 respectively.
In section 6 we consider the chargino and neutralino spectrum in the ESSM while in section 7
the potential discovery of a $Z'$ boson and new exotic particles at future
colliders are discussed. Our results are summarised in
section 8.

\section{The ESSM}
\subsection{Overview of the model}
As it is clear from the discussion in the Introduction, 
the ESSM is a low energy alternative to the MSSM or NMSSM defined as follows.
The ESSM originates from an $E_6$ GUT gauge group which is 
broken at the GUT scale to the SM gauge group together 
with an additional $U(1)_N$ gauge group which is not broken until
a scale not very far above the EW scale, giving rise 
to an observable $Z'$ gauge boson. The $U(1)_N$ gauge
group is defined such that right--handed neutrinos $N^c_i$ carry zero charges
under it. 
The matter content below the GUT scale corresponds to three
$27$--plets of $E_6$ (labelled as $27_i$)
which contain the three ordinary families of
quarks and leptons including right--handed neutrinos $N^c_i$, 
three families of candidate Higgs doublets
${H}_{1i}$, ${H}_{2i}$, three 
families of extra down-type quark singlets $D_i, \bar{D_i}$, 
and three families of extra singlets $S_i$. However only
the third family 
Higgs doublets $H_u$ and $H_d$ and singlet $S$ develop VEVs.
In addition, in order to achieve gauge coupling unification,
there is a further pair of Higgs--like doublet
supermultiplets $H'$ and $\bar{H}'$ which do not develop VEVs,
arising from an incomplete $27'+\overline{27}'$ representation.
All the extra matter described above is expected to have 
mass of order the TeV scale and may be observable at the LHC or at an ILC.
The ESSM has the following desirable features:
\begin{itemize}
\item Anomalies are cancelled 
generation by generation within each complete $27_i$ representation.
\item Gauge coupling unification is accomplished due to the complete
$27_i$ representations together with the additional pair of
Higgs--like doublets $H'$ and $\bar{H}'$
from the incomplete $27'+\overline{27}'$ representation.
\item The seesaw mechanism is facilitated due to the
right--handed neutrinos $N_i^c$ arising from the $27_i$ representations
having zero gauge charges.
\item The $\mu$-problem of the MSSM is solved since the
$\mu$-term is forbidden by the $U(1)_N$ gauge symmetry. It is 
replaced by a singlet coupling
$\hat{S}(\hat{H}_d\hat{H}_u)$
as in the NMSSM, but without the $\hat{S}^3$ term of the NMSSM
which resulted in domain wall problems. Besides, in the ESSM the would-be
Goldstone boson is eaten by the $Z'$ associated with the 
$U(1)_N$ gauge group. 
\end{itemize}

The purpose of the remainder of this section is to develop 
the theoretical aspects of the ESSM defined above, and define the
matter content, charges, couplings and symmetries of the ESSM more precisely.

\subsection{The choice of the surviving $U(1)_N$ gauge group}
Since all matter and Higgs
superfields must originate from $27$ and $\overline{27}$--plets of $E_6$, 
one cannot 
break $E_6$ in a conventional manner as the required Higgs fields are
in larger representations than the $27$.  
However, at the string scale, $E_6$ can be broken via the
Hosotani mechanism \cite{37}. Because the rank of the $E_6$ group is six
the breakdown of the $E_6$ symmetry results in several models based on
rank--5 or rank--6 gauge groups. As a consequence $E_6$ inspired SUSY
models in general may lead to low--energy gauge groups with one or
two
additional $U(1)'$ factors in comparison to the SM. Indeed $E_6$
contains the maximal subgroup $SO(10)\times U(1)_{\psi}$ while
$SO(10)$ can be decomposed in terms of the $SU(5)\times U(1)_{\chi}$
subgroup. By means of the Hosotani mechanism $E_6$ can be broken
directly to $SU(3)_C\times SU(2)_W\times U(1)_Y\times
U(1)_{\psi}\times U(1)_{\chi}$ which has rank 6. For suitable large
VEVs of the symmetry breaking Higgs fields this
rank--6 model can be reduced further to an effective rank--5 model with
only one extra gauge symmetry.  Then
an extra $U(1)'$ that appears at
low energies is a linear combination of $U(1)_{\chi}$ and
$U(1)_{\psi}$:
\be U(1)'=U(1)_{\chi}\cos\theta+U(1)_{\psi}\sin\theta\,.
\label{3}
\ee 

In general 
the right--handed neutrinos will carry non--zero
charges with respect to the extra gauge interaction $U(1)'$. It means that their
mass terms are forbidden by the gauge symmetry. The right--handed
neutrinos can gain masses only after the breakdown of the $SU(2)_W\times
U(1)_Y\times U(1)'$ symmetry. But even in this case one can expect
that the corresponding mass terms should be suppressed due to the
invariance of the low--energy effective Lagrangian under the $U(1)_L$
symmetry associated with lepton number conservation\footnote{{\noindent 
In many supersymmetric models the invariance
under the lepton and baryon $U(1)$ symmetries are caused by 
R--parity conservation which is normally imposed to prevent rapid
proton decay.}}.  If the right--handed neutrinos were lighter than a
few $\mbox{MeV}$ they would be produced prior to big bang
nucleosynthesis by the $Z'$ interactions leading to a faster expansion
rate of the Universe and to a higher $~^4He$ relic abundance \cite{21}. The current
cosmological observations of cosmic microwave background radiation and
nuclear abundances restrict the total effective number of extra
neutrino species $\Delta N_{\nu}$ to $0.3$ \cite{38}. The strength of
the interactions of right--handed neutrinos with other particles and
the equivalent number of additional neutrinos rises with a decreasing
$Z'$ boson mass. Thus cosmological and astrophysical
observations set a stringent limit on the $Z'$ mass which has to be
larger than $4.3\,\mbox{TeV}$ \cite{21}.

The situation changes dramatically if the right--handed neutrinos remain
sterile after the breakdown of the $E_6$ symmetry, i.e. have zero
charges under the surviving gauge group. The extra $U(1)'$
factor for which right--handed neutrinos transform trivially is called
$U(1)_{N}$. It corresponds to the angle $\theta=\arctan\sqrt{15}$ in
Eq.~(\ref{3}).  In this case, considered here, 
the right--handed neutrinos may be superheavy. 
Then the three known doublet neutrinos $\nu_e$,
$\nu_{\mu}$ and $\nu_{\tau}$ acquire small Majorana masses via the
seesaw mechanism. This allows for a comprehensive understanding of
the mass hierarchy in the lepton sector and neutrino oscillations
data. The successful leptogenesis in the early epoch of the Universe
is the distinctive feature of the ESSM 
with an extra $U(1)_{N}$ factor \cite{321}. Because right--handed neutrinos
are allowed to have large masses, they may decay into final states
with lepton number $L=\pm 1$, thereby creating a lepton asymmetry in the early
Universe \cite{39}. Since sphalerons violate $B+L$ but conserve $B-L$,
this lepton asymmetry subsequently 
gets converted into the present observed baryon
asymmetry of the Universe through the EW phase transition
\cite{40}.  Any other $E_6$ inspired supersymmetric extension with the
extra $U(1)$ factor would result in $B-L$ violating interactions at
$O(1)\,\mbox{TeV}$ as it is broken down to the SM. This $B-L$
violating interactions together with $B+L$ violating sphalerons would
erase any lepton or baryon asymmetry that may have been created during the
earlier epoch of the Universe. Different phenomenological aspects of
supersymmetric models with an extra $U(1)_{N}$ gauge symmetry were
studied in
\cite{29}, \cite{30}, \cite{312}, \cite{321}, \cite{331}, \cite{401}.

One of the most important issues in $U(1)'$ models is the cancellation
of the gauge and gravitational anomalies.  In $E_6$ theories the
anomalies are cancelled automatically. Therefore all models that
 are
based on the $E_6$ subgroups and contain complete representations 
 should be anomaly--free. As a result in order to make
the chosen 
supersymmetric model with the extra $U(1)_{N}$ factor anomaly--free one is
forced to augment the minimal spectrum by a number of exotics which,
together with ordinary quarks and leptons, form complete fundamental
$27$ representations of $E_6$. These decompose under the surviving
low energy gauge group as discussed in the next subsection.

\subsection{The low energy matter content of the ESSM}
The three families 
of fundamental $27_i$ representations
decompose under the $SU(5)\times U(1)_{N}$ subgroup of $E_6$ \cite{29}
as follows: 
\be 
27_i\to
\ds\left(10,\,\ds{1}\right)_i+\left(5^{*},\,\ds{2}\right)_i
+\left(5^{*},\,-\ds{3}\right)_i
+\ds\left(5,-\ds{2}\right)_i
+\left(1,\ds{5}\right)_i+\left(1,0\right)_i\,.
\label{4}
\ee 
The first and second quantities in the brackets are the $SU(5)$
representation and extra $U(1)_{N}$ charge while $i$ is a family index
that runs from 1 to 3. An ordinary SM family which contains the
doublets of left--handed quarks $Q_i$ and leptons $L_i$, right--handed
up-- and down--type quarks ($u^c_i$ and $d^c_i$) as well as right--handed
charged leptons, is assigned to the
$\left(10,\ds{1}\right)_i+
\left(5^{*},\,\ds{2}\right)_i$. These 
representations decompose under 
\be 
SU(5)\times U(1)_{N}\rightarrow 
SU(3)_C\times SU(2)_W\times U(1)_Y\times U(1)_{N}
\label{401a}
\ee
to give ordinary quarks and leptons:
\be \ba{rcl}
\left(10,\,\ds{1}\right)_i&\to&
Q_i=(u_i,\,d_i)\sim\left(3,\,2,\,\ds\frac{1}{6},\,\ds{1}\right)\,,\\[2mm]
&&
u^c_i\sim\left(3^{*},\,1,\,-\ds\frac{2}{3},\,\ds{1}\right)\,,\\[2mm]
&& e^c_i\sim\left(1,\,1,\,1,\,\ds{1}\right)\,;\\[4mm]
\left(5^{*},\,\ds{2}\right)_i&\to &
d^c_i\sim\left(3^{*},\,1,\,\ds\frac{1}{3},\,\ds{2}\right)\,,\\[2mm]
&& L_i=(\nu_i,\,
e_i)\sim\left(1,\,2,\,-\ds\frac{1}{2},\,\ds{2}\right)\,,
\ea
\label{5}
\ee
where the third quantity in the brackets is the $U(1)_Y$ hypercharge. (In Eq.~(\ref{5}) and further we omit all isospin and colour indexes
related to $SU(2)$ and $SU(3)$ gauge interactions.)

The right--handed neutrinos $N^c_i$ transform trivially under
$SU(5)\times U(1)_{N}$ by definition.
Therefore $N^c_i$ should be associated with the last term in
Eq.~(\ref{4}) $\left(1,0\right)_i$. The next--to--last term in Eq.~(\ref{4}),
$\left(1,\ds{5}\right)_i$, represents SM singlet fields $S_i$
which carry non--zero $U(1)_{N}$ charges and therefore survive
down to the EW scale.

The remaining representations in Eq.~(\ref{4})
decompose as follows: 
\be \ba{rcl}
\left(5^{*},\,-\ds{3}\right)_i&\to&
H_{1i}=(H^0_{1i},\,H^{-}_{1i})\sim
\left(1,\,2,\,-\ds\frac{1}{2},\,-\ds{3}\right)\,,\\[2mm]
&&
\overline{D}_i\sim\left(3^{*},\,1,\,\ds\frac{1}{3},\,-\ds{3}\right)\,,\\[4mm]
\left(5,\,-\ds{2}\right)_i&\to &
H_{2i}=(H^{+}_{2i},\,H^{0}_{2i})\sim
\left(1,\,2,\,\ds\frac{1}{2},\,\ds-{2}\right)\,,\\[2mm]
&&
D_i\sim\left(3,\,1,\,-\ds\frac{1}{3},\,\ds-{2}\right)\,.
\ea
\label{6}
\ee 
The pair of $SU(2)$--doublets ($H_{1i}$ and $H_{2i}$) that are
contained in $\left(5^{*},\,-\ds{3}\right)_i$ and
$\left(5,-\ds{2}\right)_i$ have 
the quantum numbers of Higgs doublets. Other components
of these exotic $SU(5)$ multiplets form extra 
colour triplet but EW singlet quarks $D_i$
and anti--quarks $\bar{D_i}$ with electric charges $-1/3$ and $+1/3$
respectively. The exotic multiplets in Eq.~(\ref{6}) form vector
pairs under the SM gauge group.

In addition to the three complete $27_i$ representations just discussed,
some components of additional $27'$ and
$\overline{27'}$ representations can and must survive to low energies,
in order to preserve gauge coupling unification.
We assume that an additional $SU(2)$ doublet and
anti--doublet $H'$ and $\bar{H}'$ originate as
incomplete multiplets of an additional 
$27'$ and $\overline{27'}$. Specifically we assume that they
originate from the $SU(2)$ doublet components of a
$\left(5^{*},\,\ds{2}\right)$ from a 
$27'$, and the corresponding 
anti--doublet from a $\overline{27'}$.
 
The low energy matter
content of the ESSM may then be summarised as: 
\be
3\left[(Q_i,\,u^c_i,\,d^c_i,\,L_i,\,e^c_i,\,N_i^c)\right]
+3(S_i)+3(H_{2i})+3(H_{1i})+3(D_i,\,\bar{D}_i)+H'+\bar{H}'\,,
\label{7}
\ee
where the right--handed neutrinos $N^c_i$ are expected to gain masses
at some intermediate scale, while the remaining matter survives
down to the EW scale near which the gauge group $U(1)_N$ is broken.

\subsection{The low energy symmetries and couplings of the ESSM}
In $E_6$ models the renormalisable part of the superpotential comes
from the $27\times 27\times 27$ decomposition of the $E_6$ fundamental
representation. The most general renormalisable superpotential which
is allowed by the $SU(3)\times SU(2)\times U(1)_{Y}\times U(1)_{N}$
gauge symmetry can be written in the following form: 
\be
W_{\rm{total}}=W_0+W_1+W_2+W_{\cancel{E_6}}\,.
\label{8}
\ee

The first, second and third terms in Eq.~(\ref{8}) represent the most
general form of the superpotential allowed by the $E_6$ symmetry.
$W_0$, $W_1$ and $W_2$ are given by 
\be \ba{rcl}
W_0&=&\lambda_{ijk}S_i(H_{1j}H_{2k})+\kappa_{ijk}S_i(D_j\overline{D}_k)+h^N_{ijk}
N_i^c (H_{2j} L_k)+ h^U_{ijk} u^c_{i} (H_{2j} Q_k)+\\[2mm]
&&+h^D_{ijk} d^c_i (H_{1j} Q_k) + h^E_{ijk} e^c_{i} (H_{1j} L_k)
\,,\\[4mm] W_1&=& g^Q_{ijk}D_{i} (Q_j Q_k)+g^{q}_{ijk}\overline{D}_i
d^c_j u^c_k\,,\\[4mm] W_2&=& g^N_{ijk}N_i^c D_j d^c_k+g^E_{ijk} e^c_i
D_j u^c_k+g^D_{ijk} (Q_i L_j) \overline{D}_k\,.  \ea
\label{10}
\ee 
The part of the superpotential (\ref{8}) coming from the $27\times
27\times 27$ decomposition of the $E_6$ fundamental representation
(i.e. $W_0+W_1+W_2$) possesses a global $U(1)$ symmetry that can be
associated with $B-L$ number conservation.  This enlarged global
symmetry is broken explicitly by most of the terms in
$W_{\cancel{E_6}}$.

The last part of the superpotential (\ref{8}) includes the $E_6$
violating set of terms:
\be
W_{\cancel{E_6}}=\frac{1}{2}M_{ij} N_i^cN_j^c+W'_0+W'_1+W'_2\,,
\label{9}
\ee 
where 
\be 
\ba{l} 
W'_0=\mu'(H'\overline{H}')+\mu'_i(\overline{H}'L_i)+h_{ij} N_i^c (H_{2j}
H')+h^{H'}_{ij} e^c_{i} (H_{1j} H')\,,\\[2mm] 
W'_1=
\ds\frac{\sigma_{ijk}}{3} N_i^c N_j^c N_k^c
+\Lambda_k N_k^c+ \lambda_{ij}
S_i(H_{1j}\overline{H}')+g^N_{ij}N^c_i(\overline{H}'L_j)\\[2mm]
+g^N_i
N^c_i(\overline{H}'H')+g^U_{ij}u^c_i(\overline{H}'Q_j)+\mu_{ij}(H_{2i}L_j)
+\mu_i (H_{2i} H')+\mu'_{ij} D_i d^c_j\,,\\[2mm] W'_2=g^{H'}_{ij} (Q_i H')
\overline{D}_j\,,\qquad\qquad\qquad i,j,k=1,2,3\,.\\[2mm] 
\ea
\label{999}
\ee 
The terms in Eq.~(\ref{9})
are invariant with respect to the SM gauge group and extra
$U(1)_{N}$ transformations but are either 
forbidden by the $E_6$ symmetry itself
or by the splitting of complete $27$ and
$\overline{27}$ representations that also breaks $E_6$.  Some of the
interactions listed in Eq.~(\ref{9}) can play a crucial role in
low--energy phenomenology. For example, Majorana 
mass terms of the right--handed
neutrinos at some intermediate scales
provide small Majorana masses for the three species of
left--handed neutrinos via the seesaw mechanism. 
Some other terms in Eqs.~(\ref{9})--(\ref{999}) (like $\mu_{ij}H_{2i}L_j$) may 
be potentially dangerous from the phenomenological point of view.

Although the $B-L$ number is conserved automatically in $E_6$
inspired SUSY models, some Yukawa interactions in Eq.~(\ref{10}) violate
baryon number resulting in rapid proton decay. The baryon and lepton
number violating operators can be suppressed by postulating the
invariance of the Lagrangian under R--parity transformations.
In the MSSM the R--parity quantum numbers are
\be
R=(-1)^{3(B-L)+2S}\,.
\label{11}
\ee 
The straightforward generalisation of the definition of 
R--parity to 
$E_6$ inspired supersymmetric models, assuming $B_{D}=1/3$ and
$B_{\overline{D}}=-1/3$, implies that $W_1$ and $W_2$ are forbidden by
the discrete symmetry (\ref{11}). In this case the rest of the
Lagrangian of the considered model, which is allowed by the $E_6$
symmetry, is invariant not only with respect to $U(1)_L$ and $U(1)_B$
but also under $U(1)_D$ symmetry 
transformations\footnote{{\noindent The term $\mu'_{ij} D_i d^c_j$ in
$W_{\cancel{E_6}}$ spoils the invariance under $U(1)_D$ symmetry but
the mechanism of its generation remains unclear.}}  
\be 
D\to e^{i\alpha} D\,,\qquad\qquad
\overline{D}\to e^{-i\alpha}\overline{D}\,.
\label{12}
\ee

The $U(1)_D$ invariance ensures that the lightest exotic quark is
stable. Any heavy stable particle would have been copiously produced
during the very early epochs of the Big Bang. Those strong or
electromagnetically interacting fermions and bosons which survive
annihilation would subsequently have been confined in heavy hadrons
which would annihilate further. The remaining heavy hadrons
originating from the Big Bang should be present in terrestrial
matter. There are very strong upper limits on the abundances of
nuclear isotopes which contain such stable relics in the mass range
from $1\,\mbox{GeV}$ to $10\,\mbox{TeV}$. Different experiments set
limits on their relative concentrations from $10^{-15}$ to $10^{-30}$
per nucleon \cite{42}. At the same time various theoretical
estimations \cite{43} show that if remnant particles would exist in
nature today their concentration is expected to be at the level of
$10^{-10}$ per nucleon. Therefore $E_6$ inspired models with stable
exotic quarks or non--Higgsinos are ruled out.

To prevent rapid proton decay in $E_6$ supersymmetric models the
definition of R--parity should be modified. There are 8 different ways
to impose an appropriate $Z_2$ symmetry resulting in baryon and
lepton number conservation \cite{44}. The requirements of successful
leptogenesis and non--zero neutrino masses single out only two ways to
do that. If $H_{1i}$, $H_{2i}$, $S_i$, $D_i$, $\overline{D}_i$ and
quark superfields ($Q_i$, $u^c_i$, $d^c_i$) are even under $Z_2$ while
lepton superfields ($L_i$, $e^c_i$, $N^c_i$) and survival components
of $27$ and $\overline{27}$ ($H'$ and $\overline{H}'$) are odd, all
terms in $W_2$ are forbidden.  Then the part of the superpotential
allowed by the $E_6$ symmetry is invariant with respect to $U(1)_B$
and $U(1)_L$ global symmetries if the exotic quarks $D_i$ and
$\overline{D_i}$ carry twice larger baryon number than the ordinary quark
fields $d^c_i$ and $Q_i$ respectively. It implies that $\overline{D_i}$
and $D_i$ are diquark and anti--diquark, i.e. $B_{D}=-2/3$ and
$B_{\overline{D}}=2/3$. This way of suppressing baryon and lepton
number violating operators will be called further Model I.  An
alternative possibility is to assume that the exotic quarks $D_i$ and
$\overline{D_i}$ as well as ordinary lepton superfields and survivors are all odd
under $Z_2$ whereas the others remain even. Then we get Model II in
which all Yukawa interactions in $W_1$ are ruled out by the discrete
$Z_2$ symmetry. Model II possesses two extra $U(1)$ global
symmetries. They can be associated with $U(1)_L$ and $U(1)_B$ if the
exotic quarks carry baryon ($B_{D}=1/3$ and $B_{\overline{D}}=-1/3$)
and lepton ($L_{D}=1$ and $L_{\overline{D}}=-1$) numbers
simultaneously. It means that $D_i$ and $\overline{D_i}$ are leptoquarks
in Model II.

In Model II the imposed $Z_2$ symmetry forbids all the terms in
the $W'_1$ part of $W_{\cancel{E_6}}$, leaving only
the mass terms for the right--handed
neutrinos, $W'_0$ and $W'_2$. The discrete symmetry postulated in the
Model I also rules out $W'_2$ but permits $\mu'_{ij} D_i d^c_j$ which
violate baryon number making possible the transition $p\to
\pi^{+}\chi^0$, where $\chi^0$ is a neutralino. In order to suppress
dangerous operators one can impose another $Z_2$ symmetry that changes
the sign of the ordinary quark superfields $Q_i$, $u^c_i$, $d^c_i$
leaving all others unchanged. Finally for the superpotentials of Model
I and II we get 
\be \ba{rcl} 
I)\quad W_{\rm{ESSM\, I}}&=& W_0+W_1+\ds\frac{1}{2}M_{ij}N_i^c N_j^c+W'_0\,,\\[3mm] 
II)\quad
W_{\rm{ESSM\, II}}&=& W_0+W_2+\ds\frac{1}{2}M_{ij}N_i^c N_j^c+W'_0+W'_2.  \ea
\label{13}
\ee

\subsection{Origin of bilinear mass terms in the ESSM}
In the superpotentials (\ref{13}) the non--Higgs doublets and the survival
components from the $27'$ can be redefined in such a way that only one
$SU(2)$ doublet $H'$ interacts with the $\overline{H}'$ from the 
$\overline{27'}$. As a result without loss of generality $\mu'_i$ in
$W'_0$ may be set to zero. Then the superpotentials of the considered
supersymmetric models include two types of bilinear terms
only. One of them, $\ds\frac{1}{2}M_{ij} N_i^c N_j^c$, determines the
spectrum of the right--handed neutrinos which are expected to be heavy
so that the corresponding mass parameters $M_{ij}$ are at intermediate
mass scales. Another one, $\mu' H'\overline{H}'$, is
characterised by the mass parameter $\mu'$ which should not be too
large otherwise it spoils gauge coupling unification
in the ESSM. On the other hand, the parameter $\mu$
cannot be too small since $\mu' H'\overline{H}'$ is solely responsible
for the mass of the charged and neutral components of
$\overline{H}'$. Therefore we typically require $\mu' \sim O(1\ {\rm{TeV}})$ 
as in the MSSM, potentially giving rise to the $\mu$ problem once again.

Within SUGRA models the appropriate term $\mu' H'\overline{H}'$ in the 
superpotentials (\ref{13}) can be induced just after the breakdown of 
local SUSY if the K\"ahler potential contains an extra term 
$(Z(H'\overline{H}')+h.c)$ \cite{45}. This mechanism is of course just 
the same one used in the MSSM to solve the $\mu$ problem. But in  
superstring inspired models the bilinear terms involving $H_d$ and $H_u$ 
are forbidden by the $E_6$ symmetry both in the K\"ahler potential and 
superpotential. As a result the mechanism mentioned above cannot be 
applied for the generation of $\mu H_d H_u$ in the ESSM superpotential.
However this mechanism may be used to give mass to the non--Higgs 
doublets $H'$ and $\overline{H}'$ from additional $27'$ and 
$\overline{27'}$ since the corresponding bilinear terms are allowed by 
the $E_6$ symmetry both in the K\"ahler potential and superpotential.

The other bilinear terms in the superpotential of the $E_6$ inspired SUSY
models responsible for right--handed neutrino masses
can be induced through the non--renormalisable interactions of
$27$ and $\overline{27}$ of the form
$\ds\frac{\kappa_{\alpha\beta}}{M_{Pl}}(27_{\alpha}\,\overline{27}_{\beta})^2$.
If the $N^c_H$ and $\overline{N}_H^c$ components of some extra $27_H$ and
$\overline{27}_H$ representations develop VEVs
along the $D$--flat direction $<N_H^c>=<\overline{N}_H^c>$ the original
gauge symmetry of the rank--6 superstring inspired model with extra
$U(1)_{\psi}$ and $U(1)_{\chi}$ reduces to $SU(3)_C\times
SU(2)_W\times U(1)_Y\times U(1)_{N}$. In this case the effective mass
terms for the right--handed neutrinos are generated automatically if
the extra $\overline{27}_H$--plet couples to the ordinary matter
representations 
\be \delta
W=\ds\frac{\kappa_{ij}}{M_{Pl}}(\overline{27}_H\, 27_i)(\overline{27}_H\,
27_j)\qquad \Longrightarrow\qquad
M_{ij}=\ds\frac{2\kappa_{ij}}{M_{Pl}}<\overline{N}_H^c>^2\,.
\label{14}
\ee
To get a reasonable pattern for the left--handed neutrino masses and
mixing the $U(1)_{\psi}$ and $U(1)_{\chi}$ gauge symmetries should be broken
down to the $U(1)_{N}$ one around the Grand Unification or Planck scale.
A similar mechanism could be applied for the generation of the $\mu$
term discussed earlier. 
However it is rather difficult to use the
same fields $N_H^c$ and $\overline{N}_H^c$ in both cases because the
values of the corresponding mass parameters are too different. In
order to obtain $\mu$ in the TeV range one should assume the existence
of additional pair of $N_H^{c'}$ and $\overline{N}_H^{c'}$ which acquire 
VEVs  of order $10^{11}\,\mbox{GeV}$.

\subsection{Yukawa couplings in the ESSM}

The superpotential (\ref{13}) of the ESSM involves a
lot of new Yukawa couplings in comparison to the SM. But only large
Yukawa couplings are significant for the study of the renormalisation
group flow and spectrum of new particles which will be analysed in the
subsequent sections. The observed mass hierarchy of quarks and charged
leptons implies that most of the Yukawa couplings in the SM and MSSM
are small. Therefore it is natural to assume some hierarchical structure 
of the Yukawa interactions of new exotic particles with ordinary
quarks and leptons as well. As discussed earlier, without loss of generality 
we can assume that only the third family Higgs doublets and singlets
$S\equiv S_3$, $H_d\equiv H_{13}$, $H_u\equiv H_{23}$ gain VEVs, and furthermore
the third family Higgs sector couples most strongly with the third family
quarks and leptons. The third family SM singlet field $S$ will also couple to 
the exotic quarks $D_i$ and $\overline{D}_i$ and $SU(2)$ non--Higgs doublets
$H_{1\alpha}$ and $H_{2\alpha}$ ($\alpha=1,2$).

Discrete and extended gauge symmetries, which were specified before, 
do not guarantee the absence of FCNCs in the ESSM. Indeed the 
considered model contains many $SU(2)$ doublets and exotic quarks 
which interact with ordinary quarks and charged leptons of different 
generations. Therefore one may expect that even in the basis
of their mass eigenstates the non--diagonal flavour transitions are not
forbidden. For example, non--diagonal flavour interactions contribute
to the amplitude of $K^0-\overline{K}^0$ oscillations and give rise to
new channels of muon decay like $\mu\to e^{-}e^{+}e^{-}$.
To suppress flavour changing processes one
can postulate $Z^{H}_2$ symmetry. If all superfields except $H_u$,
$H_d$ and $S$ are odd under $Z^{H}_2$ symmetry transformations then
only one Higgs doublet $H_d$ interacts with the down--type quarks
and charged leptons and only one Higgs doublet $H_u$ couples to up--type
quarks while the couplings of all other exotic particles to the
ordinary quarks and leptons are forbidden. This eliminates any problems
related with the non--diagonal flavour transitions in the considered model.

The most general $Z^{H}_2$ and gauge invariant part of the ESSM superpotential
that describes the interactions of the SM singlet fields $S_i$ with 
exotic quarks, $SU(2)$ Higgs and non--Higgs doublets can be written as
\be
\ba{c}
\lambda_{ijk}S_i(H_{1j}H_{2k})+\kappa_{ijk}S_i(D_j\overline{D}_k)
\longrightarrow  \lambda_i S(H_{1i}H_{2i})+\kappa_i S(D_i\overline{D}_i)+\\[3mm]
+f_{\alpha\beta}S_{\alpha}(H_d H_{2\beta})+\tilde{f}_{\alpha\beta}S_{\alpha}(H_{1\beta}H_u)\,,\\[2mm]
\ea
\label{15}
\ee 
where $\alpha,\beta=1,2$ and $i=1,2,3$\,. In Eq.~(\ref{15}) we
choose the basis of non--Higgs and exotic quark superfields so that
the Yukawa couplings of the singlet field $S$ have flavour diagonal
structure. Here we define $\lambda \equiv \lambda_3$ and $\kappa
\equiv \kappa_3$
\footnote{Note that $\kappa$ as defined here in the ESSM refers to the
coupling of the singlet $S$ to the third family exotic quarks
$D\overline{D}$, and is not related to the $\kappa$ of the NMSSM which
refers to the cubic singlet coupling $S^3$ which is absent in the
ESSM.}.  If $\lambda$ or $\kappa$ are large at the Grand Unification
scale they affect the evolution of the soft scalar mass $m_S^2$ of the
singlet field $S$ rather strongly resulting in negative values of
$m_S^2$ at low energies that triggers the breakdown of $U(1)_{N}$
symmetry. The singlet VEV must be large enough to generate
sufficiently large masses for the exotic particles to avoid conflict
with direct particle searches at present and former accelerators. This
also implies that the Yukawa couplings $\lambda_i$ and $\kappa_i$
($i\neq 3$) involving the new exotic particles although small must be
large enough. The Yukawa couplings of other SM singlets
$f_{\alpha\beta}$ and $\tilde{f}_{\alpha\beta}$ are expected to be
considerably less than $\lambda_i$ and $\kappa_i$ to ensure that only
one singlet field $S$ gains a VEV.  At the same
time $f_{\alpha\beta}$ and $\tilde{f}_{\alpha\beta}$ cannot be
negligibly small because in this case the fermion components of
superfields $S_1$ and $S_2$ becomes extremely light \footnote{When
$f_{\alpha\beta}$ and $\tilde{f}_{\alpha\beta}$ vanish singlinos
$\tilde{S}_1$ and $\tilde{S}_2$ remain massless.}.  The induced masses
of singlinos $\tilde{S}_1$ and $\tilde{S}_2$ should be larger a few
MeV otherwise the extra states could contribute to the expansion rate
prior to nucleosynthesis changing nuclear abundances.

The $Z^{H}_2$ symmetry discussed above forbids all terms in $W_1$ and
$W_2$ that would allow the exotic quarks to decay. Therefore discrete
$Z^{H}_2$ symmetry can only be approximate. In our model we allow only
the third family $SU(2)$ doublets $H_d$ and $H_u$ to have Yukawa
couplings to the ordinary quarks and leptons of the order unity. As
discussed, this is a self-consistent assumption since the large Yukawa
couplings of the third generation (in particular, the top--quark Yukawa
coupling) provides a radiative mechanism for generating the Higgs VEVs
\cite{46} which defines the third family direction. As a consequence
the neutral components of $H_u$ and $H_d$ acquire non--zero VEVs 
inducing the masses of ordinary quarks and leptons. The
Yukawa couplings of two other pairs of $SU(2)$ doublets $H_{1i}$ and
$H_{2i}$ as well as $H'$ and exotic quarks to the quarks and leptons
of the third generation are supposed to be significantly smaller
($\lesssim 0.1$) so that none of the other exotic bosons gain a VEV. 
These couplings break $Z^{H}_2$ symmetry explicitly
resulting in flavour changing neutral currents.  In order to suppress
the contribution of new particles and interactions to the
$K^0-\overline{K}^0$ oscillations and to the muon decay channel
$\mu\to e^{-}e^{+}e^{-}$ in accordance with experimental
limits, it is necessary to assume that the Yukawa couplings of exotic
particles to the quarks of the first and second generations are less
than $10^{-4}$ and their couplings to the leptons of the first two
generations are smaller than $10^{-3}$.

\section{Renormalisation group analysis}
\subsection{The approximate superpotential to be studied}
Following the discussion about the natural choice of the parameters 
in our model given at the end of section 2, we can now specify the
superpotential couplings whose RG
flow will be analysed in this section. 
In our RG analysis we shall retain only Yukawa
couplings which appear on the right--hand side of Eq.~(\ref{15}),
together with the $O(1)$ Yukawa couplings to the quarks and leptons.
We shall neglect the neutrino Yukawa couplings 
as well as the small couplings involving the
first and second family singlets in our analysis.
Then the approximate superpotential studied is given by:
\be \ba{c}
W_{0}\approx \lambda S(H_{d} H_{u})+\lambda_1 S(H_{1,1} H_{2,1})+\lambda_2
S(H_{1,2} H_{2,2})+\kappa S(D_3\overline{D}_3)+\qquad\qquad\\[3mm]
\qquad\qquad+\kappa_1 S(D_1\overline{D}_1)+\kappa_2
S(D_2\overline{D}_2)+h_t(H_{u}Q)t^c+h_b(H_{d}Q)b^c+h_{\tau}(H_{d}L)\tau^c\,,
\ea
\label{16}
\ee 
where all ordinary quark and lepton superfields appeared in
Eq.~(\ref{16}) belong to the third generation, i.e. $L=L_3$, $Q=Q_3$,
$t^c=u^c_3$, $b^c=d^c_3$ and $\tau^c=e^c_3$. Here we adopt the
notation $\lambda=\lambda_3$ and $\kappa=\kappa_3$. The obtained 
superpotential possesses the approximate $Z^{H}_2$ symmetry specified
in the previous section which ensures the natural suppression of FCNCs. 
To guarantee that
only one pair of $SU(2)$ doublets $H_u$ and $H_d$ acquires a VEV
 we impose a certain hierarchy between the couplings of
$H_{1i}$ and $H_{2i}$ to the SM singlet superfield $S$:
$\lambda\gtrsim\lambda_{1,2}$. We assume further that the
superpotential (\ref{16}) is formed near the Grand Unification or Planck
scale. But in order to compute the masses and couplings at the
EW scale one has to determine the values of gauge and Yukawa
couplings at the EW scale. The evolution of all masses and
couplings from $M_X$ to $M_Z$ is described by a system of
RG equations.  In this section we study the
behaviour of the solutions to such equations describing the gauge and
Yukawa couplings in the
framework of the ESSM.

\subsection{The mixing of $U(1)_Y$ and $U(1)_N$}
In this subsection we address a mixing phenomenon related 
with the gauge sector of models containing two $U(1)$ gauge factors. 
In the Lagrangian of any gauge extension
of the SM containing an additional $U(1)'$ gauge group
there can appear a kinetic term consistent with all
symmetries which mixes the gauge fields of the $U(1)'$ and
$U(1)_Y$ \cite{461}. Our model is not an exception in this respect. 
In the basis in
which the interactions between gauge and matter fields have the
canonical form, i.e. for instance a covariant derivative $D_{\mu}$
which acts on the scalar and fermion components of the left--handed
quark superfield given by \be
D_{\mu}=\partial_{\mu}-ig_3A_{\mu}^aT^a-ig_2W_{\mu}^b\tau^b-ig_Y
Q^{Y}_iB^{Y}_{\mu}-ig_{N}Q^{N}_iB^{N}_{\mu}\,,
\label{17}
\ee
the pure gauge kinetic part of the Lagrangian can be written as
\be
\mathcal{L}_{kin}=-\ds\frac{1}{4}\left(F^Y_{\mu\nu}\right)^2-
\frac{1}{4}\left(F^{N}_{\mu\nu}\right)^2-\frac{\sin\chi}{2}F^{Y}_{\mu\nu}F^{N}_{\mu\nu}\,-
\ds\frac{1}{4}\left(G_{\mu\nu}\right)^2-\ds\frac{1}{4}\left(W_{\mu\nu}\right)^2\,.
\label{18}
\ee 
In Eqs.~(\ref{17})--(\ref{18}) $A_{\mu}^a$, $W_{\mu}^b$,
$B^{Y}_{\mu}$ and $B^{N}_{\mu}$ represent $SU(3)$, $SU(2)$, $U(1)_Y$
and $U(1)_{N}$ gauge fields, $G_{\mu\nu}^a$, $W_{\mu\nu}^b$,
$F_{\mu\nu}^Y$ and $F_{\mu\nu}^{N}$ are field strengths for the
corresponding gauge interactions, while $g_3$, $g_2$, $g_Y$ and
$g_{N}$ are $SU(3)$, $SU(2)$, $U(1)_Y$ and $U(1)_{N}$ gauge couplings
respectively.

Because $U(1)_Y$ and $U(1)_{N}$ arise from the breaking
of the simple gauge group $E_6$ the parameter $\sin\chi$ that parametrises
the gauge kinetic term mixing is equal to zero at tree--level. 
However it arises from loop effects since 
\be
Tr\left(Q^YQ^{N}\right)=\sum_{i=~{\rm{chiral~fields}}}\left(Q^Y_i
Q^{N}_i\right)\ne 0\,.
\label{19}
\ee 
Here the trace is restricted to the states lighter than the energy
scale being considered. The complete $E_6$ multiplets do not
contribute to this trace.  Its non--zero value is due to the incomplete
$27'+\overline{27'}$ multiplets
of the original $E_6$ symmetry from which only $H'$ and $\overline{H'}$
survive to low energy in order to ensure gauge coupling unification.

The mixing in the gauge kinetic part of the Lagrangian (\ref{18}) can
be easily eliminated by means of 
a non--unitary transformation of the two $U(1)$
gauge fields \cite{291}, \cite{462}--\cite{19}: 
\be
B^Y_{\mu}=B_{1\mu}-B_{2\mu}\tan\chi\,,\qquad
B^{N}_{\mu}=B_{2\mu}/\cos\chi\,.
\label{20}
\ee 
In terms of the new gauge variables $B_{1\mu}$ and $B_{2\mu}$ the
gauge kinetic part of the Lagrangian (\ref{18}) is now diagonal
and the covariant derivative (\ref{17}) becomes \cite{461} 
\be
D_{\mu}=\partial_{\mu}-ig_3A_{\mu}^aT^a-ig_2W_{\mu}^b\tau^b-ig_1
Q^{Y}_iB_{1\mu}-i(g'_1Q^{N}_i+g_{11}Q^Y_i)B_{2\mu}\,,
\label{21}
\ee 
where the redefined gauge coupling constants, written in terms of
the original ones, are 
\be g_1=g_Y\,, \qquad
g'_1=g_{N}/\cos\chi\,,\qquad g_{11}=-g_Y\tan\chi\,.
\label{22}
\ee 
In the new Lagrangian written in terms of the new gauge variables
$B_{1\mu}$ and $B_{2\mu}$ (defined in Eq.~(\ref{20})) 
the mixing effect is concealed in the
interaction between the $U(1)_{N}$ gauge field and matter
fields. The gauge coupling constant $g'_1$ is varied from the original one
and also a new off--diagonal gauge coupling $g_{11}$ appears. The
covariant derivative (\ref{21}) can be rewritten in a more compact form
\be
D_{\mu}=\partial_{\mu}-ig_3A_{\mu}^aT^a-ig_2W_{\mu}^b\tau^b-iQ^{T}G
B_{\mu}\,,
\label{23}
\ee
where $Q^T=(Q^Y_i,\,Q^{N}_i)$, $B^{T}_{\mu}=(B_{1\mu},\,B_{2\mu})$ and $G$ is 
a $2\times 2$ matrix of new gauge couplings (\ref{22})
\be
G=\left(
\ba{cc}
g_1 & g_{11}\\[2mm]
0   & g'_1
\ea
\right)\,.
\label{24}
\ee

Now all physical phenomena can be considered by using this new
Lagrangian with the modified structure of the extra $U(1)_{N}$ interaction
(\ref{21})--(\ref{23}). In the considered approximation the gauge
kinetic mixing changes effectively the $U(1)_{N}$ charges of the
fields to 
\be
\tilde{Q}_i\equiv Q^{N}_i+Q^{Y}_i\delta ,
\label{effective}
\ee
where
$\delta=g_{11}/g'_1$ while the $U(1)_Y$ charges remain the same. As
the gauge coupling constants are scale dependent, the effective
$U(1)_{N}$ charges defined here as $\tilde{Q}_i$
are scale dependent as well. The particle spectrum now
depends on the effective $U(1)_{N}$ charges $\tilde{Q}_i$. 

In Eq.~(\ref{effective})
the correct $E_6$ normalisation of the charges should be
used, and thus the
$U(1)_Y$ hypercharges in Eqs.~(\ref{5})--(\ref{6}) should be
multiplied by a factor $\ds\sqrt{\frac{3}{5}}$,
and the $Q^{N}$ charges in Eqs.~(\ref{5})--(\ref{6}) 
should be multiplied by $\ds1/\sqrt{{40}}$.
The correctly normalised
charges of all the matter fields in the ESSM are summarised
in Table~\ref{charges}. The charges are family independent,
and the index $i$ here refers to the different multiplets
as well as the different families.

\begin{table}[ht]
  \centering
  \begin{tabular}{|c|c|c|c|c|c|c|c|c|c|c|c|c|c|}
    \hline
 & $Q$ & $u^c$ & $d^c$ & $L$ & $e^c$ & $N^c$ & $S$ & $H_2$ & $H_1$ & $D$ &
 $\overline{D}$ & $H'$ & $\overline{H'}$ \\
 \hline
$\sqrt{\frac{5}{3}}Q^{Y}_i$  
 & $\frac{1}{6}$ & $-\frac{2}{3}$ & $\frac{1}{3}$ & $-\frac{1}{2}$ 
& $1$ & $0$ & $0$ & $\frac{1}{2}$ & $-\frac{1}{2}$ & $-\frac{1}{3}$ &
 $\frac{1}{3}$ & $-\frac{1}{2}$ & $\frac{1}{2}$ \\
 \hline
$\sqrt{{40}}Q^{N}_i$  
 & $1$ & $1$ & $2$ & $2$ & $1$ & $0$ & $5$ & $-2$ & $-3$ & $-2$ &
 $-3$ & $2$ & $-2$ \\
 \hline
  \end{tabular}
  \caption{The $U(1)_Y$ and $U(1)_{N}$ charges of matter fields in the
    ESSM, where $Q^{N}_i$ and $Q^{Y}_i$ are here defined with the correct
$E_6$ normalisation factor required for the RG analysis.}
  \label{charges}
\end{table}

\subsection{The running of the gauge couplings}
At the one--loop level the full set of RG equations
describing the running of gauge and Yukawa couplings splits into two
parts.  One of them includes RG equations for the gauge couplings. In
the one--loop approximation $\beta$ functions of the gauge couplings
do not depend on the Yukawa ones. Therefore this part of the system of
RG equations can be analysed separately,
and is discussed in this subsection.

The RG flow of the Abelian gauge couplings is affected
by the kinetic term mixing as discussed in the previous subsection. 
Using the matrix notation for the
structure of $U(1)$ interactions with $G$ defined 
in Eq.~(\ref{24}) one can write down the RG
equations for the Abelian couplings in a compact form \cite{291},
\cite{462}--\cite{19} \be \ds\frac{d G}{d t}=G\times B\,,
\label{27}
\ee
where $B$ is a $2\times 2$ matrix of $\beta$ functions given by
\be
B=\left(
\ba{cc}
B_1 & B_{11} \\[2mm]
0   & B'_1
\ea
\right)=\ds\frac{1}{(4\pi)^2}
\left(
\ba{cc}
\beta_1 g_1^2 & 2g_1g'_1\beta_{11}+2g_1g_{11}\beta_1\\[2mm]
0 & g^{'2}_1\beta'_1+2g'_1 g_{11}\beta_{11}+g_{11}^2\beta_1
\ea
\right)\,.
\label{28}
\ee 
In the ESSM with $N_g=3$ the one--loop $\beta$ functions $\beta_1$,
$\beta'_1$ and $\beta_{11}$ are 
\be
\beta_1=\sum_{i}(Q^Y_i)^2=\ds\frac{48}{5}\,,\qquad
\beta'_1=\sum_i(Q^{N}_i)^2=\ds\frac{47}{5}\,, \qquad \beta_{11}=\sum_i
Q^Y_i Q^{N}_i=\mp\ds\frac{\sqrt{6}}{5}\,.
\label{29}
\ee 
The index $i$ is summed over all possible chiral
superfields and all families. 
Note that 
$\beta_1 \approx \beta'_1\gg \beta_{11}$. This implies that the
effect of $U(1)$ gauge mixing is ultimately rather small,
and furthemore that, if the (properly normalised) 
$g_Y$ and $g_N$ start out equal at the GUT scale,
then they will remain approximately equal at low energies.

The running of $SU(2)$ and $SU(3)$ couplings obey the RG equations of the
standard form: \be \frac{d g_2}{dt}=\ds\frac{\beta_2
g_2^3}{(4\pi)^2}\,,\qquad\qquad \frac{d g_3}{dt}=\frac{\beta_3
g_3^3}{(4\pi)^2}\,,
\label{25}
\ee
with $\beta$ functions
\be
\beta_2=-5+3N_g\,,\qquad\qquad \beta_3=-9+3N_g\,,
\label{26}
\ee 
where $t=\ln\left(\mu /M_X\right)$ and $\mu$ is the RG scale.
The parameter $N_g$ appeared in the
expressions for $\beta_2$ and $\beta_3$ is the number of generations
forming $E_6$ fundamental representations which the considered SUSY model
involves at low energies. As one can easily see from Eq.~(\ref{26})
$N_g=3$ is the critical value for the one--loop $\beta$ function of
strong interactions. Since by construction three complete 27--plets
survive to low energies in the ESSM $\beta_3$ is equal to zero in our
case and $SU(3)$ gauge coupling remains constant everywhere from $M_Z$
to $M_X$. Because complete 27--plets do not violate $E_6$ symmetry
each generation should give the same contribution to all $\beta$
functions. It takes place automatically in the case of $SU(2)$ and
$SU(3)$ $\beta$ functions and allows to obtain a correct normalisation
for the charges of two $U(1)'\mbox{s}$.

The RG equations for the gauge couplings
in Eqs.~(\ref{27}) and (\ref{25}) should be supplemented by the boundary
conditions.  Since we deal with an $E_6$ inspired model it seems to
be natural to assume that at high energies $E_6$ symmetry is restored
and all gauge interactions are characterised by a unique $E_6$ gauge
coupling $g_0$ which is defined as \be
g_3(M_X)=g_2(M_X)=g_1(M_X)=g'_1(M_X)=g_0\,.
\label{30}
\ee 
Also we expect that there is no mixing in the gauge kinetic
part of the Lagrangian just after the breakdown of the $E_6$ symmetry,
i.e.  
\be g_{11}(M_X)=0\,.
\label{31}
\ee

The hypothesis of gauge coupling unification (\ref{30}) permits to
evaluate the overall gauge coupling $g_0$ and the Grand Unification scale
$M_X$ using the values of $g_1(M_Z)$, $g_2(M_Z)$ and $g_3(M_Z)$ which
are fixed by LEP measurements and other experimental data \cite{47}.
The high energy scale where the unification of the gauge couplings
takes place is almost insensitive to the matter content of the
supersymmetric model. Indeed, in the one--loop approximation we have
\be
\ds\frac{1}{(4\pi)^2}\ln\frac{M_X^2}{M_Z^2}=\frac{1}{\beta_1-\beta_2}\bigg[\frac{1}{g_1^2(M_Z^2)}-\frac{1}{g_2^2(M_Z^2)}
\biggl]\,.
\label{32}
\ee 
Because the dependence of the scale $M_X$, where $U(1)_Y$ and
$SU(2)$ gauge couplings meet, on the particle content of any model
comes from the difference of the corresponding $\beta$ functions, in
which the contribution of any complete $SU(5)$ multiplets is
cancelled, the Grand Unification scale in the ESSM remains the same as
in the MSSM, i.e in the one--loop approximation $M_X\simeq 2\cdot
10^{16}\,\mbox{GeV}$. At the same time the value of the overall gauge
coupling is rather sensitive to the matter content of SUSY models.
In the ESSM the appropriate values of the $SU(2)$, $SU(3)$ and
$U(1)_Y$ gauge couplings at the EW scale can be reproduced
for $g_0\simeq 1.21$ which differs from the value of $g_0\simeq 0.72$
found in the minimal SUSY model. The growth of $g_0$ in our model is
caused by the extra exotic supermultiplets of matter.

The interesting point concerning the matter content in our model is
that $\beta_1$, $\beta'_1$, $\beta_2$ and $\beta_3$ are quite close to
their saturation limits when the gauge couplings blow up at the Grand
Unification scale. The ESSM allows to accommodate only one additional
pair of $5+\overline{5}$ representations of the usual $SU(5)$ which
form extra exotic quark and non--Higgs multiplets. Further enlargement
of the particle content leads to the appearance of the Landau pole during
the evolution of the gauge couplings from $M_Z$ to $M_X$.

Using the boundary conditions (\ref{30}) and (\ref{31}) as well as the
obtained values of $g_0$ and $M_X$ it is possible to solve the
RG equations for $g'_1$ and $g_{11}$. It turns out
that $g'_1(Q)$ is very close to $g_1(Q)$ for any value of
renormalisation scale $Q$ from $M_X$ to $M_Z$ while $g_{11}(Q)$ is
negligibly small compared to all other gauge couplings. At the
EW scale we get 
\be 
\ds\frac{g_1(M_Z)}{g'_1(M_Z)}\simeq
0.99\,,\qquad\qquad g_{11}(M_Z)\simeq 0.020\,,
\qquad\qquad g_{1}(M_Z)\simeq 0.46\,.
\label{33}
\ee
Eq.~(\ref{33}) tells us that if the (properly normalised) 
$g_Y$ and $g_N$ couplings start out equal at the GUT scale,
then they will remain approximately equal at low energies
to within an accuracy of two per cent at the one--loop level.
As previously noted, this results from
$\beta_1 \approx \beta'_1\gg \beta_{11}$
which implies that the
effect of $U(1)$ gauge mixing is small.
In the following analysis we shall continue to include the
effects of $U(1)$ gauge mixing in the correct way.      However,
it should be noted that to excellent approximation we could
take $g_1=g'_1=g_Y=g_N$ and $\tilde{Q}_i=Q^N_i$, which is 
within the accuracy of the one--loop result.

\subsection{The running of the Yukawa couplings}
The running of the Yukawa couplings appearing in the superpotential
in Eq.~(\ref{16}) obey the following system of differential 
equations\footnote{See also \cite{29}--\cite{291}.}:
\be
\ba{rcl}
\ds\frac{dh_t}{dt}&=&\ds\frac{h_t}{(4\pi)^2}\biggl[\lambda^2+6h_t^2+h_b^2-\ds\frac{16}{3}g_3^2-3g_2^2-\ds\frac{13}{15}g_1^2-
2\left(\tilde{Q}_2^2+\tilde{Q}_{Q}^2+\tilde{Q}_{u}^2\right)g^{'2}_1 \biggr]\,,\\[3mm]
\ds\frac{dh_b}{dt}&=&\ds\frac{h_b}{(4\pi)^2}\biggl[\lambda^2+h_t^2+6h_b^2+h_{\tau}^2-\ds\frac{16}{3}g_3^2-3g_2^2-\ds\frac{7}{15}g_1^2-
2\left(\tilde{Q}_1^2+\tilde{Q}_{Q}^2+\tilde{Q}_{d}^2\right)g^{'2}_1\biggr]\,,\\
[3mm]
\ds\frac{dh_{\tau}}{dt}&=&\ds\frac{h_{\tau}}{(4\pi)^2}\biggl[\lambda^2+3h_b^2+4h_{\tau}^2-3g_2^2-\frac{9}{5}g_1^2-
2\left(\tilde{Q}_1^2+\tilde{Q}_L^2+\tilde{Q}_e^2\right)g^{'2}_1\biggr]\,,\qquad\qquad\\[3mm]
\ds\frac{d\lambda_i}{dt}&=&\ds\frac{\lambda_i}{(4\pi)^2}\biggl[2\lambda_i^2+2\Sigma_{\lambda}+3\Sigma_{\kappa}+
(3h_t^2+3h_b^2+h_{\tau}^2)\delta_{i3}-\\[3mm]
&&\ds-3g_2^2-\frac{3}{5}g_1^2-2\left(\tilde{Q}_S^2+\tilde{Q}_2^2+\tilde{Q}_1^2\right)g^{'2}_1\biggr]\,,\\[3mm]
\ds\frac{d\kappa_i}{dt}&=&\ds\frac{\kappa_i}{(4\pi)^2}\biggl[2\kappa^2_i+2\Sigma_{\lambda}+3\Sigma_{\kappa}-\ds\frac{16}{3}g_3^2-
\frac{4}{15}g_1^2-2\left(\tilde{Q}_S^2+\tilde{Q}_{D}^2+\tilde{Q}_{\overline{D}}^2\right)g^{'2}_1
\biggr]\,,
\ea
\label{34}
\ee
where 
$\Sigma_{\lambda}=\lambda_1^2+\lambda_2^2+\lambda_3^2$ 
and $\Sigma_{\kappa}=\kappa_1^2+\kappa_2^2+\kappa_3^2$
and where the index $i=1,2,3$.

The couplings $h_t$, $h_b$ and $h_{\tau}$ in Eq.~(\ref{34}) determine
the running masses of the fermions of the third generation at the
EW scale 
\be m_t(M_t)=\ds\frac{h_t(M_t)
v}{\sqrt{2}}\sin\beta, \ \ m_b(M_t)=\ds\frac{h_b(M_t)
v}{\sqrt{2}}\cos\beta, \ \ m_{\tau}(M_t)=\ds\frac{h_{\tau}(M_t)
v}{\sqrt{2}}\cos\beta\,,
\label{35}
\ee 
which are generated after EWSB. In
Eq.~(\ref{35}) $m_{\tau}$, $m_t$ and $m_b$ are the running masses
of the $\tau$--lepton, top-- and bottom--quark respectively, $M_t$ is a top
quark pole mass, $v=\sqrt{v_1^2+v_2^2}=246\,\mbox{GeV}$, while
$\tan\beta=v_2/v_1$, where $v_2$ and $v_1$ are the VEVs
 of the Higgs doublets $H_u$ and $H_d$.  Since the running
masses of the fermions of the third generation are known,
Eq.~(\ref{35}) can be used to derive the Yukawa couplings $h_\tau(M_t)$,
$h_t(M_t)$ and $h_{b}(M_t)$ for each particular value of
$\tan\beta$ establishing boundary conditions for the renormalisation
group equations (\ref{34}). In this paper we restrict our analysis to
moderate values of $\tan\beta\ll m_t(M_t)/m_b(M_t)$ for which
$b$--quark and $\tau$--lepton Yukawa couplings are much smaller than $h_t$
and thus can be safely neglected.

The boundary conditions for the Yukawa couplings of the SM singlet
field $S$ to the Higgs doublets and exotic particles are unknown.
These couplings give rise to the masses of the exotic quarks and
non--Higgsinos after the breakdown of gauge symmetry. Since none of the
exotic particles or Higgs bosons have been found yet $\lambda_i$ and
$\kappa_i$ should be considered as free parameters in our model.
There are two different assumptions regarding these couplings that
look rather natural and allow one to reduce the number of new parameters.
One of them implies that the masses of the exotic particles mimic the
hierarchy observed in the sector of ordinary quarks and charged
leptons. Then non--observation of new exotic states may be related
with the considerable hierarchy of VEVs of the
singlet field $S$ and Higgs doublets. In this case $\lambda_1$,
$\lambda_2$, $\kappa_1$ and $\kappa_2$ are tiny and can be simply
ignored.

Although the suggested pattern is quite simple and natural it does not permit to tell anything about the masses
of the exotic particles of the first two generations because the corresponding Yukawa couplings are set to zero from the beginning.
In the meantime one has to ensure that exotic fermions gain large enough masses to avoid any conflict with direct new particle searches
at present and former colliders. Therefore it is worth to incorporate all Yukawa couplings of exotic particles to the SM singlet field $S$
in our analysis of RG flow. Because the Yukawa interactions of extra coloured singlets $D_i$ and $\overline{D}_i$
in the superpotential (\ref{16}) and as a consequence the corresponding RG equations are symmetric with respect to the generation index
$i$ exchange, i.e. $1\leftrightarrow 2$, $2\leftrightarrow 3$ and $3\leftrightarrow 1$, there is a solution of the RG equations
when all $\kappa_i$ are equal to each other. Moreover the solutions for $\kappa_i(Q)$ tend to converge to each other during the evolution
of these couplings from the Grand Unification to EW scale. Thus the choice $\kappa_1(M_t)=\kappa_2(M_t)=\kappa(M_t)$ is
well motivated by the RG flow.

Similar results can be obtained for $\lambda_1$ and $\lambda_2$. However the running of $\lambda_1(Q)=\lambda_2(Q)$ differs from the evolution
of $\lambda(Q)$ because the top--quark and its superpartners give significant contributions to the renormalisation of $\lambda(Q)$. Nevertheless the
difference between the RG flow of $\lambda_1(Q)=\lambda_2(Q)$ and $\lambda(Q)$ is not so appreciable in comparison with the
running of $\kappa_i(Q)$ and $h_t(Q)$. The couplings of the Yukawa interactions involving quark superfields renormalise by virtue of the strong
interactions that push their values up considerably at low energies. Therefore it seems to be reasonable to ignore the difference between the evolution
of $\lambda_i(Q)$ in first approximation and consider separately the limit in which $\lambda_1(M_t)=\lambda_2(M_t)=\lambda(M_t)$.

At first we consider the limit when the Yukawa couplings of the exotic quarks and non--Higgses of the first two generation are negligibly
small. Then the system of the RG equations can be rewritten in the suggestive form
\be
\ba{rcl}
8\pi^2 \ds\frac{d}{dt}\left[\frac{\lambda^2}{h_t^2}\right]&=&\biggl(3\lambda^2+3\kappa^2-3h_t^2+\ds\frac{16}{3}g_3^2+\frac{4}{15}g_1^2+\\[3mm]
&&+2(\tilde{Q}_{Q}^2+\tilde{Q}_{u}^2-\tilde{Q}_S^2-\tilde{Q}_1^2)\biggr)\left[\ds\frac{\lambda^2}{h_t^2}\right]\,,\\[3mm]
8\pi^2 \ds\frac{d}{dt}\left[\frac{\kappa^2}{h_t^2}\right]&=&\biggl(5\kappa^2+\lambda^2-6h_t^2+3g_2^2+\ds\frac{3}{5}g_1^2+\\[3mm]
&&+2(\tilde{Q}_2^2+\tilde{Q}_{Q}^2+\tilde{Q}_{u}^2-\tilde{Q}_S^2-\tilde{Q}_{D}^2-\tilde{Q}_{\overline{D}}^2)\biggr)
\left[\ds\frac{\kappa^2}{h_t^2}\right]\,.
\ea
\label{36}
\ee
If we ignore gauge couplings the system of differential equations (\ref{36}) has two fixed points
\be
\ba{rl}
(I)\qquad&\ds\frac{\lambda^2}{h_t^2}=1\,,\qquad\qquad \ds\frac{\kappa^2}{h_t^2}=0\,;\\[5mm]
(II)\qquad&\ds\frac{\lambda^2}{h_t^2}=0\,,\qquad\qquad \ds\frac{\kappa^2}{h_t^2}=\frac{6}{5}\,.
\ea
\label{37}
\ee 
Of these fixed points only the last one is infrared stable.  The
presence of the infrared stable fixed point $(II)$ means that if we
start with random boundary conditions for the couplings $\lambda$,
$\kappa$ and $h_t$ in the gaugeless ($g_0=0$) limit
the solutions of the RG equations (\ref{36}) tend
to converge towards the values which respect the ratios $(II)$. This
is illustrated in Fig.~1a where the running $\lambda/h_t$ versus
$\kappa/h_t$ for $h_t(M_X)=10$, $g_0=0$ and regular distribution of
boundary conditions for $\lambda(M_X)$ and $\kappa(M_X)$ at the Grand
Unification scale is shown. A point in the plane $\lambda/h_t -
\kappa/h_t$ will flow rapidly towards the valley, that corresponds to
the invariant line which connects fixed points $(I)$ and $(II)$, and
then more slowly along it to the stable fixed point $(II)$. The
properties of invariant lines and surfaces were reviewed in detail in
\cite{48}.

So far we have neglected the effects of gauge couplings. However their role is quite important especially at low energies where the gauge
coupling of the strong interactions is larger than the Yukawa ones. As one can see from Fig.~1b the inclusion of gauge couplings
spoils the valley along which the solutions of RG equations flow to the fixed points $(II)$. But the convergence of
$h_t(Q)$, $\lambda(Q)$ and $\kappa(Q)$ to the quasi--fixed point, which becomes more close to unity, increases.

Similar analysis can be performed in the case when all Yukawa
couplings of exotic particles have non--zero values.  As before in the
gaugeless limit there is only one stable fixed point of the
RG equations (\ref{34}) which corresponds to \be
\frac{\lambda^2}{h_t^2}=0\,,\qquad\qquad
\ds\frac{\kappa^2}{h_t^2}=\frac{\kappa^2_{1,\,2}}{h_t^2}=\frac{\lambda^2_{1,\,2}}{h_t^2}
=\frac{2}{5}\,.
\label{38}
\ee 
Turning the gauge couplings on induces a certain hierarchy between
the Yukawa couplings of exotic quarks and non--Higgses. Due to the growth
of the gauge coupling of strong interactions at low energies the
Yukawa couplings of the top-- and exotic quarks tend to dominate over
$\lambda_i$. It shifts the position of the quasi--fixed point where the
solutions of the RG equations are focused 
\be
\frac{\lambda^2}{h_t^2}\to 0\,,\qquad
\frac{\lambda^2_{1,\,2}}{h_t^2}\simeq 0.26\,,\qquad
\ds\frac{\kappa^2}{h_t^2}=\frac{\kappa^2_{1,\,2}}{h_t^2}\simeq 0.66\,.
\label{39}
\ee

In spite of their attractiveness fixed points cannot provide a
complete description of the RG flow of Yukawa
couplings.  Indeed, the strength of the attraction of the solutions to the
RG equations (\ref{37}) towards the invariant line
and fixed point is governed by $h_t(M_X)$. The larger $h_t(M_X)$
the faster the stable fixed point is reached. However at small values
of $h_t(M_X)=0.2-0.4$ the convergence of $h_t(Q)$, $\lambda(Q)$ and
$\kappa(Q)$ to the fixed points is extremely weak. Therefore even at
moderate values of $\tan\beta=1.5-3$ the values of the Yukawa
couplings at the EW scale may be far away from the stable
fixed point.

In this context it is worth to study the limits on the values of the
Yukawa couplings imposed by the perturbative RG
flow.  The growth of Yukawa couplings at the EW scale entails
the increase of their values at the Grand Unification scale
resulting in the appearance of the Landau pole. Large values of the
Yukawa couplings spoil the applicability of perturbation theory at
high energies so that the one--loop RG equations
cannot be used for an adequate description of the evolution of gauge and
Yukawa couplings at high scales $Q\sim M_X$. The requirement of
validity of perturbation theory up to the Grand Unification scale
restricts the interval of variations of Yukawa couplings at the
EW scale. In the simplest case when
$\kappa_1=\kappa_2=\lambda_1=\lambda_2=0$ the assumption that
perturbative physics continues up to the scale $M_X$ sets an upper
limit on the low energy value of $\kappa(M_t)$, evaluated at the
top mass $M_t$, for each fixed set of
$\lambda(M_t)$ and $h_t(M_t)$ (or $\tan\beta$). With decreasing
(increasing) $\lambda(M_t)$ the maximal possible value of
$\kappa(M_t)$, which is consistent with perturbative gauge coupling
unification, increases (decreases) forming a demarcating line that
restricts the allowed range of the parameter space in the $\kappa/h_t
- \lambda/h_t$ plane for each particular value of $\tan\beta$.

For $\tan\beta=2$ the corresponding limits on the 
low energy Yukawa couplings evaluated at $M_t$ are
shown in Fig.~2a. Outside the permitted region the solutions of the
RG equations blow up before the Grand Unification
scale and perturbation theory is not valid at high energies.  The
allowed range for the Yukawa couplings varies when $\tan\beta$
changes. When $\tan\beta$ tends to its lower bound caused by the
applicability of perturbation theory, which is around unity in our
model, the permitted part of the parameter space narrows in the
direction of $\lambda/h_t$ so that only a small interval of variations
of $\lambda/h_t$ is allowed. At large $\tan\beta$ the allowed range for
the Yukawa couplings enlarges. The typical pattern of the
RG flow of the ratios of Yukawa couplings from the
permitted region of the parameter space is presented in Fig.~2b. It
differs significantly from the analogous pattern obtained in the
top--bottom approach (see Fig.~1) because the chosen value of top--quark
Yukawa coupling, which corresponds to $\tan\beta=2$, is quite far from
the quasi--fixed point.  The peculiar feature of both patterns is that
the ratio of $\lambda/h_t$ changes during the evolution more strongly than
$\kappa/h_t$. Owing to this most trajectories in Fig.~2b are
almost parallel to the axis $\lambda/h_t$.

A similar pattern for the RG flow can be found for
most values of $\tan\beta$ and does not change much after the inclusion
of the Yukawa couplings of exotic particles of the first two
generations. But the introduction of $\kappa_1$, $\kappa_2$, $\lambda_1$
and $\lambda_2$ makes more rigorous the restrictions on the Yukawa
couplings of the third generation. In Fig.~2c we plot the allowed
range of the Yukawa couplings in the case when $\lambda_1$ and
$\lambda_2$ are still negligibly small while
$\kappa=\kappa_1=\kappa_2$.  It is easy to see that in this case the
upper bounds on $\kappa/h_t$ are more stringent than in
Fig.~2a. Turning $\lambda_1$ and $\lambda_2$ on so that
$\lambda(M_t)=\lambda_1(M_t)=\lambda_2(M_t)$ reduces the limit on
$\lambda/h_t$ (see Fig.~2d). The narrowing of the allowed range of the
parameter space is not an unexpected effect. New Yukawa couplings appear
in the right--hand side of the differential equations (\ref{34}) with
positive sign. As a consequence they increase the growth of the Yukawa
couplings of the third generation and perturbation theory becomes
inapplicable for lower values of $\lambda(M_t)$ and $\kappa(M_t)$.

Because the allowed range of the Yukawa couplings always shrinks when
additional Yukawa couplings are introduced one can find an absolute
upper limit on the value of $\lambda(M_t)$ as
function of $h_t(M_t)$ or $\tan\beta$ by setting all other Yukawa
couplings to zero. The dependence of this upper limit
$\lambda_{\max}$ on $\tan\beta$ is shown in Fig.~3. 
The upper bound on $\lambda(M_t)$ grows with
increasing $\tan\beta$ because the top--quark Yukawa coupling
decreases. The value of $\lambda_{\max}$ vanishes at $\tan\beta\simeq
1$ when the top--quark Yukawa coupling attains a fixed point (see also
Fig.~1b). At large $\tan\beta$ the upper bound on $\lambda(M_t)$
approaches the saturation limit where $\lambda_{\max}\simeq 0.84$. The
restrictions on $\lambda(M_t)$ and other Yukawa couplings obtained in
this section are extremely useful for the analysis of the Higgs particle
spectrum which we are going to consider next.

\section{Electroweak symmetry breaking and Higgs sector}
\subsection{The Higgs potential and its minimisation}
The sector responsible for EWSB in the ESSM
includes two Higgs doublets $H_u$ and $H_d$ as well as the SM singlet
field $S$. The interactions between them are defined by the structure of
the gauge interactions and by the superpotential in Eq.~(\ref{16}).
Including soft SUSY breaking terms, and radiative corrections,
the resulting Higgs effective potential is the sum of four pieces: 
\be \ba{rcl}
V&=&V_F+V_D+V_{soft}+\Delta V\, ,\\[3mm]
V_F&=&\lambda^2|S|^2(|H_d|^2+|H_u|^2)+\lambda^2|(H_d H_u)|^2\,,\\[3mm]
V_D&=&\ds\frac{g_2^2}{8}\left(H_d^\dagger \sigma_a H_d+H_u^\dagger \sigma_a
H_u\right)^2+\frac{{g'}^2}{8}\left(|H_d|^2-|H_u|^2\right)^2+\\[3mm]
&&+\ds\frac{g^{'2}_1}{2}\left(\tilde{Q}_1|H_d|^2+\tilde{Q}_2|H_u|^2+\tilde{Q}_S|S|^2\right)^2\,,\\[3mm]
V_{soft}&=&m_{S}^2|S|^2+m_1^2|H_d|^2+m_2^2|H_u|^2+\biggl[\lambda
A_{\lambda}S(H_u H_d)+h.c.\biggr]\,, \ea
\label{40}
\ee 
where $g'=\sqrt{3/5} g_1$ is the low energy (non-GUT
normalised) gauge coupling and $\tilde{Q}_1$,
$\tilde{Q}_2$ and $\tilde{Q}_S$ are effective $U(1)_{N}$ charges of
$H_d$, $H_u$ and $S$ respectively.  Here $H_d^T=(H_d^0,\,H_d^{-})$,
$H_u^T=(H_u^{+},\,H_u^{0})$ and $(H_d
H_u)=H_u^{+}H_d^{-}-H_u^{0}H_d^{0}$.  At tree--level the Higgs
potential in Eq.~(\ref{40}) is described by the sum of the first three
terms. The structure of the $F$--terms $V_F$ is exactly the same as in
the NMSSM without the self--interaction of the singlet superfield. 
However the $D$--terms in $V_D$ contain a new ingredient: the
terms in the expression for $V_D$ proportional to ${g'_1}^2$
represent $D$--term contributions due to the
extra $U(1)_{N}$ which are not present in the MSSM or NMSSM.
The soft SUSY breaking terms are collected
in $V_{soft}$.

The term $\Delta V$ represents the contribution
of loop corrections to the Higgs effective potential. In the MSSM
the dominant contribution to $\Delta V$ comes from the
loops involving the top--quark and its superpartners because of their
large Yukawa coupling. However the ESSM contains
many new exotic supermultiplets and the RG analysis 
described in the previous section revealed that the Yukawa
couplings of the exotic $D$--quarks to the SM singlet field $S$
can be large at the EW scale. 
Therefore the contribution of $D$--quarks
and their superpartners to $\Delta V$ can be enhanced as well. Keeping
only leading one--loop corrections to the Higgs effective potential
in Eq.~(\ref{40}) from the top-- and exotic quarks and their superpartners we
find 
\be 
\ba{c} 
\Delta
V=\ds\frac{3}{32\pi^2}\left[m_{\tilde{t}_1}^4\left(
\ln\frac{m_{\tilde{t}_1}^2}{Q^2}-\frac{3}{2}\right)+
m_{\tilde{t}_2}^4\left(\ln\frac{m_{\tilde{t}_2}^2}{Q^2}-\frac{3}{2}\right)
-2m_t^4\left(\ln\frac{m_t^2}{Q^2}-\frac{3}{2}\right)\right.+\qquad\qquad
\label{401b}
\ea
\ee
$$
\ba{c}
+\left.\sum_{i=1,2,3}\left\{\ds m_{\tilde{D}_{1,i}}^4\left(
\ln\frac{m_{\tilde{D}_{1,i}}^2}{Q^2}-\frac{3}{2}\right)+
m_{\tilde{D}_{2,i}}^4\left(\ln\frac{m_{\tilde{D}_{2,i}}^2}{Q^2}-\frac{3}{2}\right)
-2\mu_{Di}^4\left(\ln\frac{\mu_{Di}^2}{Q^2}-\frac{3}{2}\right)\right\}
\right],
\ea
$$ 
where $\mu_{Di}=\kappa_i<S>=\ds\frac{\kappa_i s}{\sqrt{2}}$ are
masses of exotic quarks, while $m_{\tilde{t}_1}$, $m_{\tilde{t}_2}$,
$m_{\tilde{D}_{1,i}}$ and $m_{\tilde{D}_{2,i}}$ are the masses of the
superpartners of the top and $D$--quarks which are given by \be \ba{c}
m^2_{\tilde{t}_1,\tilde{t}_2}=\ds\frac{1}{2}\biggl[m^2_Q+m^2_U+2+2m_t^2\pm\sqrt{(m_Q^2-m_U^2)^2+
4 m_t^2\biggl(A_t-\ds\frac{\lambda s
}{\sqrt{2}\tan\beta}\biggr)^2}\biggr]\,,\\[3mm]
m^2_{\tilde{D}_{1,i},\tilde{D}_{2,i}}=\ds\frac{1}{2}\biggl[m^2_{Di}+m^2_{\overline{D}i}+2\mu_{Di}^2\pm
\sqrt{(m_{Di}^2-m_{\overline{D}i}^2)^2+4\biggl(A_{\kappa
i}\mu_{Di}-\ds\frac{\kappa_i\lambda}{2}v_1v_2\biggr)^2} \biggr]\,.
\ea
\label{402}
\ee

The couplings $g_2,\,g',\,g'_1$ and $\lambda$ in the scalar potential
(\ref{40}) do not violate SUSY. Moreover the gauge couplings
$g_2$ and $g'$ are well known \cite{47}. The value of the extra $U(1)_{N}$
coupling $g'_1$ and the effective $U(1)_{N}$ charges of $H_d$, $H_u$
and $S$ can be determined assuming gauge coupling unification (see
Eq.~(\ref{33})). The Yukawa coupling $\lambda$ cannot be fixed as
directly as the gauge couplings. But as we discussed in the previous
section the requirement of validity of perturbation theory up to
the GUT scale leads to an upper bound $\lambda \leq \lambda_{\max}$.

A set of soft SUSY breaking parameters in the tree--level Higgs boson
potential includes the soft masses $m_1^2,\,m_2^2,\,m_{S}^2$ and the trilinear
coupling $A_{\lambda}$. The part of the scalar potential (\ref{40})
which contains soft SUSY breaking terms $V_{soft}$ coincides with the
corresponding one in the NMSSM when the 
NMSSM parameters $\kappa$ and $A_{\kappa}$
vanish. Since the only complex phase (of $\lambda A_{\lambda}$) that
appears in the tree--level scalar potential (\ref{40}) can easily be
absorbed by a suitable redefinition of the Higgs fields,
CP--invariance is preserved in the Higgs sector of the considered
model at tree--level. The inclusion of loop corrections draws into
the analysis many other soft SUSY breaking parameters which define
masses of different superparticles. Some of these parameters can be
complex creating potential sources of CP--violation.

At the physical minimum of the scalar potential (\ref{40}) the Higgs
fields develop VEVs
\be
<H_d>=\ds\frac{1}{\sqrt{2}}\left(
\begin{array}{c}
v_1\\ 0
\end{array}
\right) , \qquad
<H_u>=\ds\frac{1}{\sqrt{2}}\left(
\begin{array}{c}
0\\ v_2
\end{array}
\right) ,\qquad
<S>=\ds\frac{s}{\sqrt{2}}.
\label{41}
\ee 
The vacuum configuration (\ref{41}) is not the most general
one. Because of the $SU(2)$ invariance of the Higgs potential
(\ref{40}) one can always make $<H_u^{+}>=0$ by virtue of a suitable
gauge rotation. Then the requirement $<H_d^{-}>=0$, which is a
necessary condition to preserve $U(1)_{em}$ associated with
electromagnetism in the physical vacuum, is equivalent to requiring
the squared mass of the physical charged scalar to be positive. It
imposes additional constraints on the parameter space of the model.

The equations for the extrema of the Higgs boson potential in the
directions (\ref{41}) in field space read: 
\be \ba{rcl}
\ds\frac{\partial V}{\partial s}&=&\ds m_{S}^2 s-\frac{\lambda
A_{\lambda}}{\sqrt{2}}v_1v_2+\frac{\lambda^2}{2}(v_1^2+v_2^2)s+\\[2mm]
&&+\ds\frac{g^{'2}_1}{2}\biggl(\tilde{Q}_1v_1^2+\tilde{Q}_2v_2^2+\tilde{Q}_S
s^2\biggr)\tilde{Q}_S s+\ds\frac{\partial\Delta V}{\partial
s}=0\,,\\[2mm] \ds\frac{\partial V}{\partial v_1}&=&\ds
m_1^2v_1-\frac{\lambda A_{\lambda}}{\sqrt{2}}s v_2
+\frac{\lambda^2}{2}(v_2^2+s^2)v_1+\frac{\bar{g}^2}{8}\biggl(v_1^2-v_2^2)\biggr)v_1+\\[2mm]
&&+\ds\frac{g^{'2}_1}{2}\biggl(\tilde{Q}_1v_1^2+\tilde{Q}_2v_2^2+\tilde{Q}_Ss^2\biggr)\tilde{Q}_1
v_1+\ds\frac{\partial\Delta V}{\partial v_1}=0\,,\\[2mm] 
\label{42}
\ds\frac{\partial V}{\partial v_2}&=&\ds m_2^2v_2-\frac{\lambda A_{\lambda}}{\sqrt{2}}s v_1+
\frac{\lambda^2}{2}(v_1^2+s^2)v_2+\frac{\bar{g}^2}{8}\biggl(v_2^2-v_1^2\biggr)v_2+\\[2mm]
&&+\ds\frac{g^{'2}_1}{2}\biggl(\tilde{Q}_1v_1^2+\tilde{Q}_2v_2^2+\tilde{Q}_Ss^2\biggr)\tilde{Q}_2 v_2+\ds\frac{\partial\Delta V}{\partial v_2}=0\,,
\ea
\ee
where $\bar{g}=\sqrt{g_2^2+g'^2}$. Instead of $v_1$ and $v_2$ it is more convenient to use $\tan\beta$ and
$v$ defined above. To simplify the analysis of the Higgs spectrum it is worth to express the soft masses $m_1^2,\,m_2^2,\,m_{S}^2$
in terms of $s,\, v$, $\tan\beta$ and other parameters. Because from precision measurements we know that $v=246\,\mbox{GeV}$
the tree--level Higgs masses and couplings depend on four variables only:
\be
\lambda\,,\qquad s\,,\qquad \tan\beta\,,\qquad A_{\lambda}\,.
\label{43}
\ee

\subsection{$Z$-$Z'$ mixing}
Initially the sector of EWSB involves ten
degrees of freedom. However four of them are massless Goldstone modes
which are swallowed by the $W^{\pm}$, $Z$ and $Z'$ gauge bosons. The
charged $W^{\pm}$ bosons gain masses via the interaction with the
neutral components of the Higgs doublets just in the same way as in
the MSSM so that $M_W=\ds\frac{g_2}{2}v$. Meanwhile the mechanism of
the neutral gauge boson mass generation differs significantly. Letting
$Z'$ be the gauge boson associated with $U(1)_{N}$, i.e.  \be
Z'_{\mu}=B_{2\mu}\,,\qquad\qquad
Z_{\mu}=W^3_{\mu}\cos\theta_W-B_{1\mu}\sin\theta_{W}\,,
\label{44}
\ee
the $Z-Z'$ mass squared matrix is given by
\be
M^2_{ZZ'}=\left(
\ba{cc}
M^2_{Z}  & \Delta^2\\[2mm]
\Delta^2 & M^2_{Z'}
\ea
\right)\,,
\label{45}
\ee
where
\be
\ba{c}
M_Z^2=\ds\frac{\bar{g}^2}{4}v^2\,,\qquad\qquad \Delta^2=\ds\frac{\bar{g}g'_1}{2}v^2\biggl(
\tilde{Q}_1\cos^2\beta-\tilde{Q}_2\sin^2\beta\biggr)\,,\\[3mm]
M^2_{Z'}=g^{'2}_1 v^2\biggl(\tilde{Q}_1^2\cos^2\beta+\tilde{Q}_2^2\sin^2\beta\biggr)+g^{'2}_1\tilde{Q}^2_S s^2\,.
\ea
\label{46}
\ee
The eigenvalues of this matrix are
\be
M^2_{Z_1,\,Z_2}=\ds\frac{1}{2}\left[M_Z^2+M^2_{Z'}\mp\sqrt{(M_Z^2-M_{Z'}^2)^2+4\Delta^4}\right]\,.
\label{47}
\ee
The eigenvalues $M^2_{Z_1}$ and $M^2_{Z_2}$ correspond to the mass eigenstates $Z_1$ and $Z_2$ which are linear
superpositions of $Z$ and $Z'$
\be
\ba{c}
Z_1=Z\cos\alpha_{ZZ'}+Z'\sin\alpha_{ZZ'}\,,\qquad\qquad Z_2=-Z\sin\alpha_{ZZ'}+Z'\cos\alpha_{ZZ'}\,,\\[2mm]
\alpha_{ZZ'}=\ds\frac{1}{2}\arctan\left(\frac{2\Delta^2}{M^2_{Z'}-M_Z^2}\right)\,.
\ea
\label{48}
\ee
Phenomenological constraints typically require the mixing angle $\alpha_{ZZ'}$ to be less than $2-3\times 10^{-3}$ \cite{16} and the
mass of the extra neutral gauge boson to be heavier than 
$500-600\,\mbox{GeV}$ \cite{15}. A suitable mass hierarchy and mixing between $Z$ and $Z'$ are
maintained if the field $S$ acquires a
large VEV $s \gtrsim 1.5\,\mbox{TeV}$. Then the mass of the lightest neutral
gauge boson $Z_1$ is very close to $M_Z$ whereas the mass of $Z_2$ is set by 
the VEV
 of the singlet field
$M_{Z_2}\simeq M_{Z'}\approx g'_1\tilde{Q}_S\, s$.

\subsection{Charged Higgs}
Due to electric--charge conservation the charged components of the
Higgs doublets are not mixed with neutral Higgs fields.  They form a
separate sector whose spectrum is described by a $2\times 2$ mass
matrix. Its determinant has zero value leading to the appearance of
two Goldstone states 
\be G^{-}=H_d^{-}\cos\beta-H_u^{+*}\sin\beta\,,
\label{49}
\ee
which are absorbed into the longitudinal degrees of freedom of the $W^{\pm}$ 
gauge boson. Their orthogonal linear combination
\be
H^{+}=H_d^{-*}\sin\beta+H_u^{+}\cos\beta
\label{50}
\ee
gains mass
\be
m^2_{H^{\pm}}=\ds\frac{\sqrt{2}\lambda A_{\lambda}}{\sin 2\beta}s-\frac{\lambda^2}{2}v^2+\frac{g^2}{2}v^2+\Delta_{\pm}\,.
\label{51}
\ee
In the leading one--loop approximation the corrections to the charged Higgs boson mass $\Delta_{\pm}$ in the ESSM
are almost the same as in the MSSM where the parameter $\mu$ has to be replaced by $\ds\frac{\lambda s}{\sqrt{2}}$.
The explicit expressions for the leading one--loop corrections to $m^2_{H^{\pm}}$ in the MSSM can be found in \cite{49}.

\subsection{CP--odd Higgs}
The imaginary parts of the neutral components of the Higgs doublets
and imaginary part of the SM singlet field $S$ compose the CP--odd (or pseudoscalar)
Higgs sector of the considered model. This sector includes two
Goldstone modes $G_0,G'$
which are swallowed by the $Z$ and $Z'$ bosons after 
EWSB, leaving only one physical CP--odd
Higgs state $A$. In the field basis $(A,\,G',\,G_0)$ one has
\be
\ba{l} A=P_S\sin\varphi+P\cos\varphi\,,\\[2mm]
G'=P_S\cos\varphi-P\sin\varphi\,,\\[2mm] G_0=\sqrt{2}(Im\,
H_d^0\cos\beta-Im\, H_u^0\sin\beta)\,,
\ea
\label{52}
\ee 
where
\be
\ba{l} 
P=\sqrt{2}(Im\,H_d^0\sin\beta+Im\, H_u^0\cos\beta)\,,\\[2mm] 
P_S= \sqrt{2}Im\, S\,,
\qquad \tan\varphi=\ds\frac{v}{2s}\sin2\beta .
\ea
\label{52'}
\ee 
Two massless pseudoscalars $G_0$ and $G'$ decouple from the rest of
the spectrum whereas the physical CP--odd Higgs boson $A$ acquires mass
\be 
m^2_{A}=
\ds\frac{\sqrt{2}\lambda A_{\lambda}}{\sin 2\varphi}v+\Delta_A\,,
\label{53}
\ee
where $\Delta_A$ is the contribution of loop corrections. In the leading one--loop approximation the expressions for the mass of the
pseudoscalar Higgs boson in the ESSM and PQ symmetric NMSSM coincide. The CP--odd Higgs sector of the NMSSM and one--loop
corrections to it were studied in \cite{50}. In phenomenologically acceptable models, in which the singlet VEV
 is much larger than $v$, $\varphi$ goes to zero and the physical pseudoscalar is predominantly the superposition of the imaginary parts of
the neutral components of the Higgs doublets, i.e. $P$.

\subsection{CP--even Higgs}
The CP--even Higgs sector involves $Re\,H_d^0$, $Re\,H_u^0$ and $Re\,
S$. In the field space basis $(h,\,H,\,N)$ rotated by an angle $\beta$
with respect to the initial one 
\be \ba{c} Re\,H_d^0=(h \cos\beta- H
\sin\beta+v_1)/\sqrt{2}\,, \\[2mm] Re\,H_u^0=(h \sin\beta+ H
\cos\beta+v_2)/\sqrt{2}\,, \\[2mm] Re\,S=(s+N)/\sqrt{2}\,, \ea
\label{54}
\ee
the mass matrix of the Higgs scalars takes the form \cite{51}:
\be
M^2=
\left(
\ba{ccc}
\ds\frac{\partial^2 V}{\partial v^2}&
\ds\frac{1}{v}\frac{\partial^2 V}{\partial v \partial\beta}&
\ds\frac{\partial^2 V}{\partial v \partial s}\\[0.3cm]
\ds\frac{1}{v}\frac{\partial^2 V}{\partial v \partial\beta}&
\ds\frac{1}{v^2}\frac{\partial^2 V}{\partial^2\beta}&
\ds\frac{1}{v}\frac{\partial^2 V}{\partial s \partial\beta}\\[0.3cm]
\ds\frac{\partial^2 V}{\partial v \partial s}&
\ds\frac{1}{v}\frac{\partial^2 V}{\partial s \partial\beta}&
\ds\frac{\partial^2 V}{\partial^2 s}
\ea
\right)=\left(
\ba{ccc}
M_{11}^2 & M_{12}^2 & M_{13}^2\\
M_{21}^2 & M_{22}^2 & M_{23}^2\\
M_{31}^2 & M_{32}^2 & M_{33}^2
\ea
\right)\,.
\label{55}
\ee
Taking second derivatives of the Higgs boson effective potential and substituting $m_1^2$, $m_2^2$, $m_{S}^2$
from the minimisation conditions (\ref{42}) one obtains:
\be
\ba{rcl}
M_{11}^2&=&\ds\frac{\lambda^2}{2}v^2\sin^22\beta+\ds\frac{\bar{g}^2}{4}v^2\cos^22\beta+g^{'2}_1 v^2(\tilde{Q}_1\cos^2\beta+
\tilde{Q}_2\sin^2\beta)^2+\Delta_{11}\,,\\
M_{12}^2&=&M_{21}^2=\ds\left(\frac{\lambda^2}{4}-\ds\frac{\bar{g}^2}{8}\right)v^2
\sin 4\beta+\ds\frac{g^{'2}_1}{2}v^2(\tilde{Q}_2-\tilde{Q}_1)\times\\
&&\times(\tilde{Q}_1\cos^2\beta+\tilde{Q}_2\sin^2\beta)\sin 2\beta+\Delta_{12}\, ,\\
M_{22}^2&=&\ds\frac{\sqrt{2}\lambda A_{\lambda}}{\sin 2\beta}s+\left(\frac{\bar{g}^2}{4}-\ds\frac{\lambda^2}{2}\right)v^2
\sin^2 2\beta+\ds\frac{g^{'2}_1}{4}(\tilde{Q}_2-\tilde{Q}_1)^2 v^2 \sin^22\beta+\Delta_{22}\,,\\
M_{23}^2&=&M_{32}^2=-\ds\frac{\lambda A_{\lambda}}{\sqrt{2}}v\cos 2\beta+\ds\frac{g^{'2}_1}{2}(\tilde{Q}_2-\tilde{Q}_1)\tilde{Q}_S
v s\sin 2\beta+\Delta_{23}\,,\\
M_{13}^2&=&M_{31}^2=-\ds\frac{\lambda A_{\lambda}}{\sqrt{2}}v\sin 2\beta+\lambda^2 v s+g^{'2}_1(\tilde{Q}_1\cos^2\beta+
\tilde{Q}_2\sin^2\beta)\tilde{Q}_S v s+\Delta_{13}\,,\\
M_{33}^2&=&\ds\frac{\lambda A_{\lambda}}{2\sqrt{2}s}v^2\sin 2\beta+g^{'2}_1\tilde{Q}_S^2s^2+\Delta_{33}\,.
\ea
\label{56}
\ee
In Eq.~(\ref{56}) $\Delta_{ij}$ represents the
 contribution of loop corrections which in the leading one--loop approximation are rather similar to the
ones calculated in the NMSSM. The one--loop corrections to the mass matrix of the NMSSM CP--even Higgs sector were analysed in
\cite{50}, \cite{52}.

When the SUSY breaking scale $M_S$ and VEV
of the singlet field are considerably larger than the
EW scale the mass matrix (\ref{55})--(\ref{56}) has a
hierarchical structure. Therefore the masses of the heaviest Higgs
bosons are closely approximated by the diagonal entries $M_{22}^2$ and
$M_{33}^2$ which are expected to be of the order of $M_S^2$ or even
higher. All off--diagonal matrix elements are relatively small
$\lesssim M_S M_Z$. As a result the mass of one CP--even Higgs boson
(approximately given by $H$)
is governed by $m_A$ while the mass of another one
(predominantly the $N$ singlet field) is set by $M_{Z'}$.  Since the minimal
eigenvalue of the mass matrix (\ref{55})--(\ref{56}) is always less
than its smallest diagonal element at least one Higgs scalar in the
CP--even sector (approximately $h$) remains light 
even when the SUSY breaking
scale tends to infinity, i.e. $m^2_{h_1}\lesssim M_{11}^2$.

The direct Higgs searches at LEP set stringent limits on the parameter
space of supersymmetric extensions of the SM.  In order to establish
the corresponding restrictions on the parameters of the ESSM we need
to specify the couplings of the neutral Higgs particles to the
$Z$ boson. In the rotated field basis $(h,\,H,\,N)$ the trilinear part
of the Lagrangian, which determines the interaction of the neutral
Higgs states with the $Z$ boson, is simplified: \be
L_{AZH}=\ds\frac{\bar{g}}{2}
M_{Z}Z_{1\mu}Z_{1\mu}h+\frac{\bar{g}}{2}Z_{1\mu}
\biggl[H(\partial_{\mu}A)-(\partial_{\mu}H)A\biggr]~.
\label{57}
\ee 
Here we assume that the mixing between $Z$ and $Z'$ is negligibly
small and can be safely ignored so that $Z_{1}\simeq Z$. In the
considered case only one CP--even component $h$ couples to a pair of
$Z$ bosons while another one $H$ interacts with the pseudoscalar $A$ and
$Z_1$.  The coupling of $h$ to the $Z_1$ pair is exactly the same as
in the SM. In the Yukawa interactions with fermions the first
component of the CP--even Higgs basis also manifests itself as  a
SM--like Higgs boson.

The couplings of the Higgs scalars to a $Z_1$ pair ($g_{ZZi}$,
$i=1,2,3$) and to the Higgs pseudoscalar and $Z$ boson ($g_{ZAi}$)
appear because of the mixing of $h$ and $H$ with other components of
the CP--even Higgs sector. Following the traditional notations we
define the normalised $R$--couplings as: $g_{ZZh_i}=R_{ZZi}\times$
SM coupling; $g_{ZAh_i}=\ds\frac{\bar{g}}{2}R_{ZAi}$.  The
absolute values of all $R$--couplings vary from zero to unity.

The components of the CP--even Higgs basis are related to the physical CP--even Higgs eigenstates by virtue of a unitary transformation:
\be
\left(
\begin{array}{c}
h\\ H\\ N
\end{array}
\right)=
U^{\dagger}
\left(
\begin{array}{c}
h_1 \\ h_2\\ h_3
\end{array}
\right)\,.
\label{58}
\ee
Combining the Lagrangian (\ref{57}) and relations (\ref{58}) the normalised $R$--couplings may be written in terms of the mixing matrix elements
according to
\be
R_{ZZi}=U^\dagger_{hh_i}~,\qquad\qquad R_{ZAi}=U^\dagger_{Hh_i}\,.
\label{59}
\ee
If all fundamental parameters are real the CP--even Higgs mass matrix (\ref{55})--(\ref{56}) is symmetric and the unitary transformation
(\ref{58}) reduces to an orthogonal one. The orthogonality of the mixing matrices $U$ results in sum rules:
\be
\sum_i R_{ZZi}^2=1~,\qquad \sum_i R_{ZAi}^2=1~,\qquad \sum_i R_{ZZi}R_{ZAi}=0\, .
\label{60}
\ee
The conditions (\ref{60}) allow to eliminate three $R$--couplings. As a result, in the limit $\alpha_{ZZ'}\to 0$ the interactions of the
neutral Higgs particles with a $Z$ boson are described by three independent $R$--couplings. The dependence of spectrum and couplings of the
Higgs bosons on the parameters of the ESSM will be examined in the following section.

\section{Higgs phenomenology}

\subsection{Higgs masses and couplings}

\subsubsection{The MSSM limit $\lambda \rightarrow 0$, $s\rightarrow \infty$}

First of all we consider the spectrum and couplings of the Higgs
bosons in the ESSM. Let us start from the MSSM limit of the ESSM when
$\lambda \to 0$, $s\rightarrow \infty$ with 
$\mu_{eff}\sim \lambda s$ held fixed in order to 
give an acceptable chargino mass and EWSB.
From the first minimisation
conditions (\ref{42}) it follows that such solution can be obtained
for very large and negative values of $m_S^2$ only.

As $s\rightarrow \infty$ the CP--even Higgs state, which is
predominantly a singlet field, $Z'$ boson and all exotic quarks and
non--Higgsinos become very heavy and decouple from the rest of the particle
spectrum. Then by means of a small unitary transformation the CP--even
Higgs mass matrix in Eq.~(\ref{55})
reduces to the block diagonal form \cite{55}
\be M'^2\simeq\left( \ba{ccc} \ds
M_{11}^2-\ds\frac{M_{13}^4}{M_{33}^2} &
M_{12}^2-\ds\frac{M^2_{13}M^2_{32}}{M^2_{33}} & 0\\[3mm]
M_{21}^2-\ds\frac{M_{23}^2 M_{31}^2}{M_{33}^2} &
M_{22}^2-\ds\frac{M_{23}^4}{M_{33}^2} & 0\\[3mm] 0 & 0 & \ds
M_{33}^2+\ds\frac{M_{13}^4}{M_{33}^2}+\ds\frac{M_{23}^4}{M_{33}^2} \ea
\right)\,.
\label{64}
\ee
For small values of $\lambda$ the top--left $2\times 2$ submatrix in Eq.~(\ref{64}) reproduces the mass matrix of the CP--even Higgs
sector in the MSSM. So at tree--level we find
\be
\ba{rcl}
m_{H^{\pm}}^2 &\simeq & m_A^2+m_W^2\,,\qquad\qquad\qquad m_A^2=\ds\frac{\sqrt{2}\lambda A_{\lambda}}{\sin 2\beta}s\,,\\[3mm]
m_{h_1, h_2}^2 &\simeq &\ds\frac{1}{2}\left[m_A^2+M_Z^2\mp \sqrt{(m_A^2+M_Z^2)^2-4m_A^2M_Z^2\cos^2 2\beta}\right]\,,\\[3mm]
m_{h_3}^2 &\simeq & g^{'2}_1\tilde{Q}_S^2s^2.
\ea
\label{65}
\ee
In Eq.~(\ref{65}) the terms of  $O(\lambda^2 v^2)$ are
omitted, and $s^2\gg v^2$ is assumed.

Since the enlargement of $s$ leads to the growth of the mass of the
singlet dominated Higgs state $m_{h_3}$, which is very close to
$M_{Z'}$, the mixing between $N$ and neutral components of the Higgs
doublets diminishes when $\lambda$ tends to zero. Thus in the MSSM
limit of the ESSM the couplings of the heaviest CP--even Higgs boson
to the quarks, leptons and gauge bosons vanish and the MSSM sum rules
for the masses and couplings of the two lightest Higgs scalars and
pseudoscalar are recovered. As in the minimal SUSY model the masses of
MSSM--like Higgs bosons are defined by $m_A$ and $\tan\beta$. They
grow if $m_A$ increases and at large values of $m_A$ ($m_A^2>>M_Z^2$)
the mass of the lightest CP--even Higgs boson attains its theoretical
upper bound which is determined by the $Z$ boson mass at 
tree--level, i.e. $m_{h_1}\le M_Z |\cos 2\beta|$ \cite{56}.

\subsubsection{$\lambda \gtrsim g_1$}

When $\lambda\gtrsim g'_1\approx g_1\approx 0.46$ the qualitative
pattern of the spectrum of the Higgs bosons is rather similar to the
one which arises in the PQ symmetric NMSSM
\cite{55}, \cite{57}. We first give an analytic discussion of the
spectrum at tree--level, then discuss the spectrum numerically including
one--loop radiative corrections.

Assuming that in
the allowed part of the parameter space $M_{22}^2\gg M_{33}^2\gg
M_{11}^2$ the perturbation theory method yields \be \ba{rcl} m^2_{h_3}
&\simeq & M_{22}^2+\ds\frac{M_{23}^4}{M_{22}^2}\,,\\[3mm] m^2_{h_2}
&\simeq &
M_{33}^2-\ds\frac{M_{23}^4}{M_{22}^2}+\ds\frac{M_{13}^4}{M_{33}^2}\,,\\[3mm]
m^2_{h_1} &\simeq & M_{11}^2-\ds\frac{M_{13}^4}{M_{33}^2}\,.  \ea
\label{66}
\ee
Here we neglect all terms suppressed by inverse powers of $m_A^2$ or $M_{Z'}^2$, i.e. $O(M_Z^4/m_A^2)$ and $O(M_Z^4/M_{Z'}^2)$.

At tree--level the masses of the Higgs bosons can written as
\be
\ba{rclcrcl}
m_A^2&=&\ds\frac{2\lambda^2 s^2 x}{\sin^2 2\beta}+O(M_Z^2)\,,&\qquad\qquad & m^2_{H^{\pm}}&=&m_A^2+O(M_Z^2)\,,\\[3mm]
m^2_{h_3}&=&m_A^2+O(M_Z^2)\,,&\qquad\qquad & m^2_{h_2}&=&g^{'2}_1\tilde{Q}_S^2s^2+O(M_Z^2)\,,
\ea
\label{67}
\ee
\be
\ba{rcl}
m^2_{h_1}&\simeq &\ds\frac{\lambda^2}{2}v^2\sin^22\beta+\ds\frac{\bar{g}^2}{4}v^2\cos^22\beta+g^{'2}_1 v^2\biggl(\tilde{Q}_1\cos^2\beta+
\tilde{Q}_2\sin^2\beta\biggr)^2-\\[3mm]
&&-\ds\frac{\lambda^4 v^2}{g^{'2}_1Q_S^2}\biggl(1-x+\ds\frac{g^{'2}_1}{\lambda^2}\biggl(\tilde{Q}_1\cos^2\beta+Q_2\sin^2\beta\biggr)Q_S\biggr)^2
+O(M_Z^4/M_{Z'}^2)\,,
\ea
\label{68}
\ee
where
$$
x=\ds\frac{A_{\lambda}}{\sqrt{2}\lambda s}\sin 2\beta\,.
$$
As evident from the explicit expression for $m^2_{h_1}$ given above at $\lambda^2\gg g^{2}_1$ the last term in Eq.~(\ref{68}) dominates and
the mass of the lightest Higgs boson tends to be negative if the auxiliary variable $x$ is not close to unity. In this case the vacuum stability
requirement constrains the variable $x$ around unity. As a consequence $m_A$ is confined in the vicinity of $\mu\, \tan\beta$ and is much larger
than the masses of the $Z'$ and lightest CP--even Higgs boson. At so large values of $m_A$ the masses of the heaviest CP--even, CP--odd and charged
states are almost degenerate around $m_A$.

In Fig.~4 we plot masses and couplings of the Higgs bosons as a
function of $m_A$. As a
representative example we fix $\tan\beta=2$ and VEV
 of the singlet field $s=1.9\,\mbox{TeV}$, that
corresponds to $M_{Z'}\simeq 700\,\mbox{GeV}$ which is quite close to
the current limit on the $Z'$ boson mass. For our numerical study we
also choose the maximum possible value of $\lambda(M_t)\simeq 0.794$
which does not spoil the validity of perturbation theory up to the
Grand Unification scale. In order to obtain a realistic spectrum, we
include the leading one--loop corrections from the top and stop
loops. The contributions of these corrections to $m_A^2$,
$m^2_{H^{\pm}}$ and mass matrix of the CP--even Higgs states
(\ref{55})--(\ref{56}) depend rather strongly on the soft masses of
the the superpartners of the top--quark ($m_Q^2$ and $m_U^2$) and the 
stop mixing
parameter $X_t=A_t-\ds\frac{\lambda s }{\sqrt{2}\tan\beta}$. Here and
in the following we set $m_Q=m_U=M_{S}=700\,\mbox{GeV}$ while the stop mixing
parameter is taken to be $\sqrt{6}\,M_S$ in order to enhance
stop--radiative effects.

From Fig.~4a it becomes clear that the mass of
the lightest Higgs scalar changes considerably when $m_A$ varies. At
$m_A$ below $2\,\mbox{TeV}$ or above $3\,\mbox{TeV}$ the mass squared
of the lightest Higgs boson tends to be negative. A negative eigenvalue
of the mass matrix (\ref{55})--(\ref{56}) means that the considered
vacuum configuration ceases to be a minimum and turns into a saddle
point. Near this point there is a direction in field space along
which the energy density decreases generating instability of the given
vacuum configuration.
The requirement of stability of the physical vacuum therefore
limits the range of
variations of $m_A$ from below and above. Together with the
experimental lower limit on the mass of the $Z'$ boson it maintains the mass
hierarchy in the spectrum of the Higgs particles seen in
Figs.~4b and 4c.  Relying on this mass hierarchy one can diagonalise the
$3\times 3$ mass matrix of the CP--even Higgs sector. 

The numerical results in Figs.~4a--4c
confirm the analytic tree--level results 
discussed earlier.
The numerical analysis reveals that the masses of the two
heaviest CP--even, CP--odd and charged Higgs states grow when the
VEV of the SM singlet field (or $M_{Z'}$)
increases. The masses of the heaviest scalar, pseudoscalar and charged
Higgs fields also rise with increasing $\lambda$, $m_A$ and
$\tan\beta$ while the mass of the second lightest Higgs scalar is
almost insensitive to variations of these parameters.
The growth of the masses of heavy Higgs bosons caused by the
increase of $\tan\beta$ or $s$ does not affect much the lightest
Higgs scalar mass which lies below $200\,\mbox{GeV}$ (see Fig.~4a).

Turning now to a discussion of the couplings,
the hierarchical structure of the mass matrix of the CP--even Higgs
sector for $\lambda\gtrsim g_1$ allows one to get approximate solutions
for the Higgs couplings to the $Z$ boson. They are given by \be
\ba{lclcl} |R_{ZZ1}|\simeq
1-\ds\frac{1}{2}\left(\frac{M^2_{13}}{M_{33}^2}\right)^2\,,&\,\,\,
&|R_{ZZ2}|\simeq\ds\frac{|M^2_{13}|}{M^2_{33}}\,,&\,\,\, &
|R_{ZZ3}|\simeq \ds\frac{|M_{12}^2|}{M_{11}^2}\,,\\[3mm]
|R_{ZA1}|\simeq
\ds\biggl|\ds\frac{M^2_{12}}{M_{22}^2}-\ds\frac{M_{23}^2
M_{13}^2}{M_{22}^2 M_{33}^2}\biggr|\,,&\,\,\, & |R_{ZA2}|\simeq
\ds\frac{|M_{23}^2|}{M_{22}^2}\,,&\,\,\, &|R_{ZA3}|\simeq
1-\ds\frac{1}{2}\left(\frac{M^2_{13}}{M_{33}^2}\right)^2\,.  \ea
\label{69}
\ee 
The obtained approximate formulae for the Higgs couplings
(\ref{69}) indicate that\\ $R_{ZZ1}\gg R_{ZZ2}\gg R_{ZZ3}$ and
$R_{ZA3}\gg R_{ZA2}\gg R_{ZA1}$. 

The analytic discussion of the couplings is confirmed 
by the numerical results for 
$R_{ZZi}$ and $R_{ZAi}$ shown in Figs.~4d and 4e where the
results of our numerical analysis including leading one--loop
corrections to the CP--even Higgs mass matrix from the top--quark and
its superpartners are presented. From Eq.~(\ref{69}) as well as from
Figs.~4d and 4e one can see that the couplings of the second lightest
Higgs boson to a $Z$ pair and to the Higgs pseudoscalar and $Z$ are always
suppressed. They are of $O(M_Z/M_{Z'})$ and $O(M_Z/m_A)$
respectively which is a manifestation of the singlet dominated structure
of the wave function of the second lightest Higgs scalar. The heaviest
CP--even Higgs boson is predominantly a superposition of neutral
components of Higgs doublets $H$. This is a reason why its relative
coupling to the pseudoscalar and $Z$ is so close to unity (see
Eq.~(\ref{57})). The main contribution to the wave function of the
lightest Higgs scalar gives the first component of the CP--even Higgs
basis $h$. Due to this the relative coupling of the lightest CP--even
Higgs boson to $Z$ pairs tends to unity in the permitted range of the
parameter space. Because mixing between $H$ and $h$ is extremely small
the couplings $R_{ZZ3}$ and $R_{ZA1}$ are almost negligible, i.e. they
are of $O(M_Z^2/m_A^2)$.

\subsubsection{$\lambda \lesssim g_1$}

With decreasing $\lambda$ the qualitative pattern of the Higgs
spectrum changes significantly. In Fig.~5 the masses of the Higgs
particles and their couplings to $Z$ are examined as a function of $m_A$
for $\lambda(M_t)=0.3$. The values of $\tan\beta$ and $M_{Z'}$ are
taken to be the same as in Fig.~4. 
For $\lambda\lesssim g'_1\approx g_1\approx 0.46$
the allowed range
of $m_A$ enlarges because mixing between the first and third
components of the CP--even Higgs basis reduces. In particular the
mass squared of the lightest Higgs boson remains positive even when
$m_A\sim M_Z$, as shown in Fig.~5a.
Therefore the lower bound on the Higgs pseudoscalar
mass disappears so that charged, CP--odd and second lightest CP--even
Higgs states may have masses in the $200-300\,\mbox{GeV}$ range (see
Figs.~5b and 5c). But the requirement of vacuum stability still prevents
having very high values of $m_A$ (or $x$). Indeed from Eq.~(\ref{68})
it is obvious that very large values of $x$ (or $m_A$) pulls the
mass squared of the lightest Higgs boson below zero destabilising the
vacuum.  This sets upper limits on the masses of charged and
pseudoscalar Higgs bosons.

At least one scalar in the Higgs spectrum is always heavy since it has
almost the same mass as the $Z'$ boson, which must be heavier than
$600\,\mbox{GeV}$. The mass of this CP--even Higgs state is determined
by the VEV of the singlet field and does not
change much if the other parameters $\lambda$, $\tan\beta$ and $m_A$
vary. As before the masses of the other CP--even, CP--odd and charged
Higgs fields grow when $m_A$ rises providing the degeneracy of the
corresponding states at $m_A\gg M_Z$. The growth of $\tan\beta$ and
$s$ enlarges the allowed range of $m_A$ increasing the upper limit on
the pseudoscalar mass. The permitted interval for $m_A$ is also
expanded when $\lambda$ diminishes.

The couplings of the neutral Higgs bosons to $Z$ depend rather
strongly on the value of the pseudoscalar mass. At small values of
$m_A$ the second lightest Higgs scalar gains a relatively low
mass. Because of this the mixing between $H$ and $h$ is large and
relative couplings of the lightest Higgs scalars to $Z$ pairs and to the
pseudoscalar and $Z$ are of the order of unity (see Figs.~5d and 5e). The
couplings of the heaviest CP--even Higgs state to other bosons and
fermions are tiny in this case since it is predominantly a singlet
field. If the lightest Higgs scalar and pseudoscalar had low masses and
large couplings to the $Z$ they could be produced at
LEPII. Non--observation of these particles at LEP rules out most
parts of the ESSM parameter space for $m_A\lesssim 200\,\mbox{GeV}$.

When $m_A$ is much larger than $M_Z$ but is less than $M_{Z'}$ the heaviest
Higgs scalar state is still singlet dominated which makes its couplings
to the observed particles negligibly small. The hierarchical structure
of the CP--even Higgs mass matrix ensures that the lightest and second
lightest Higgs scalars are predominantly composed of the first and
second components of the CP--even Higgs basis respectively. Therefore
$R_{ZZ1}$ and $R_{ZA2}$ are very close to unity while $R_{ZZ2}$ and
$R_{ZA1}$ are suppressed. When $m_A$ approaches the $Z'$ boson mass
the mixing between $S$ and $H$ becomes large. This leads to 
appreciable values of the $R_{ZA2}$ and $R_{ZA3}$ couplings as 
displayed in Fig.~5e. However both of these relative couplings may be
simultaneously large only in a very narrow part of the parameter space
where $m_A\simeq M_{Z'}$. At the same time the relative couplings of
the heaviest Higgs scalars to $Z$ pairs are still much less than unity
in this range of parameters because the mixing between the first and the
other components of the CP--even Higgs basis remains very small (see
Fig.~5d). Further increasing $m_A$ mimics the mass hierarchy of the
CP--even Higgs sector appeared at $\lambda\gtrsim g_1$. As a result
the pattern of the Higgs couplings is rather similar to the one shown
in Figs.~4d and 4e.

\subsection{Upper bound on the lightest CP--even Higgs boson mass}

It is apparent from Figs.~4b and 5b, as
well as our analytic considerations, that at some value of $m_A$
(or $x$) the lightest CP--even Higgs boson mass attains its maximum value.
This coincides with the theoretical upper bound on $m_{h_1}$ given by
the first element of the mass 
$\sqrt{M_{11}^2}$. In this subsection we shall obtain an absolute
upper bound on the lightest CP--even Higgs boson mass in the ESSM,
and compare it to similar bounds obtained in the MSSM and NMSSM.

\subsubsection{Tree--level upper bound}
At tree--level the lightest
Higgs scalar mass is obtained from Eq.~(\ref{68})
where the first three terms on the 
right--hand side are positive definite, while the fourth term
is always negative, and the upper bound therefore corresponds to 
taking this term to be zero.
The contribution of the extra $U(1)_{N}$ $D$--term to the upper
limit on $m_{h_1}$ may be closely approximated as \be g^{'2}_1
v^2\biggl(\tilde{Q}_1\cos^2\beta+\tilde{Q}_2\sin^2\beta\biggr)^2\simeq
\left(\frac{M_Z}{2}\right)^2\biggl(1+\frac{1}{4}\cos2\beta\biggr)^2\,.
\label{70}
\ee
Using this approximation, the tree--level upper bound 
on the lightest CP--even Higgs boson is given by:
\be m_{h_1}^2\lesssim
\ds\frac{\lambda^2}{2}v^2\sin^22\beta+M_Z^2\cos^22\beta+\ds\frac{M_Z^2}{4}\biggl(1+\frac{1}{4}\cos2\beta\biggr)^2\, .
\label{710}
\ee 
The first and second terms are similar to the 
tree--level terms in the NMSSM \cite{111}.  The extra $U(1)_{N}$ effect appears
through the third term in Eq.~(\ref{68}) which is a contribution coming
from the additional $U(1)_{N}$ $D$--term in the Higgs scalar potential
\cite{58}. 

At tree--level the theoretical restriction on the lightest Higgs
mass in the ESSM depends on $\lambda$ and $\tan\beta$ only.  As it was
noticed in section 3 the requirement of validity of perturbation
theory up to the Grand Unification scale constrains the interval of
variations of the Yukawa coupling $\lambda$ for each value of
$\tan\beta$. The allowed range of $\lambda$ as a function of
$\tan\beta$ was shown in Fig.~3. Using the results of the analysis of
the RG flow in the ESSM one can obtain the maximum
possible value of the lightest Higgs scalar for each particular choice
of $\tan\beta$.

In Fig.~6a we plot maximum values of the square roots of different
contributions in Eq.~(\ref{710})
to the tree--level upper limit on $m_{h_1}^2$ versus
$\tan\beta$. It is clear that at moderate values of $\tan\beta \sim 1-3$
the term $\lambda^2 v^2 /2\sin^2 2\beta$ from the
$F$-term involving the singlet field
dominates. With increasing $\tan\beta$ it falls quite rapidly
and becomes negligibly small as $\tan\beta\gtrsim 15$. In contrast
the contribution of the $SU(2)$ and $U(1)_Y$
$D$--terms grows when $\tan\beta$ becomes larger. At $\tan\beta\gtrsim 4$ it
exceeds $\lambda^2 v^2 /2\sin^2 2\beta$ and gives the
dominant contribution to the tree--level upper bound on $m_{h_1}$. As
one can see from Fig.~6a the $D$--term of the extra $U(1)_{N}$ gives the
second largest contribution to the tree--level theoretical restriction
on $m_{h_1}$ at very large and low values of $\tan\beta$, when
$\tan\beta$ is less than $1.6$ or larger than $8$. In the part of the ESSM
parameter space where the upper limit on $m_{h_1}$ reaches its
absolute maximum value its contribution is the smallest one. According
to Eq.~(\ref{710}) the square root of the $U(1)_{N}$ $D$--term
contribution to $m^2_{h_1}$ varies from $45$ to $34\,\mbox{GeV}$ when
$\tan\beta$ changes from $1.1$ to $14$.

The resulting {tree--level} upper bound on the mass of the lightest
Higgs particle in the ESSM is presented in Fig.~6b,
and compared to the corresponding bounds in the
MSSM and NMSSM.
In the ESSM the bound attains a maximum value of
$130\,\mbox{GeV}$ at $\tan\beta=1.5-1.8$. Remarkably, we find that in
the interval of $\tan\beta$ from $1.2$ to $3.4$ the absolute maximum
value of the mass of the lightest Higgs scalar in the ESSM is larger
than the experimental lower limit on the SM--like Higgs boson even at
tree--level. Therefore non--observation of the Higgs boson at LEP
does not cause any trouble for the ESSM, even at tree--level.

The upper bound
on the mass of the lightest CP--even Higgs scalar in the NMSSM
exceeds the corresponding limit in the MSSM because of the extra
contribution to $m^2_{h_1}$ induced by the additional $F$--term in the
Higgs scalar potential of the NMSSM. The size of this contribution,
which is described by the first term in Eq.~(\ref{710}), is determined
by the Yukawa coupling $\lambda$ whose interval of variations is
constrained by the applicability of perturbation theory at high
energies.  The upper limit on the coupling $\lambda$ caused by the validity
of perturbation theory in the NMSSM is more stringent than in the
ESSM due to the presence of exotic
$5+\overline{5}$--plets of matter in the particle spectrum of the
ESSM. Indeed extra $SU(5)$ multiplets of matter change the running of
the gauge couplings so that their values at the intermediate scale
rise when the number of new supermultiplets increases. Since $g_i(Q)$
occurs in the right--hand side of the differential equations
(\ref{34}) with negative sign the growth of the gauge couplings
prevents the appearance of the Landau pole in the evolution of the
Yukawa couplings. It means that for each value of the top--quark
Yukawa coupling (or $\tan\beta$) at the EW scale the maximum
allowed value of $\lambda(M_t)$ rises when the number of
$5+\overline{5}$--plets increases. The increase of $\lambda(M_t)$
is accompanied by the growth of the theoretical restriction on the
mass of the lightest CP--even Higgs particle. For instance, it was shown that
the introduction of four pairs of $5+\overline{5}$ supermultiplets in
the NMSSM raised the two--loop upper limit on the lightest Higgs boson
mass from $135\,\mbox{GeV}$ to $155\,\mbox{GeV}$ \cite{581}.  This is
also a reason why the tree--level theoretical restriction on $m_{h_1}$
in the Next-to-Minimal SUSY model is considerably less than in the
ESSM at moderate values of $\tan\beta$.

At large $\tan\beta\gg 10$ the contribution of the $F$--term of the SM
singlet field to $m_{h_1}^2$ vanishes. Therefore with increasing 
$\tan\beta$ the upper bound on the lightest Higgs boson mass in the
NMSSM approaches the corresponding limit in the minimal SUSY model.
In the ESSM the theoretical restriction on the mass of the lightest
Higgs scalar also diminishes when $\tan\beta$ rises. But even at very
large values of $\tan\beta$ the tree--level upper limit on $m_{h_1}$ in
the ESSM is still $6-7\,\mbox{GeV}$ larger than the ones in the MSSM
and NMSSM because of the $U(1)_{N}$ $D$--term contribution.

\subsubsection{One--loop upper bound}

So far we have discussed the bounds at tree--level.
Now we shall include radiative corrections in our discussion.
It is well known that 
the inclusion of loop corrections from the top--quark and its
superpartners increases the bound on the lightest
Higgs boson mass in the ESSM substantially. In the ESSM and in the
NMSSM these corrections are nearly the same as in the 
MSSM. The leading one--loop and two--loop corrections to the lightest
Higgs boson mass in the MSSM were calculated and studied in \cite{59}
and \cite{60}--\cite{61} respectively. However, in contrast with the
MSSM and NMSSM, the ESSM contains extra supermultiplets of exotic
matter. Because it is not clear {\it a priori}\, if the corrections
induced by the loops involving new particles affect the mass of the
lightest Higgs scalar considerably, we include in our analysis leading
one--loop corrections to $m_{h_1}^2$ from the exotic quarks since
their couplings to the singlet Higgs field tend to be large at low
energies enhancing radiative effects. In the leading approximation the
upper bound on the lightest Higgs boson mass in the ESSM can be
written as 
\be m_{h_1}^2\lesssim
\ds\frac{\lambda^2}{2}v^2\sin^22\beta+M_Z^2\cos^22\beta+\ds\frac{M_Z^2}{4}\biggl(1+\frac{1}{4}\cos2\beta\biggr)^2+\Delta^t_{11}+\Delta^D_{11}\,,
\label{71}
\ee 
where the third term represents the $U(1)_{N}$ $D$--term contribution
while $\Delta^t_{11}$ and $\Delta^D_{11}$ are one--loop corrections
from the top--quark and $D$--quark supermultiplets respectively. When
$m_{Di}^2=m_{\overline{D}i}^2=M_S^2$ the contribution of one--loop
corrections to $m^2_{h_1}$ from the superpartners of $D$--quarks reduces
to 
\be \Delta^{D}_{11}=\ds\sum_{i=1,2,3}\frac{3 \lambda^2\kappa_i^2
v^2}{32\pi^2}\,\sin^22\beta\,\,\ln\biggl[\frac{m_{D_{1,i}}m_{D_{2,i}}}{Q^2}\biggr]\,.
\label{72}
\ee

In Figs.~7a and 7b we explore the dependence of one--loop upper bound on
the lightest Higgs boson mass on the Yukawa couplings $\kappa_i$ and
$\lambda$. We consider two different cases when $\kappa_1=\kappa_2=0$,
$\kappa_3=\kappa$ (see Fig.~7a) and $\kappa_1=\kappa_2=\kappa_3=\kappa$
(see Fig.~7b). To simplify our analysis the soft masses of the
superpartners of exotic and top--quarks are set to be equal, i.e.
$m_Q^2=m_U^2=m_{Di}^2=m^2_{\overline{D}i}=M_S^2$. In order to enhance
the contribution of loop effects we assume maximal mixing in the stop
sector ($X_t=\sqrt{6} M_{S}$) and minimal mixing between the
superpartners of exotic quarks $D$ and $\overline{D}$, i.e. $A_{\kappa
i}=0$. As before we keep $M_S=M_{Z'}=700\,\mbox{GeV}$ and
$\tan\beta=2$. Then the theoretical restriction on the mass of the
lightest Higgs scalar (\ref{71}) is defined by the couplings $\lambda$
and $\kappa$ only.

In the plane $\lambda/h_t - \kappa/h_t$ the set of points that
results in the same upper limit on $m_{h_1}$ forms a line. For any
choice of $\lambda$ and $\kappa$ lying below the line the lightest
Higgs particle has a mass which is less than the theoretical restriction
that corresponds to this line. Curvature of the line characterises the
dependence of the upper bound on the lightest Higgs boson mass on the
Yukawa coupling $\kappa$. If $\kappa$ is zero the one--loop contribution of
the exotic squarks to $m^2_{h_1}$ vanishes. When $\kappa$ grows the
exotic squark contribution to $m_{h_1}^2$ and the upper limit on the
mass of the lightest Higgs scalar rise. As a consequence the same
theoretical restriction on $m_{h_1}$ is obtained for smaller values of
$\lambda(M_t)$. But from Figs.~7a and 7b one can see that the
increase of $\kappa(M_t)$ within the allowed range of the parameter
space does not lead to the appreciable decrease of $\lambda(M_t)$ that
should compensate the growth of exotic squark contribution to
$m_{h_1}^2$. It means that the contribution of the exotic squarks is
always much smaller than the first term in Eq.~(\ref{71}). Numerically
the increase of the lightest Higgs boson mass caused by the inclusion
of the exotic squark contribution does not exceed a few $\mbox{GeV}$.

\subsubsection{Two--loop upper bound}

We also include in our analysis leading two--loop
corrections to $m^2_{h_1}$ from the top--quark and its superpartner.
In the two--loop leading--log approximation the upper bound on the
lightest Higgs boson mass in the ESSM can be written in the following
form 
\be \ba{c}
m_h^2\lesssim\biggl[\ds\frac{\lambda^2}{2}v^2\sin^22\beta+M_Z^2\cos^22\beta+\ds\frac{M_Z^2}{4}\biggl(1+\frac{1}{4}\cos2\beta\biggr)^2\biggr]
\times\qquad\qquad\\[5mm]
\times\left(1-\ds\frac{3h_t^2}{8\pi^2}l\right)+\ds\frac{3 h_t^4 v^2
\sin^4\beta}{8\pi^2}\left\{\ds\frac{1}{2}U_t+l+
\ds\frac{1}{16\pi^2}\biggl(\frac{3}{2}h_t^2-8g_3^2\biggr)(U_t+l)l\right\}+\Delta^D_{11}\,,
\ea
\label{73}
\ee
$$
U_t=2\ds\frac{X_t^2}{M_S^2}\biggl(1-\frac{1}{12}\frac{X_t^2}{M_S^2}\biggr)\,,\qquad\qquad\qquad l=\ln\biggl[\ds\frac{M_S^2}{m_t^2}\biggr]\,.
$$
Here we keep one--loop leading--log corrections from the exotic squarks. Eq.~(\ref{73}) is a simple generalisation of the approximate expressions
for the theoretical restriction on the mass of the lightest Higgs particle obtained in the MSSM \cite{61} and NMSSM \cite{62}.
The inclusion of leading two--loop corrections reduces the upper limit on $m_{h_1}$ significantly and nearly compensates the growth of the
theoretical restriction on $m_{h_1}$ with increasing SUSY breaking scale $M_S$ which is caused by one--loop corrections.

The dependence of the two--loop upper bound (\ref{73}) on $\lambda$
and $\kappa$ for two different choices of the Yukawa couplings of
exotic quarks described above is examined in Figs.~7c and 7d. After the
incorporation of two--loop corrections the line that corresponds to the
$160\,\mbox{GeV}$ upper limit on the lightest Higgs boson mass lies
beyond the permitted range of the parameter space while in the
one--loop approximation even larger values of $m_{h_1}$ are
allowed. The distortion of the lines, which represent different
theoretical restrictions on the mass of the lightest Higgs scalar in
the ESSM, still remains negligible. It demonstrates the fact that the
exotic squark contribution to $m^2_{h_1}$ is much less than the
leading two--loop corrections from the top--quark and its superpartners
to $m^2_{h_1}$.

From Fig.~7 it follows that for each given value of $\tan\beta$ the
mass of the lightest Higgs particle attains its maximum when
$\lambda(M_t)\to\lambda_{\max}$ and $\kappa\to 0$. Nevertheless the
dependence of the upper limit (\ref{73}) on $\kappa$ is rather weak so
that the theoretical restriction on the lightest Higgs boson mass for
$\kappa=0$ and for $\kappa=g'_1$ are almost identical. The upper limit
on $m_{h_1}$ is very sensitive to the choice of $\lambda$ and
$\tan\beta$. Therefore at the last stage of our analysis we explore
the dependence of the two--loop upper bound (\ref{73}) on $\tan\beta$
keeping $\kappa=0$ (i.e. $\Delta^D_{11}=0$) and relying on the results
of our study of the RG flow summarised in Fig.~3.

The dependence of the two--loop theoretical restrictions on $m_{h_1}$
on $\tan\beta$ shown in Figs.~8a and 8b resembles the tree--level one.
But the interval of variations of the upper bound on $m_{h_1}$
shrinks. In Figs.~8a and 8b we consider maximal ($X_t=\sqrt{6} M_{S}$)
and minimal ($X_t=0$) mixing in the stop sector respectively. Again at
moderate values of $\tan\beta=1.6-3.5$ the upper bound on the lightest
Higgs boson mass in the ESSM is considerably higher than in the MSSM
and NMSSM because of the enhanced contribution of the $F$--term of the
SM singlet field to $m^2_{h_1}$. Although the two--loop theoretical
restriction on $m_{h_1}$ in the ESSM reduces with increasing 
$\tan\beta$ it still remains $4-5\,\mbox{GeV}$ larger than the
corresponding limits in the MSSM and NMSSM owing to the $U(1)_{N}$
$D$--term contribution. This contribution is especially important in
the case of minimal mixing between the superpartners of the top
quark. In the considered case the two--loop theoretical restriction on
$m_{h_1}$ in the MSSM and NMSSM is less than the experimental limit on
the SM--like Higgs boson mass set by LEPII. As a result the scenario
with $X_t=0$ is ruled out in the MSSM. The contribution of an extra
$U(1)_{N}$ $D$--term to $m_{h_1}^2$ raises the upper bound (\ref{73}) at
large $\tan\beta\gtrsim 10$ slightly above the existing LEP limit thus relaxing
the constraints on the ESSM parameter space (see Fig.~8b).

The growth of $X_t$ from $0$ to $\sqrt{6} M_S$ increases the
theoretical restriction on the lightest Higgs boson mass in the ESSM
by $10-20\,\mbox{GeV}$. The upper limit on $m_{h_1}$ is most
sensitive to the choice of $X_t$ at low and large values of
$\tan\beta$ where the growth of the corresponding theoretical
restriction reaches $20$ and $15\,\mbox{GeV}$ respectively. At the
same time the absolute maximum value of the lightest Higgs boson mass
rises by $10\,\mbox{GeV}$ only. In total leading one--loop and
two--loop corrections modify the maximum possible value of the mass
of the lightest Higgs scalar by $20\,\mbox{GeV}$ by increasing it up to
about $150\,\mbox{GeV}$.

Note that the quoted upper limits for the ESSM, as well as the 
MSSM and NMSSM, are sensitive to the value of the top--quark mass,
and the SUSY breaking scale, and depend on the precise form
of the two--loop approximations used. Here we have used an analytic
approximation of the two--loop effects which slightly underestimates
the full two--loop corrections. We have also taken the SUSY scale to 
be given by $700\,\mbox{GeV}$. The upper bounds quoted here
may therefore be further increased by several GeV by making slightly 
different assumptions. The main point we wish to make is that the
upper bound on the lightest CP--even Higgs scalar
in the ESSM is always significantly larger than
in the NMSSM, as well as the MSSM.

\section{Charginos and Neutralinos}

\subsection{Chargino and neutralino states in the ESSM}

After EWSB all superpartners of the gauge and Higgs bosons get non--zero masses. Since the supermultiplets
of the $Z'$ boson and SM singlet Higgs field $S$ are electromagnetically neutral they do not contribute any extra particles to the chargino
spectrum. Consequently the chargino mass matrix and its eigenvalues remain the same as in the MSSM, namely
\be
\ba{rcl}
m^2_{\chi^{\pm}_{1,\,2}}&=&\ds\frac{1}{2}\biggl[M_2^2+\mu_{eff}^2+2 M^2_{W}\pm\biggl.\\[2mm]
&&\biggr.\qquad\qquad\qquad\sqrt{(M_2^2+\mu^2_{eff}+2M^2_{W})^2-4(M_2\mu_{eff}-M^2_{W}\sin 2\beta)^2}
\biggr]\,,
\ea
\label{61}
\ee
where $M_2$ is the $SU(2)$ gaugino mass and $\mu_{eff}=\ds\frac{\lambda s}{\sqrt{2}}$. Unsuccessful LEP searches for SUSY particles including data
collected at $\sqrt{s}$ between $90\,\mbox{GeV}$ and $209\,\mbox{GeV}$ set a $95\%$ CL lower limit on the chargino mass of about $100\,\mbox{GeV}$
\cite{53}. This lower bound constrains the parameter space of the ESSM restricting the absolute values of the effective $\mu$-term and $M_2$
from below, i.e. $|M_2|$, $|\mu_{eff}|\ge 90-100\,\mbox{GeV}$.

In the neutralino sector of the ESSM there are two extra neutralinos besides the four MSSM ones. One of them is
an extra gaugino coming from the $Z'$ vector supermultiplet. The other one is an additional Higgsino $\tilde{S}$ (singlino) which is
a fermion component of the SM singlet superfield $S$. The Higgsino mass terms in the Lagrangian of the ESSM are induced by the trilinear
interaction $\lambda S(H_d H_u)$ in the superpotential (\ref{16}) after the breakdown of gauge symmetry. Because of this their values are
determined by the coupling $\lambda$ and VEVs  of the Higgs fields. The mixing between gauginos and Higgsinos is
proportional to the corresponding gauge coupling and VEV that the scalar partner of the considered Higgsino gets. Taking
this into account one can obtain a $6\times 6$ neutralino mass matrix that in the interaction basis
$(\tilde{B},\,\tilde{W}_3,\,\tilde{H}^0_1,\,\tilde{H}^0_2,\,\tilde{S},\,\tilde{B}')$ reads
\be
M_{\tilde{\chi}^0}=
\left(
\ba{cccccc}
M_1                  & 0                   & -\ds\frac{1}{2}g'v_1 & \ds\frac{1}{2}g'v_2  & 0                                & 0\\[2mm]
0                    & M_2                 & \ds\frac{1}{2}g v_1  & -\ds\frac{1}{2}g v_2 & 0                                & 0\\[2mm]
-\ds\frac{1}{2}g'v_1 & \ds\frac{1}{2}g v_1 & 0                    & -\mu_{eff}           & -\ds\frac{\lambda v_2}{\sqrt{2}} & \tilde{Q}_1 g'_1
v_1\\[2mm]
\ds\frac{1}{2}g'v_2  & -\ds\frac{1}{2}gv_2 & -\mu_{eff}           &  0                   & -\ds\frac{\lambda v_1}{\sqrt{2}} & \tilde{Q}_2 g'_1
v_2\\[2mm]
0                    & 0                   & -\ds\frac{\lambda v_2}{\sqrt{2}}  & -\ds\frac{\lambda v_1}{\sqrt{2}} & 0       & \tilde{Q}_S g'_1 s
\\[2mm]
0                    & 0                   & \tilde{Q}_1 g'_1 v_1 & \tilde{Q}_2 g'_1 v_2 & \tilde{Q}_S g'_1 s & M'_1
\ea
\right)\,,
\label{62}
\ee
where $M_1$, $M_2$ and $M'_1$ are the soft gaugino masses for $\tilde{B}$, $\tilde{W}_3$ and $\tilde{B}'$ respectively. In Eq.~(\ref{62})
we neglect the Abelian gaugino mass mixing $M_{11}$ between $\tilde{B}$ and $\tilde{B}'$ that arises at low energies as a result of the kinetic term
mixing even if there is no mixing in the initial values of the soft SUSY breaking gaugino masses near the Grand Unification or Planck scale
\cite{462}. The top--left $4\times 4$ block of the mass matrix (\ref{62}) contains the neutralino mass matrix of the MSSM where the parameter $\mu$
is replaced by $\mu_{eff}$. The lower right $2\times 2$ submatrix represents extra components of neutralinos in the considered model.
The neutralino sector in $E_6$ inspired SUSY models was studied recently in \cite{24}, \cite{29}, \cite{32}, \cite{341}--\cite{342},
\cite{401}, \cite{54}--\cite{231}.

As one can see from Eqs.~(\ref{61})--(\ref{62}) the masses of charginos and neutralinos depend on $\lambda$, $s$, $\tan\beta$, $M_1,\,M'_1$ and $M_2$.
In SUGRA models with uniform gaugino masses at the Grand Unification scale the RG flow yields a relationship between
$M_1,\,M'_1$ and $M_2$ at the EW scale:
\be
M'_1\simeq M_1\simeq 0.5 M_2\,.
\label{63}
\ee
This reduces the parameter space in the neutralino sector of the ESSM drastically. It allows to
study the spectrum of chargino and neutralino as a function of only one gaugino mass, for example $M_1$, for each set of $\lambda$, $s$ and $\tan\beta$.

\subsection{Chargino and neutralino spectrum}

The qualitative pattern of chargino and neutralino masses is determined by the Yukawa coupling $\lambda$, depending on whether $\lambda$ is less
than $g'_1$ or not. Because in the MSSM limit of the ESSM, when $\lambda\to 0$, the phenomenologically acceptable solution
implies that $s\gtrsim M_Z/\lambda$, the extra $U(1)_{N}$ gaugino $\tilde{B}'$ and singlino $\tilde{S}$ decouple from the rest of the spectrum
forming two eigenstates $(\tilde{B}'\pm\tilde{S})/\sqrt{2}$ with mass $M_{Z'}= \tilde{Q}_S g'_1 s$\,. Mixing between new
neutralino states and other gauginos and Higgsinos vanishes in this case rendering the neutralino sector in the ESSM indistinguishable
from MSSM at the LHC and ILC. Here it is worth to emphasise that the direct observation of extra neutralino states in the ESSM is unlikely to occur
in the nearest future anyway. Since off--diagonal entries of the bottom right $2\times 2$ submatrix of the neutralino mass matrix (\ref{62})
are controlled by the $Z'$ boson mass new neutralinos are always very heavy $(\sim 1\,\mbox{TeV})$ preventing the distinction between the
ESSM and
MSSM neutralino sectors.

When $\lambda>g'_1$ the typical pattern of the spectrum of neutralinos and charginos changes. In Figs.~9a and 9b we examine
the dependence of the neutralino and chargino masses on $M_1$ assuming the unification of the soft gaugino mass parameters at the scale $M_X$.
As a representative example we fix $\lambda\simeq 0.794$, $\tan\beta=2$ and $M_{Z'}=700\,\mbox{GeV}$ ($s\simeq 1.9\,\mbox{TeV}$). We restrict
our consideration to the most attractive part of the parameter space, in which the lightest chargino is accessible  at future
colliders, i.e. $|M_1|\lesssim 300\,\mbox{GeV}$. In order to get the spectrum of neutralinos we diagonalise the mass matrix (\ref{62})
numerically. As a consequence we obtain a set of positive and negative eigenvalues of this matrix which are presented in Fig.~9a. However the physical
meaning is only their absolute values.

In the considered part of the parameter space the heaviest chargino and neutralinos are almost degenerate with mass $|\mu_{eff}|$. They are formed
by the neutral and charged superpartners of the Higgs bosons. As one can see from Figs.~9a and 9b the masses of the heaviest chargino and neutralinos
are almost insensitive to the choice of the gaugino masses if $|M_1|\lesssim 300\,\mbox{GeV}$. The $U(1)_{N}$ gaugino $\tilde{B}'$ and
singlino $\tilde{S}$ compose two other heavy neutralino eigenstates whose masses are closely approximated as
\be
|m_{\chi^0_{3,4}}|\simeq\frac{1}{2}\biggl[\sqrt{M^{'2}_1+4M^2_{Z'}}\mp M'_1\biggr]\,.
\label{74}
\ee
The four heaviest neutralinos and chargino gain masses beyond $500\,\mbox{GeV}$ range so that their observation at the LHC and ILC looks rather
problematic. The masses of the heaviest neutralino and chargino states rise with increasing VEV of the SM singlet field and
are practically independent of $\tan\beta$.

At low energies heavy neutralinos decouple and the spectrum of the two lightest ones is described by the $2\times 2$ mass matrix
\be
M'_{\tilde{\chi}^0}\simeq
\left(
\ba{cc}
M_1-\ds\frac{g^{'2}v^2}{4\mu}\sin 2\beta  & \ds\frac{gg'v^2}{4\mu}\sin 2\beta \\[2mm]
\ds\frac{gg'v^2}{4\mu}\sin 2\beta         & M_2-\ds\frac{g^{2}v^2}{4\mu}\sin 2\beta
\ea
\right)\,,
\label{75}
\ee
whose eigenvalues are
\be
\ba{rcl}
|m_{\chi^0_{1,2}}|&\simeq&\biggl|M_1+M_2-\ds\frac{M_Z^2}{\mu}\sin 2\beta\mp\\[3mm]
&\mp&\sqrt{\left(M_1+M_2-\ds\frac{M_Z^2}{\mu}\sin 2\beta\right)^2-4\left(M_1M_2-\ds\frac{M_Z^2}{\mu}\tilde{M}\right)}\biggr|\,,
\ea
\label{76}
\ee
where
$$
\tilde{M}=M_2\sin^2\theta_W+M_1\cos^2\theta_W\,.
$$
The superpartners of charged $SU(2)$ gauge bosons form the lightest chargino state with mass
\be
|m_{\chi^{\pm}_1}|=\biggl|M_2-\frac{M^2_W}{\mu}\sin 2\beta\biggr|\,.
\label{77}
\ee
The numerical analysis and our analytic consideration show that the masses of the lightest neutralinos and chargino do not change
much when $\lambda$, $s$ or $\tan\beta$ vary unless $M_1$ and $M_2$ are quite small. The second lightest neutralino and the lightest
chargino are predominantly superpartners of the $SU(2)$ gauge bosons. Their masses are governed by $|M_2|$. The lightest neutralino state is
basically bino, $\tilde{B}$, whose mass is set by $|M_1|$.

In Figs.~10a and 10b we explore the spectrum of neutralino and chargino in the case when $\lambda$ is less than $g'_1$.
The Yukawa coupling $\lambda$ is taken to be $0.3$ while the other parameters remain the same as in Fig.~9. Now the two heaviest neutralinos are
mixtures of $\tilde{B}'$ and $\tilde{S}$. They get masses larger than $500\,\mbox{GeV}$ as before. However unlike in the large $\lambda$ limit the other
four neutralinos and both charginos can be light enough. Therefore they may be observed in the nearest future. The composition of the
wave functions of the lightest neutralinos and charginos depends on the choice of the parameters of the model. For the set of $\lambda$, $\tan\beta$
and $s$ chosen in Fig.~10 the lightest neutralino is still predominantly bino, $\tilde{B}$. Till $|M_1|$ is less than $200\,\mbox{GeV}$, i.e.
$|M_2|<|\mu_{eff}|$, the second lightest neutralino and the lightest chargino are basically formed by the superpartners of $SU(2)$ gauge bosons.
When $|M_1|>200\,\mbox{GeV}$ ($|M_2|>|\mu_{eff}|$) the wave functions of the second lightest neutralino and the lightest chargino are Higgsino
dominated.

Obvious disadvantage of the considered scenarios with $\lambda<g'_1$ and $\lambda>g'_1$ in the ESSM is that any pattern of the masses and
couplings of the lightest neutralino and chargino, which can be observed at the LHC and ILC, may be reproduced in the framework of the
minimal SUSY model. This is a consequence of the stringent lower bound on the mass of the $Z'$ boson set by Tevatron.

\section{$Z'$ and exotic phenomenology}

\subsection{Masses and couplings of new states}

The presence of a $Z'$ gauge boson and exotic multiplets of matter in the particle spectrum
is a very peculiar feature that permits to distinguish $E_6$ inspired supersymmetric models
from the MSSM or NMSSM. At tree--level the masses of these new particles
are determined by the VEV of the singlet field $S$ that remains a free
parameter in the considered models. Therefore the $Z'$ boson mass and the masses of exotic
quarks and non--Higgses cannot be predicted. But collider experiments \cite{16}--\cite{15} and
precision EW tests \cite{17} set stringent limits on the $Z'$ mass and $Z-Z'$ mixing.
The lower bounds on the $Z'$ mass from direct searches at the Fermilab Tevatron
$(p\overline{p}\to Z'\to l^{+}l^{-})$ \cite{15} are model dependent but are typically around
$500-600\,\mbox{GeV}$ unless couplings of ordinary particles to $Z'$ are suppressed such as in
leptophobic models \cite{19}, \cite{18}. Similarly, bounds on the mixing angle are around
$(2-3)\times 10^{-3}$ \cite{16}. As has been already mentioned, even more stringent constraints
on the $Z'$ mass and mixing follow from nucleosynthesis and astrophysical observations. They
imply that the equivalent number of additional neutrinos with full--strength weak interactions
$\Delta N_{\nu}$ is less than $0.3$ (for a recent review, see \cite{20}). This requires
$M_{Z'}\gtrsim 4.3\,\mbox{TeV}$ \cite{21}. However these restrictions cannot be applied
to the $Z'$ gauge boson in the ESSM because right--handed neutrinos here
are expected to be superheavy and do not change the effective number
of neutrino species at low energies.

The analysis performed in \cite{22} revealed that $Z'$ boson in the $E_6$ inspired models
can be discovered at the LHC if its mass is less than $4-4.5\,\mbox{TeV}$. At the same time
the determination of the couplings of the $Z'$ should be possible up to $M_{Z'}\sim 2-2.5\,\mbox{TeV}$
\cite{23}. Possible $Z'$ decay channels in $E_6$ inspired supersymmetric models were studied in
\cite{231}.


The restrictions on the masses of exotic particles are not so rigorous as the experimental
bounds on the mass of the $Z'$ boson. The most stringent constraints come from the non--observation of
exotic colour states at HERA and Tevatron. But most searches imply that exotic quarks, i.e
leptoquarks or diquarks, have integer--spin. So they are either scalars or vectors.
Because of this new coloured objects can be coupled directly to either a pair of quarks or to quark
and lepton. Moreover it is usually assumed that leptoquarks and diquarks have appreciable couplings to
the quarks and leptons of the first generation. Experiments at LEP, HERA and Tevatron excluded such
leptoquarks if their masses were less than $290\,\mbox{GeV}$ \cite{63} whereas CDF and D0 ruled out
diquarks with masses up to $420\,\mbox{GeV}$ \cite{64}. The production of diquarks at the LHC was
studied recently in \cite{65}.

In the ESSM the exotic squarks and non--Higgses are expected to be heavy since their masses are
determined by the SUSY breaking scale. Moreover, their couplings to the quarks and leptons
of the first and second generation should be rather small to avoid processes with non--diagonal
flavour transitions. As a result the production of exotic squarks and non--Higgses will be
very strongly suppressed or even impossible at future colliders. However the exotic fermions
(quarks and non--Higgsinos) can be relatively light in the ESSM since their masses are
set by the Yukawa couplings $\kappa_i$ and $\lambda_i$ that may be small. This happens,
for example, when the Yukawa couplings of the exotic particles have hierarchical structure
similar to the one observed in the ordinary quark and lepton sectors. Then $Z'$ mass lie beyond
$10\,\mbox{TeV}$ and the only manifestation of the considered model may be the presence of
light exotic quark or non--Higgses in the particle spectrum.

The new exotic particles consist of vector--like multiplets with respect to the SM gauge group.
Hence their axial couplings to the SM gauge bosons go to zero in the limit of no $Z-Z'$ mixing
reducing the contribution of additional particles to the electroweak observables measured at LEP.
The contribution of $Z'$ gauge boson to the electroweak observables is also negligibly small
because it is supposed to be very heavy so that the mixing between $Z$ and $Z'$ almost vanishes.

In order to amplify the signal coming from the presence of additional particles we shall assume
further that the three families of exotic quarks and two generations of exotic non--Higgsinos, whose masses
are determined by extra Yukawa couplings, are considerably lighter than the $Z'$ and in this case they
can possibly be observed at the LHC and ILC. Since the analysis of the RG flow
performed in section 3 revealed that the exotic quarks as well as non--Higgsino Yukawa couplings
tend to be equal we keep
$\mu_{D1}=\mu_{D2}=\mu_{D3}=\ds\frac{\kappa(M_t)}{\sqrt{2}}s$ and
$\mu_{H1}=\mu_{H2}=\ds\frac{\lambda_{1,2}(M_t)}{\sqrt{2}}s$ ($\lambda_1(M_t)=\lambda_2(M_t)$) at
the EW scale. Moreover to simplify our numerical studies we fix the masses of exotic quarks
and non--Higgsinos to be equal to $300\,\mbox{GeV}$.

Before proceeding to discuss the scope of future colliders in
testing the features of the ESSM in the $Z'$ and exotic sectors,
we should remind the reader that we assume the mixing angle
between the $Z$ and
$Z'$ to be extremely small. In this case, any vertex
involving a $Z'$ boson and fermions can be written as
\begin{equation}
-\ds\frac{ig'_1\gamma_{\mu}}{2}\biggl(g_V-g_A\gamma_5\biggr)\,.
\end{equation}
For, e.g. $M_Z'=1.5\,\mbox{TeV}$, the vector $g_V(M_{Z'})$ and axial $g_A(M_{Z'})$
couplings take the following values:
\begin{itemize}
\item  charged leptons:\qquad $g_V=\tilde{Q}_{e_L}-\tilde{Q}_{e^c_L}\simeq 0.1081$,\qquad $g_A=\tilde{Q}_{e^c_L}+\tilde{Q}_{e_L}\simeq 0.4910$\,;
\item neutrinos:\qquad $g_V=g_A=\tilde{Q}_{\nu_L}=0.2996$\,;
\item $u$--quarks:\qquad $g_V=\tilde{Q}_{u_L}-\tilde{Q}_{u^c_L}\simeq 0.0278$,\qquad $g_A=\tilde{Q}_{u^c_L}+\tilde{Q}_{u_L}\simeq 0.2996$\,;
\item $d$--quarks:\qquad $g_V=\tilde{Q}_{d_L}-\tilde{Q}_{d^c_L}\simeq -0.1637$,\qquad $g_A=\tilde{Q}_{d^c_L}+\tilde{Q}_{d_L}\simeq 0.4910$\,;
\item exotic $D$--quarks:\qquad
  $g_V=\tilde{Q}_{D}-\tilde{Q}_{\bar{D}}\simeq 0.1359$,\qquad
  $g_A=\tilde{Q}_{\bar{D}}+\tilde{Q}_{D}\simeq-0.7906$\,;
\item fermion partners of extra non--Higgs fields (non--Higgsinos):
$$
g_V=\tilde{Q}_{H_1}-\tilde{Q}_{H_2}\simeq -0.1915\,,\qquad g_A=\tilde{Q}_{H_2}+\tilde{Q}_{H_1}\simeq-0.7906\,;
$$
\item fermion partners of extra singlet fields (singlinos):
$$
g_V=g_A=\tilde{Q}_{S}\simeq 0.7906\,.
$$
\end{itemize}


As for additional couplings, the interaction of exotic quarks and non--Higgsinos with the neutral
SM gauge bosons $\gamma$ and $Z$ takes the form
$$
-ieQ_{em}\gamma_{\mu}\,,\qquad
-\ds\frac{i\overline{g}}{2}g_V\gamma_{\mu}\,,
$$
respectively, where $Q_{em}$ and $g_V$ are given as follows:
\begin{itemize}
\item exotic $D$--quarks: $Q^D_{em}=-1/3\,,\qquad g_V=-2Q^D_{em}\sin^2\theta_{W}$\,;
\item charged non--Higgsinos:  $Q^E_{em}=-1\,,\qquad g_V=2T_3^f-2Q_{em}^f\sin^2\theta_W\simeq-\cos2\theta_W$\,;
\item neutral non--Higgsinos: $Q^N_{em}=0\,,\qquad g_V=1$\,.
\end{itemize}
For completeness and to fix our conventions, we also quote the couplings of the $\gamma$ and $Z$ bosons 
to ordinary quarks and leptons, as
\begin{itemize}
\item ordinary fermions (quarks and leptons): $g_V=T_3^f-2Q_{em}^f\sin^2\theta_W\,,\qquad g_A=T_3^f$\,.
\end{itemize}

\subsection{Phenomenology at future colliders}

Fig.~11 shows the differential distribution in invariant mass of the lepton pair $l^+l^-$ (for one species
of lepton $l=e,\mu$) in Drell-Yan production at the LHC, assuming the SM only as well as the latter
augmented, in turn, by a $Z'$ field ($M_{Z'}=1.5$ TeV) with and without light exotic quarks
or non--Higgsinos ($\mu_{Di}=\mu_{Hi}=300$ GeV) separately\footnote{Recall that we assume three identical generations of
the former but only two of the latter. Besides, we allow the existence of only one at a time of either.}.
This distribution is promptly measurable at the CERN collider with a high resolution and would 
enable one to not only confirm the existence of a $Z'$ state but also to
establish the possible presence as well as nature of additional exotic matter, by simply fitting
to the data the width of the $Z'$ resonance \cite{23}.  In fact for our choice of $\mu_{Di}$, $\mu_{Hi}$ and $M_{Z'}$ the
$Z'$ total width varies from $\approx19$ GeV (in case of no exotic matter) to $\approx 25$ GeV (in case of light exotic
$D$--quarks) and to $\approx 21$ GeV (in case of light non--Higgsinos). (Also notice the different
normalisation around the $Z'$ resonance of the three curves in Fig.~11, as this scales
like $\sim\Gamma(Z'\to l^+l^-)/\Gamma(Z'\to\rm{anything})$.)
Clearly, in order to perform such an
exercise, the $Z'$ couplings to ordinary matter ought to have been previously established elsewhere, as a
modification of the latter may well lead
to effects similar to those induced by the additional matter present in our model. (Recall that in our model
$Z'$ couplings to SM particles and exotic matter are simultaneously fixed.)

However, if exotic particles of the nature described here do exist at such low scales, they could possibly be
accessed through direct pair hadroproduction. In fact, the corresponding fully inclusive
cross sections are in principle sufficient to such a purpose in the case of
exotic $D$--quarks (up to masses of the TeV order) while  this statement is presumably true for non--Higgsinos
only up to masses of few hundred GeV. (Notice that the former are generated via gluon-induced QCD interactions while
the latter via quark--induced EW ones.) This should be manifest as a close inspection of Fig.~12.

In practice, detectable final states do depend on the underlying nature of the exotic particles.
The lifetime and decay modes of the latter are determined by the operators that break the
$Z_2^{H}$ symmetry. When $Z_2^H$ is broken significantly exotic fermions can produce a remarkable 
signature\footnote{ If $Z_2^{H}$ is only slightly broken exotic quarks and non--Higgsinos may live for a long time.
Then exotic quarks will form compound states with ordinary quarks. It means that at future colliders it may be
possible to study the spectroscopy of new composite scalar leptons or baryons. Also one can observe 
quasi--stable charged colourless fermions with zero lepton number.}. Since according to our initial assumptions the 
$Z_2^{H}$ symmetry is mostly broken by the operators involving quarks and leptons of the third generation
the exotic quarks decay either via
$$
\overline{D}\to t+\tilde{b}\,,\qquad\qquad\qquad \overline{D}\to b+\tilde{t}\,,
$$
if exotic quarks $\overline{D}_i$ are diquarks or via
$$
\ba{c}
D\to t+\tilde{\tau}\,,\qquad\qquad\qquad D\to \tau+\tilde{t}\,,\\
D\to b+\tilde{\nu}_{\tau}\,,\qquad\qquad\qquad D\to \nu_{\tau}+\tilde{b}\,,\\
\ea
$$
if exotic quarks of type $D$ are leptoquarks. Because in general sfermions decay into
corresponding fermion and neutralino one can expect that each diquark will decay further
into $t$-- and $b$--quarks while a leptoquark will produce a $t$--quark and $\tau$--lepton
in the final state with rather high probability.  Thus 
 the presence of light exotic quarks in the
particle spectrum could result in an appreciable enhancement of the cross section of
either $pp\to t\overline{t}b\overline{b}+X$ and $pp\to b\overline{b}b\overline{b}+X$
if exotic quarks are diquarks or $pp\to t\overline{t}\tau^+{\tau^-}+X$ and
consequently $pp\to b\overline{b}\tau^+{\tau^-}+X$ if new quark states are leptoquarks\footnote{It 
is worth to remind here that the production cross sections
of $pp\to t\overline{t}b\overline{b}+X$ and $pp\to t\overline{t}\tau^+{\tau^-}+X$
in the SM are suppressed {\sl at least} by a factor $\left(\ds\frac{\alpha_s}{\pi}\right)^2$ and
$\left(\ds\frac{\alpha_W}{\pi}\right)^2$ respectively as compared
to the cross section of $t\overline{t}$ pair production (and, similarly, for $t$--quarks replaced by 
$b$--quarks).}.
In compliance with our initial assumptions non--Higgsinos decay predominantly 
into either quarks and squarks or leptons and sleptons of the third 
generation as well, i.e.
$$
\ba{c}
\tilde{H}^0\to t+\tilde{\overline{t}}\,,\qquad\qquad\qquad \tilde{H}^0\to \overline{t}+\tilde{t}\,,\\
\tilde{H}^0\to b+\tilde{\overline{b}}\,,\qquad\qquad\qquad \tilde{H}^0\to \overline{b}+\tilde{b}\,,\\
\tilde{H}^0\to \tau+\tilde{\overline{\tau}}\,,\qquad\qquad\qquad \tilde{H}^0\to \overline{\tau}+\tilde{\tau}\,,\\
\tilde{H}^{-}\to b+\tilde{\overline{t}}\,,\qquad\qquad\qquad \tilde{H}^{-}\to \overline{t}+\tilde{b}\,,\\
\tilde{H}^{-}\to \tau+\tilde{\overline{\nu}_{\tau}}\,,\qquad\qquad\qquad \tilde{H}^{-}\to \overline{\nu}_{\tau}+\tilde{\tau}\,.
\ea
$$
If we assume again that a sfermion decays predominantly into the corresponding fermion and neutralino
then also the production of non--Higgsinos should lead to a significant enlargement of the cross sections of
$Q\bar{Q}Q^{(')}\bar{Q}^{(')}$ and $Q\bar{Q}\tau^+{\tau^-}$ production, where $Q$ is a
heavy quark of the third generation, that allows to identify these particles if they are light enough.

As each $t$--quark decays
into a $b$--quark whereas a $\tau$--lepton gives one charged lepton $l$
in the final state with a probability of $35\%$, both these scenarios would generate an excess in the $b$--quark production cross section.
In this respect SM data samples which should be altered by the presence of exotic $D$--quarks or non--Higgsinos
are those involving $t\bar t$ production and decay as well as direct $b\bar b$ production. For this
reason, Fig.~13 also shows the cross sections for these two genuine SM processes alongside those for the
exotica.
Detailed LHC analyses will be required to establish the feasibility of extracting the excess due to
the light exotic particles predicted by our model. 
However, Fig.~13 should clearly make the point that
-- for the discussed parameter configuration -- one is in a favourable position in this respect, as the
rates for the exotica times their BRs into the aforementioned
decay channels are typically larger than the expected four--body cross sections involving
heavy quarks and/or leptons. 

The situation will experimentally be much easier at a future ILC. Here, under the same assumptions as
above concerning their decay patterns, both species of exotic particles should contribute to the inclusive
hadronic cross section, see Fig.~14. Assuming, again, the mass choice $\mu_{Di}=\mu_{Hi}=300$ GeV, the onset
at 600 GeV of the exotic pair production threshold would clearly be visible above the SM continuum (with or
without a much heavier $Z'$, again, with $M_{Z'}=1.5$ TeV). The rise of the hadronic cross section at
$\sqrt s=2\mu_{Di}=2\mu_{Hi}$ would be different, depending on the kind of exotic particles being generated,
owning to the different EW charges involved and the fact that three generations of light $D$'s can be
allowed in our model as opposed to only two in the case of light $\widetilde H$'s (as already intimated).
Furthermore, the line shape of the $Z'$ resonance would be different too, depending on whether one or
the other kind of exotic matter is allowed. Both the enormous luminosity and extremely clean environment
of an ILC, joined with a significant degree of control on the beam energy spread, should allow one to
explore in detail all such possible features of the hadronic cross section. In fact, as the actual value
of the ILC beam energy has yet to be fixed and our illustrative choice for $\mu_{Di}$ and $\mu_{Hi}$ may
not correspond to what nature has chosen, we present in Fig.~15 the mass dependence of the pair production
cross section for our exotic states at two reference collider energies, of 700 GeV and identical to
the $Z'$ mass. While the scope for exotic $D$--quark production at the ILC has probably little to add
to what could be obtained earlier at the LHC, a TeV scale $e^+e^-$ linear collider is definitely crucial
in increasing the discovery reach in mass for non--Higgsinos beyond the limits obtainable at the CERN
hadronic machine.

\section{Conclusions}
In this paper we have made a
comprehensive study of the theory and
phenomenology of a low energy supersymmetric standard model
originating from a string-inspired $E_6$ grand unified gauge group,
which we called Exceptional Supersymmetric Standard Model, or ESSM for short.
The ESSM considered here
is based on the low energy SM gauge group together with
an extra $Z'$ corresponding to an extra $U(1)_{N}$ gauge
symmetry under which right--handed neutrinos have zero charge.
This allows right--handed neutrinos to gain large Majorana
masses, resulting in the conventional (high-scale) seesaw
mechanism for neutrino masses. The extra $U(1)_{N}$ gauge
symmetry survives to the TeV scale, and 
forbids the term $\mu H_d H_u$ in the
superpotential, but permits the term $\lambda S(H_u H_d)$,
where $S$ is a low energy singlet that carries $U(1)_{N}$ 
charge and breaks the gauge symmetry when it develops
its VEV, giving rise to a
massive $Z'$ and an effective $\mu$ term.
Therefore the $\mu$ problem of the MSSM is solved in a
similar way to that in the NMSSM, but without the accompanying
problems of singlet tadpoles or domain walls since there is
no $S^3$ term, and the would-be Goldstone boson is eaten by the $Z'$.

The low energy matter content of the ESSM corresponds to three $27$
representations of the $E_6$ symmetry group, to ensure anomaly
cancellation, plus an additional pair of Higgs--like doublets as
required for high energy gauge coupling unification. 
The ESSM is therefore a low energy alternative to the MSSM or NMSSM.
The ESSM involves extra matter beyond the MSSM contained in
three $5+5^*$ representations of $SU(5)$, plus a total of
three $SU(5)$ singlets which carry $U(1)_{N}$ charges.
Thus there are three families of new exotic charge 1/3 quarks and 
non--Higgs multiplets predicted in the ESSM, in addition to the $Z'$.

As in the MSSM, the gauge symmetry of the ESSM does not forbid
baryon and lepton number violating interactions that result in rapid 
proton decay. The straightforward generalisation of R--parity,
assuming that the exotic quarks carry the same baryon number as the ordinary 
ones, guarantees not only proton stability but also the stability of
the lightest exotic quark.  The presence of heavy stable exotic quarks,
that should survive annihilation, is ruled out by different
experiments.  Therefore the R--parity definition in the ESSM 
has to be modified. There are two different ways to impose
an appropriate $Z_2$ symmetry that lead to two different versions of
the ESSM where baryon and lepton number is conserved.
ESSM version I implies
that exotic quarks have twice larger baryon number than the ordinary quark
fields. In the ESSM version II exotic quarks carry baryon and lepton numbers
simultaneously.

Because the supermultiplets of exotic matter interact with the quark,
lepton and Higgs superfields the Lagrangian of the ESSM includes many
new Yukawa couplings. In general these couplings give rise to the
processes with non--diagonal flavour transitions that have not been
observed yet. In order to suppress flavour changing processes and to
provide the correct breakdown of gauge symmetry we assumed a
hierarchical structure of the Yukawa interactions, and imposed an
approximate $Z_2^H$ symmetry under which all superfields are odd
except Higgs doublets ($H_u$ and $H_d$) and singlet field $S$.  With
these assumptions only one SM singlet field $S$ may have appreciable
couplings with exotic quarks and $SU(2)$ doublets $H_{1i}$ and
$H_{2i}$ and the couplings are flavour diagonal. It also follows that
only one pair of $SU(2)$ Higgs doublets $H_d$ and $H_u$ have Yukawa
couplings to the ordinary quarks and leptons of order unity. The
Yukawa couplings of other exotic particles to the quarks and leptons
of the first two generations must be less than $10^{-4}$ and $10^{-3}$
respectively in order to suppress FCNCs.  We would like to emphasise
that from the perspective of REWSB
 these assumptions are completely natural. Without loss of
generality it is always possible to work in a basis where only one
family of singlets $S$ and Higgs doublets $H_d$ and $H_u$ have VEVs
and the remaining states do not (the non--Higgs).  Then REWSB makes it
natural that the so defined Higgs fields have large couplings to third
family quarks and leptons, while the non--Higgs fields have small
couplings.

We have analysed the RG flow of the gauge and
Yukawa couplings in the framework of the ESSM taking into account
kinetic term mixing between $U(1)_Y$ and $U(1)_{N}$. Imposing the gauge
coupling unification at high energies we have found that the gauge
coupling of the extra $U(1)_{N}$ is very close to the $U(1)_{Y}$ gauge
coupling while the off--diagonal gauge coupling which describes the mixing
between $U(1)_Y$ and $U(1)_{N}$ is negligibly small. Since by
construction extra exotic quarks and non--Higgses fill in complete $SU(5)$
representations the Grand Unification scale remains almost the same as
in the MSSM. At the same time the overall gauge coupling $g_0$ that
characterises gauge interactions above the scale $M_X$ is considerably
larger than in the MSSM: $g_0\simeq 1.21$. The increase of the gauge
couplings at the Grand Unification and intermediate scales in the ESSM
is caused by the extra supermultiplets of exotic matter.

The growth of $g_i(Q)$ relaxes the restrictions on the Yukawa
couplings coming from the validity of perturbation theory up to the
scale $M_X$ as compared with the MSSM. In particular
$h_t$ and $\lambda$ can take larger values at the EW scale
than in the constrained MSSM and NMSSM. If the top--quark Yukawa coupling
is large at the Grand Unification scale, i.e. $h_{t}(M_X)\gtrsim 1$,
the solutions of the RG equations for the Yukawa
couplings in the gaugeless limit approach the invariant line and along
this line are attracted to the quasi--fixed point where $\lambda(M_t)$
is going to zero. After the inclusion of gauge couplings, the valley
along which the solutions of RG equations flow to
the fixed point disappear but their convergence to the fixed point
becomes even stronger. The analysis of the RG flow
shows that the Yukawa couplings of top-- and exotic quarks tend to
dominate over the Yukawa couplings of non--Higgs supermultiplets that 
are considerably larger than $\lambda$ at the EW scale. 
However the solutions for the Yukawa couplings are concentrated near 
the fixed points only when $h_t(M_X)$ is large enough that corresponds 
to $\tan\beta\simeq 1-1.1$. At moderate and large values of $\tan\beta$ 
the values of the Yukawa couplings at the EW scale may be 
quite far from the stable fixed point. As a result for values of
$\tan\beta\gtrsim 1.5$ the coupling $\lambda(M_t)$ can be comparable 
with the top--quark Yukawa coupling.

We have used the above theoretical restrictions on $\tan\beta$ and
$\lambda$ for the analysis of the Higgs, neutralino and chargino
sectors of the ESSM. Although the particle content of the ESSM
involves many particles with similar quantum number only
one singlet field $S$ and two Higgs doublets $H_u$ and $H_d$ acquire
VEVs breaking the $SU(2)\times U(1)_Y\times U(1)_N$
symmetry.  Since the Higgs sector of the ESSM contains only one new
field $S$ and one additional parameter compared to the MSSM it can be
regarded as the simplest extension of the Higgs sector of the MSSM. As in
the MSSM, the ESSM Higgs sector does not provide extra
sources for the CP--violation at tree--level. The ESSM Higgs
spectrum includes three CP--even, one CP--odd and two charged
states. The singlet dominated CP--even state is always almost
degenerate with the $Z'$ gauge boson. The masses of another CP--even and
charged Higgs fields are set by the mass of pseudoscalar state
$m_A$. The lightest CP--even Higgs boson is confined around the
EW scale. The superpartners of the $Z'$ boson and singlet field
$S$ also contribute to the ESSM neutralino spectrum while the number
of states in the chargino sector remain the same as in the MSSM. The
masses of extra states in the neutralino sector are governed by
$M_{Z'}$.

The qualitative pattern of the Higgs, neutralino and chargino spectra
in the ESSM is determined by the Yukawa coupling $\lambda$. When
$\lambda<< g_1$ (the MSSM limit of the ESSM) new states in the Higgs
and neutralino sectors become very heavy and decouple from the rest of
the spectrum making them indistinguishable from the MSSM ones. In the
case when $\lambda\gtrsim g_1$ the lightest Higgs scalar can
be heavier than in the MSSM and NMSSM.  In this case the vacuum
stability requirement constrains $m_A$ so that the heaviest CP--even,
CP--odd and charged states lie beyond the $\mbox{TeV}$ range. It means
that in this case only the lightest Higgs scalar can be discovered at
the LHC and ILC. We have found that the mass of the lightest Higgs
particle does not exceed $150\,\mbox{GeV}$.  

If $\lambda\gtrsim g_1$ then the heaviest chargino and neutralino 
are formed by the neutral and charged superpartners of the Higgs doublets 
$H_u$ and $H_d$. Extra neutralino states are lighter than the heaviest 
one but still too heavy to be observed in the near future. The lightest 
chargino is predominantly superpartner of $W^{\pm}$ gauge bosons while the
lightest neutralino state is basically bino. We have obtained the
approximate solutions for the masses and couplings of the Higgs
particles as well as for the masses of the lightest neutralino and
chargino.

As we have already mentioned the ESSM predicts the existence of many
new exotic quarks and non--Higgsinos. They compose vector--like multiplets of
matter with respect to the SM gauge group, so that the axial couplings
of the SM gauge bosons to exotic particles vanish. As a consequence
their contributions to the EW observables measured at LEP are
suppressed by inverse powers of their masses.  The contribution of the
$Z'$ is also negligibly small since the latter is supposed to be very heavy
and practically does not mix with the $Z$ boson. At the same time
$Z'$, exotic quarks and non--Higgsinos can be produced directly at future
colliders if they are light enough.  The lifetime of new exotic
particles is defined by the extent to which the $Z_2^H$ symmetry is
broken. If $Z_2^{H}$ was exact the lightest exotic quark would be
absolutely stable. Since we have assumed that $Z_2^H$ is mainly broken
by the operators involving quarks and leptons of the third generation
the exotic quarks decay into either two heavy quarks $Q\bar Q$ or a heavy
quark and a lepton $Q\tau (\nu_{\tau})$, where $Q$ is either a $b$-- or 
$t$--quark.  If exotic quarks are light enough they will be intensively
produced at the LHC.  In the case when $Z_2^{H}$ is broken
significantly this results in the growth of the cross section of either
$pp\to Q\bar{Q}Q^{(')}\bar{Q}^{(')}+X$ or $pp\to Q\bar{Q}l^{+}l^{-}+X$, with
$l=e,\mu$.  If the
violation of the $Z_2^{H}$ invariance is extremely small then a set of new
baryons or composite leptons containing quasi--stable exotic quarks
could be discovered at the LHC. As compared with the exotic quarks
the production of non--Higgsinos will be rather suppressed at the
LHC. In contrast, at an ILC the production rates 
of exotic quarks and non--Higgsinos can be comparable allowing their
simultaneous observation. The $Z'$ gauge boson has to be detected at
the LHC if it has a mass below $4-4.5\,\mbox{TeV}$. 

The ESSM can in principle be derived from a rank--6 model which 
naturally arises after the breakdown of the $E_6$ symmetry via the Hosotani 
mechanism near the string or Grand Unification scale $M_X$. The discovery 
at future colliders of the exotic particles and extra $Z'$ boson 
predicted by the ESSM would therefore 
represent a possible indirect signature of an underlying 
$E_6$ gauge structure at high energies, and provide circumstantial
evidence for superstring theory.

\section*{Acknowledgements}

\vspace{-2mm} The authors are grateful to A.~Djouadi, J.~Kalinowski, D.~J.~Miller 
and P.~M.~Zerwas for valuable comments and remarks. RN would also like to thank 
E.~Boos, D.~I.~Kazakov and M.~I.~Vysotsky for fruitful discussions. The authors acknowledge
support from the PPARC grant PPA/G/S/2003/00096, the 
NATO grant PST.CLG.980066 and the EU network MRTN 2004--503369.


\newpage
\noindent
{\Large \bf Figure captions}
\vspace{5mm}

\noindent {\bf Fig.~ 1.}\, (a) RG flow of a set of points in the $(\lambda/h_t)$--$(\kappa/h_t)$ plane
in the gaugeless limit ($g_0=0$). (b) The running of $\left(\ds\frac{\lambda(\mu)}{h_t(\mu)}\right)$ versus
$\left(\ds\frac{\kappa(\mu)}{h_t(\mu)}\right)$. The gauge couplings
are included. In both cases the energy scale $\mu$ is varied
from $M_X$ to $M_t$, $h_t(M_X)=10$, while the Yukawa couplings of the exotic quark and non--Higgs supermultiplets of the first and second
generations are set to zero. Different trajectories correspond to different initial conditions for $\lambda$ and
$\kappa$ at the scale $M_X$.

\vspace{5mm}
\noindent {\bf Fig.~ 2.}\, (a) The allowed range of the parameter space in the $(\kappa/h_t)$--$(\lambda/h_t)$ plane
for $\kappa_1=\kappa_2=\lambda_1=\lambda_2=0$ and $\tan\beta=2$
where the couplings are evaluated at the top mass $\mu=M_t$.
(b) RG flow of $\left(\ds\frac{\kappa(\mu)}{h_t(\mu)}\right)$
versus $\left(\ds\frac{\lambda(\mu)}{h_t(\mu)}\right)$ for $\tan\beta=2$. The Yukawa couplings of the exotic quark and non--Higgs supermultiplets 
of the first and second generations are taken to be zero. Different trajectories correspond to different initial conditions at the EW scale.
The solutions of the RG equations flow from the left to the right when the RG scale changes
from $M_t$ to $M_X$. (c) Upper limit on $(\kappa/h_t)$ versus $(\lambda/h_t)$ for  $\kappa_1=\kappa_2=\kappa$, $\lambda_1=\lambda_2=0$ and
$\tan\beta=2$. (d) The allowed part of the parameter space in the $(\kappa/h_t)$--$(\lambda/h_t)$ plane for $\kappa_1=\kappa_2=\kappa$,
$\lambda_1(M_t)=\lambda_2(M_t)=\lambda(M_t)$ and $\tan\beta=2$.

\vspace{5mm}
\noindent {\bf Fig.~ 3.}\, Upper limit on $\lambda$ versus $\tan\beta$.

\vspace{5mm}
\noindent {\bf Fig.~ 4.}\, 
Higgs masses and couplings for
$\lambda(M_t)=0.794$, $\tan\beta=2$, $M_{Z'}=M_S=700\,\mbox{GeV}$ and
$X_t=\sqrt{6}M_S$.  
(a) The dependence of the lightest Higgs boson mass on
$m_A$.  
(b) One--loop masses of the CP--even Higgs bosons
versus $m_A$. Solid, dashed and dashed--dotted lines correspond to the
masses of the lightest, second lightest and heaviest Higgs scalars
respectively.  
(c) One--loop masses
of the CP--odd, heaviest CP--even and charged Higgs bosons versus
$m_A$. Dotted, dashed--dotted and solid lines correspond to the masses
of the charged, heaviest scalar and pseudoscalar states.
(d) Absolute values of the relative couplings $R_{ZZh_i}$ of
the Higgs scalars to $Z$ pairs. Solid, dashed and dashed--dotted curves
represent the dependence of the couplings of the lightest, second
lightest and heaviest CP--even Higgs states to Z pairs on $m_A$.  
(e) Absolute values of the relative couplings $R_{ZAh_i}$ of the CP--even
Higgs bosons to the Higgs pseudoscalar and $Z$ as a function of
$m_A$. The notations are the same as in Fig.~4c.

\vspace{5mm}
\noindent {\bf Fig.~ 5.}\, 
Higgs masses and couplings for
$\lambda(M_t)=0.3$, $\tan\beta=2$, $M_{Z'}=M_S=700\,\mbox{GeV}$ and
$X_t=\sqrt{6}M_S$.  
(a) The dependence of the lightest Higgs boson mass on $m_A$.
(b) One--loop masses of the Higgs scalars versus
$m_A$.  
(c) One--loop masses of the CP--odd,
heaviest CP--even and charged Higgs states versus $m_A$. 
(d) Absolute values of the relative couplings $R_{ZZh_i}$ of the
CP--even Higgs bosons to $Z$ pairs. 
(e) Absolute values of the relative
couplings $R_{ZAh_i}$ of the Higgs scalars to the Higgs pseudoscalar
and $Z$ as a function of $m_A$. 
The notations are the same as in Fig.~4.

\vspace{5mm}
\noindent {\bf Fig.~ 6.}\, (a) Different contributions to the tree--level upper bound on $m_{h_1}$ in the ESSM versus $\tan\beta$. Solid line
represents the tree--level theoretical restriction on the lightest Higgs boson mass in the MSSM: $M_Z|\cos 2\beta|$. Dash--dotted line is a
contribution of extra $U(1)_{N}$ $D$--term: $g^{'}_1 v\biggl|\tilde{Q}_1\cos^2\beta+\tilde{Q}_2\sin^2\beta\biggr|$. Dotted line is the maximum
possible contribution of the $F$--term corresponding to the SM singlet field $S$: $\ds\frac{\lambda}{\sqrt{2}}v\sin 2\beta$.
(b) tree--level upper bound on the lightest Higgs boson mass as a function of $\tan\beta$. The solid, lower and upper dotted lines correspond to the
theoretical restrictions on $m_{h_1}$ in the MSSM, NMSSM and ESSM respectively.

\vspace{5mm}
\noindent {\bf Fig.~ 7.}\, (a) One--loop upper bound on the lightest Higgs boson mass as a function of the Yukawa couplings in the
$(\lambda/h_t)$--$(\kappa/h_t)$ plane for $\kappa_1(M_t)=\kappa_2(M_t)=0$ and $\kappa_3(M_t)=\kappa$. (b) One--loop upper limit on $m_{h_1}$ versus
Yukawa couplings in the $(\lambda/h_t)$--$(\kappa/h_t)$ plane for $\kappa_1(M_t)=\kappa_2(M_t)=\kappa_3(M_t)=\kappa$. (c) Two--loop upper limit on
$m_{h_1}$ as a function of the Yukawa couplings in the $(\lambda/h_t)$--$(\kappa/h_t)$ plane for $\kappa_1(M_t)=\kappa_2(M_t)=0$ and
$\kappa_3(M_t)=\kappa$. (d) Two--loop upper bound on the mass of the lightest Higgs scalar versus Yukawa couplings in the $(\lambda/h_t)$--$(\kappa/h_t)$
plane for $\kappa_1(M_t)=\kappa_2(M_t)=\kappa_3(M_t)=\kappa$. Thick, solid, dash--dotted and dashed lines in Figs.~7a--7d correspond to
$m_h=160,\,150$, $140,\,130\,\mbox{GeV}$ respectively. The dotted line represent the allowed range of the parameter space for
$\lambda_1(M_t)=\lambda_2(M_t)=0$. Theoretical restrictions on the lightest Higgs boson mass are obtained for $\tan\beta=2$,
$m_Q^2=m_U^2=m_{Di}^2=m^2_{\overline{D}i}=M_S^2$, $X_t=\sqrt{6} M_S$ and $M_S=700\,\mbox{GeV}$.

\vspace{5mm}
\noindent {\bf Fig.~ 8.}\, (a) The dependence of the two--loop upper bound on the lightest Higgs boson mass on $\tan\beta$ for
$m_t(M_t)=165\,\mbox{GeV}$, $m_Q^2=m_U^2=M_S^2$, $X_t=\sqrt{6} M_S$ and $M_S=700\,\mbox{GeV}$. The solid, lower and upper dotted lines
represent the theoretical restrictions on $m_{h_1}$ in the MSSM, NMSSM and ESSM respectively. (b) Two--loop upper bound on the mass of
the lightest Higgs particle in the MSSM, NMSSM and ESSM versus $\tan\beta$ for $X_t=0$. Other parameters and notations are the same as in Fig.~8a.

\vspace{5mm}
\noindent {\bf Fig.~ 9.}\, (a) Neutralino spectrum in the ESSM as function of $M_1$ for $\lambda(M_t)=0.794$, $\tan\beta=2$ and
$M_{Z'}=700\,\mbox{GeV}$. (Only the absolute values of the neutralino masses have physical meaning.)
(b) Chargino masses versus $M_1$. The parameters are the same as in Fig.~9a.

\vspace{5mm}
\noindent {\bf Fig.~ 10.}\, (a) The dependence of neutralino masses in the ESSM on $M_1$ for $\lambda(M_t)=0.3$, $\tan\beta=2$ and
$M_{Z'}=700\,\mbox{GeV}$. (Only the absolute values of the neutralino masses have physical meaning.)
(b) Chargino spectrum as function of $M_1$. The set of parameters is the same as in Fig.~10a.

\vspace{5mm}
\noindent {\bf Fig.~ 11.}\, Differential cross section in the final state invariant mass, denoted by
$M_{l^+l^-}$, 
at the LHC for Drell-Yan production ($l=e$ or $\mu$ only) in presence of a $Z'$
with and without the (separate) contribution of exotic $D$--quarks or non--Higgsinos $\tilde{H}$ (both via EW interactions), with $\mu_{Di}=\mu_{Hi}=300$ GeV.
Here, $M_{Z'}=1.5$ TeV. 

\vspace{5mm}
\noindent {\bf Fig.~ 12.}\, Cross section at the LHC for pair production of exotic $D$--quarks (via QCD interactions)
as well as non--Higgsinos $\tilde{H}$ (via EW interactions), as a function of their (common) mass, denoted by
$M_F$. Here, $M_{Z'}=1.5$ TeV.

\vspace{5mm}
\noindent {\bf Fig.~ 13.}\, Differential cross section in the final state invariant mass, denoted by
$M_{FF}$, 
at the LHC for pair production of $b$--, $t$-- and exotic $D$--quarks (all via QCD interactions)
as well as non--Higgsinos $\tilde{H}$ (via EW interactions), with  $\mu_{Di}=\mu_{Hi}=300$ GeV. Note the rescaling
of the rates for the first and last process. Here, $M_{Z'}=1.5$ TeV.

\vspace{5mm}
\noindent {\bf Fig.~ 14.}\, Energy-dependent hadronic cross section at a future ILC in the SM with an additional $Z'$, with and without the (separate) contribution 
of exotic $D$--quarks or non--Higgsinos $\tilde{H}$ (both via EW interactions), with  $\mu_{Di}=\mu_{Hi}=300$ GeV.  Here, $M_{Z'}=1.5$ TeV. 

\vspace{5mm}
\noindent {\bf Fig.~ 15.}\, Cross section at a future ILC for pair production  of exotic $D$--quarks and of non--Higgsinos $\tilde{H}$
(both via EW interactions), as a function of their (common) mass, , denoted by
$M_F$, for two collider energies. Here, $M_{Z'}=1.5$ TeV.

\newpage
\vspace{0mm}
\hspace{-1cm}{\large $\ds\frac{\lambda(\mu)}{h_t(\mu)}$}\\
\begin{center}
{\hspace*{-20mm}\includegraphics[height=90mm,keepaspectratio=true]{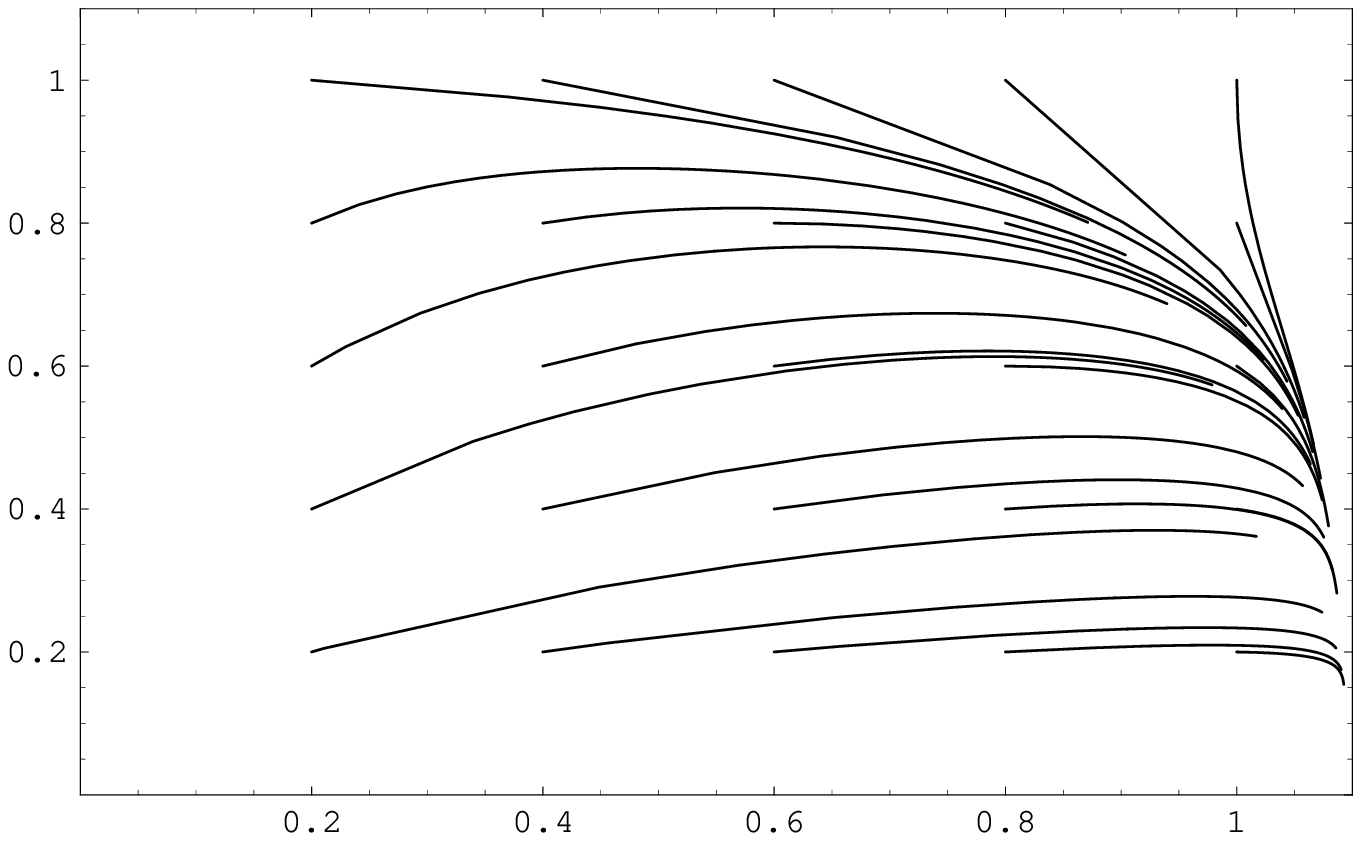}}\\
{\large $\kappa(\mu)/h_t(\mu)$}\\[2mm]
{\large\bfseries Fig.~1a}\\ \vspace{0cm}
\end{center}
\hspace{-1cm}{\large $\ds\frac{\lambda(\mu)}{h_t(\mu)}$}\\
\begin{center}
{\hspace*{-20mm}\includegraphics[height=90mm,keepaspectratio=true]{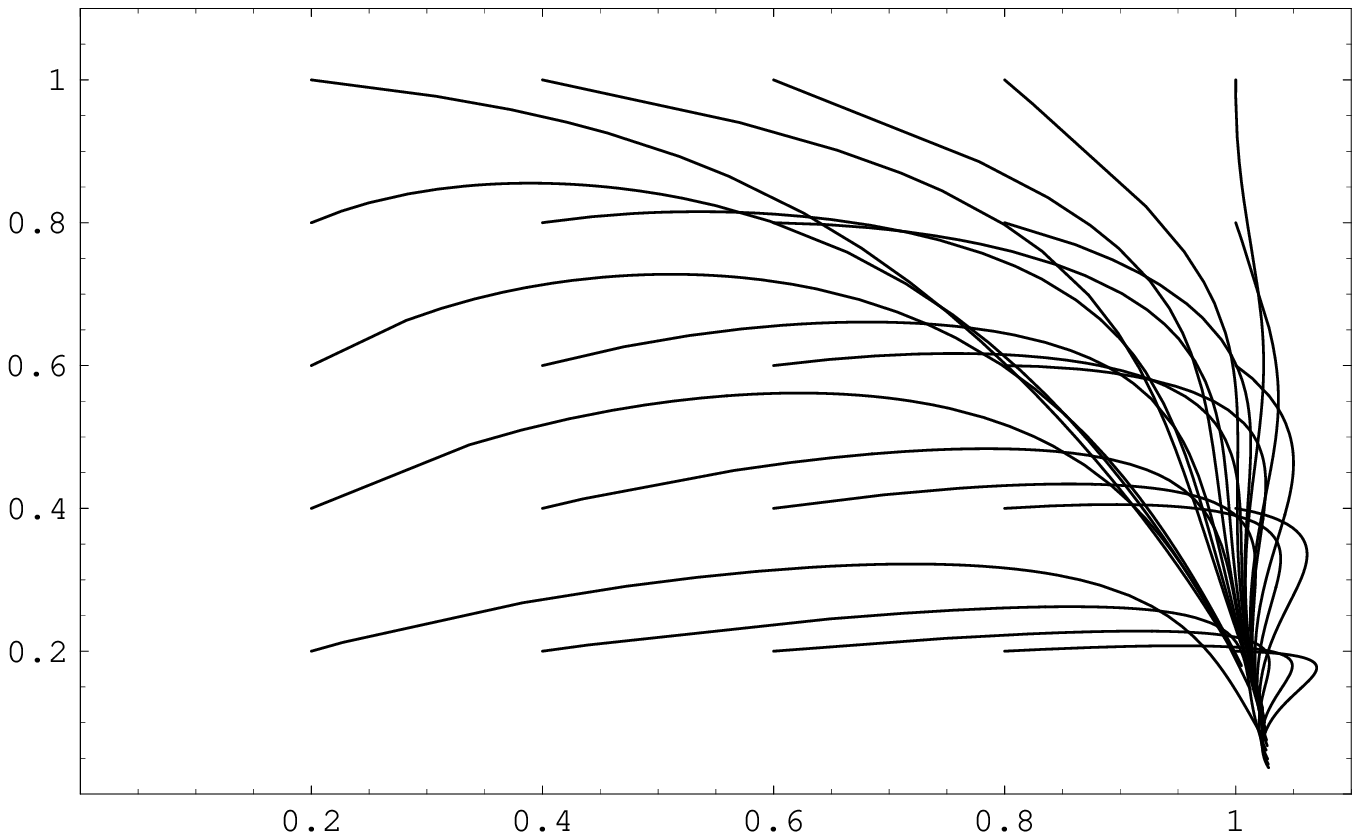}}\\
{\large $\kappa(\mu)/h_t(\mu)$}\\[2mm]
{\large\bfseries Fig.~1b}
\end{center}

\newpage
\vspace{0mm}
\hspace{-1cm}{\large $\ds\frac{\kappa(M_t)}{h_t(M_t)}$}\\
\begin{center}
{\hspace*{-20mm}\includegraphics[height=90mm,keepaspectratio=true]{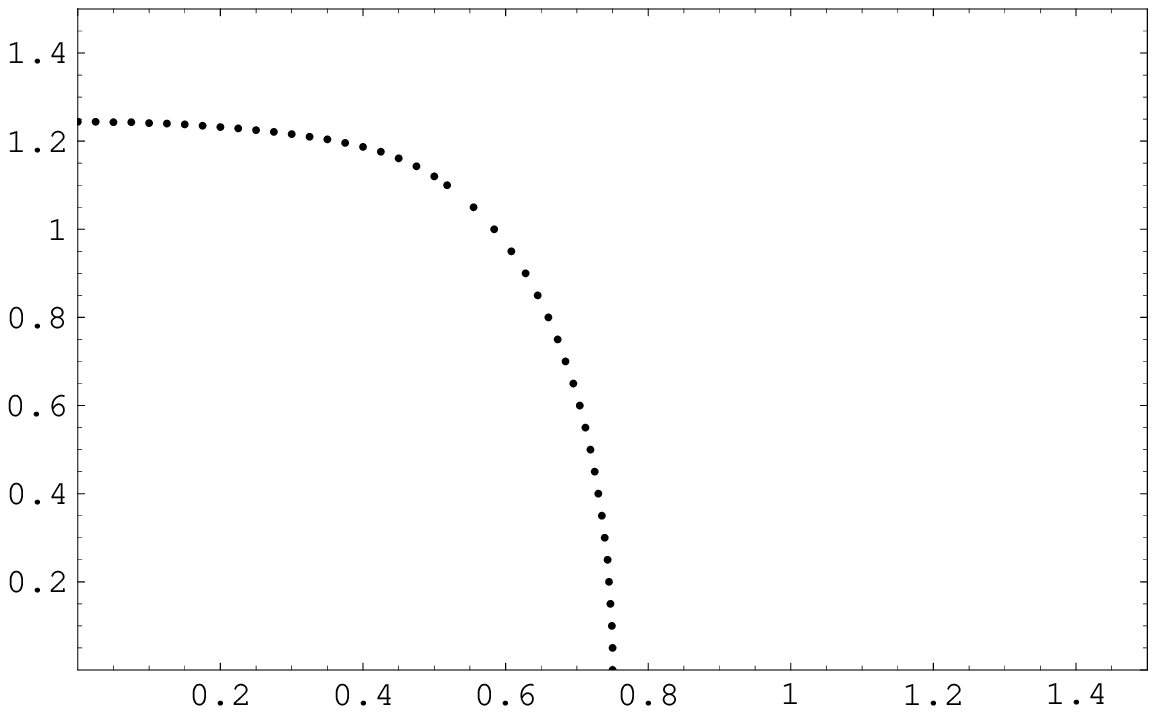}}\\
{\large $\lambda(M_t)/h_t(M_t)$}\\[2mm]
{\large\bfseries Fig.~2a}\\ \vspace{0cm}
\end{center}
\hspace{-1cm}{\large $\ds\frac{\kappa(\mu)}{h_t(\mu)}$}\\
\begin{center}
{\hspace*{-20mm}\includegraphics[height=90mm,keepaspectratio=true]{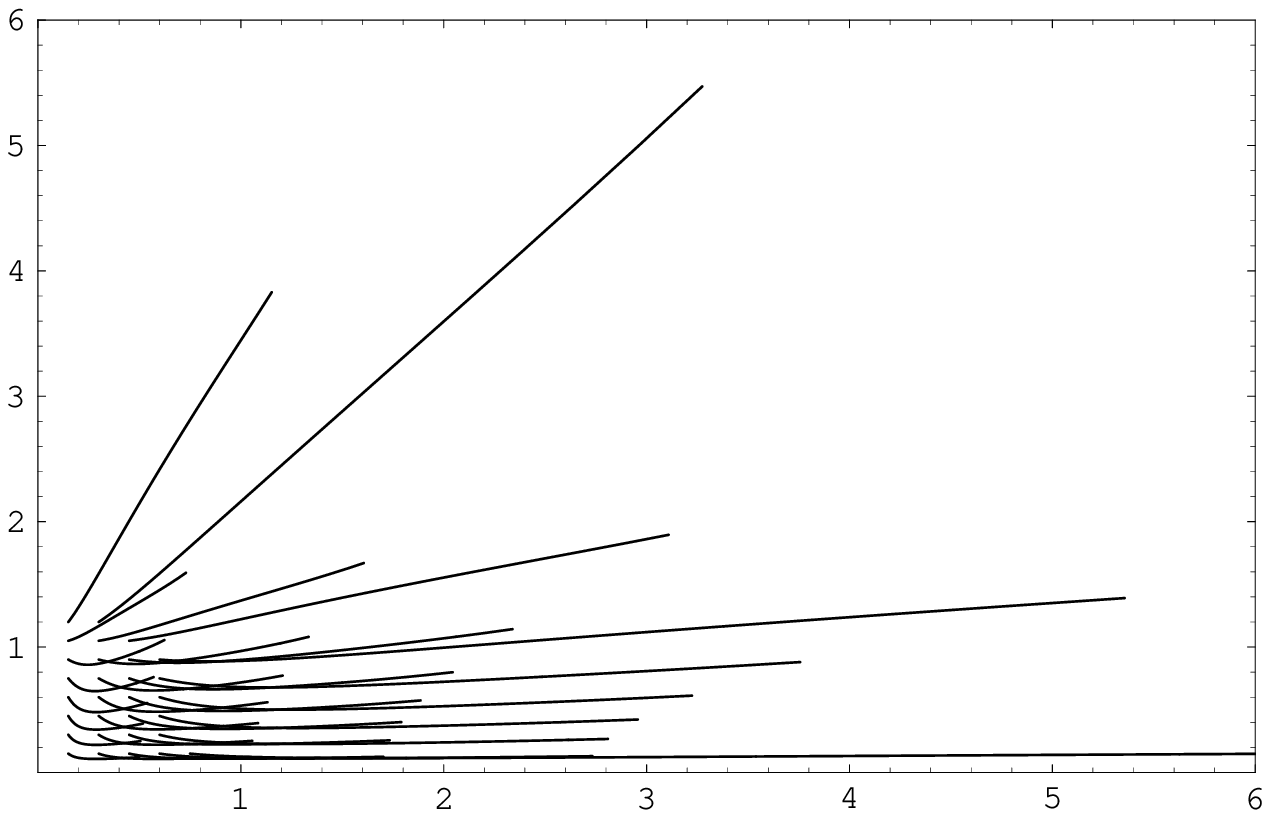}}\\
{\large $\lambda(\mu)/h_t(\mu)$}\\[2mm]
{\large\bfseries Fig.~2b}
\end{center}

\newpage
\vspace{0mm}
\hspace{-1cm}{\large $\ds\frac{\kappa(M_t)}{h_t(M_t)}$}\\
\begin{center}
{\hspace*{-20mm}\includegraphics[height=90mm,keepaspectratio=true]{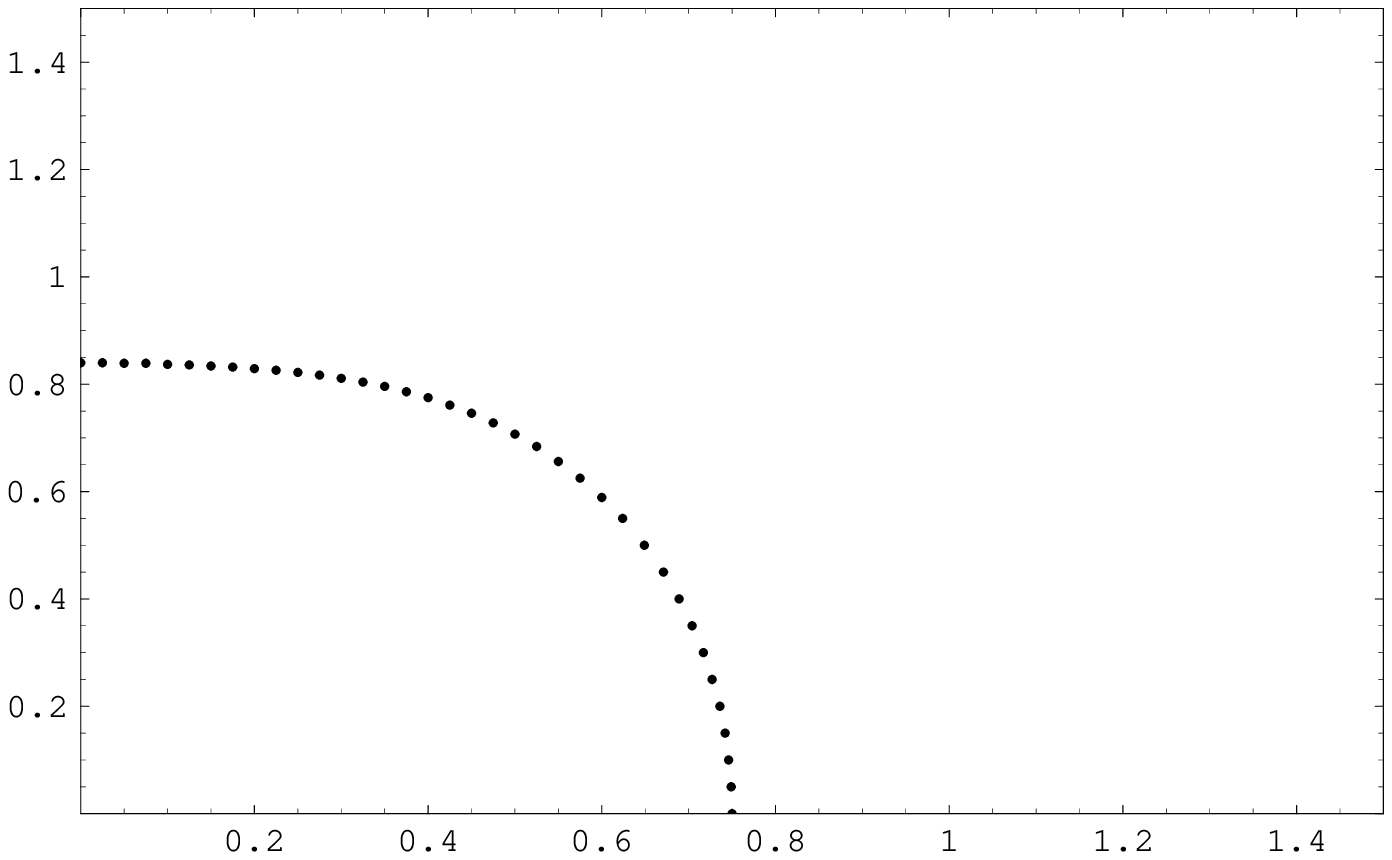}}\\
{\large $\lambda(M_t)/h_t(M_t)$}\\[2mm]
{\large\bfseries Fig.~2c}\\ \vspace{0cm}
\end{center}
\hspace{-1cm}{\large $\ds\frac{\kappa(M_t)}{h_t(M_t)}$}\\
\begin{center}
{\hspace*{-20mm}\includegraphics[height=90mm,keepaspectratio=true]{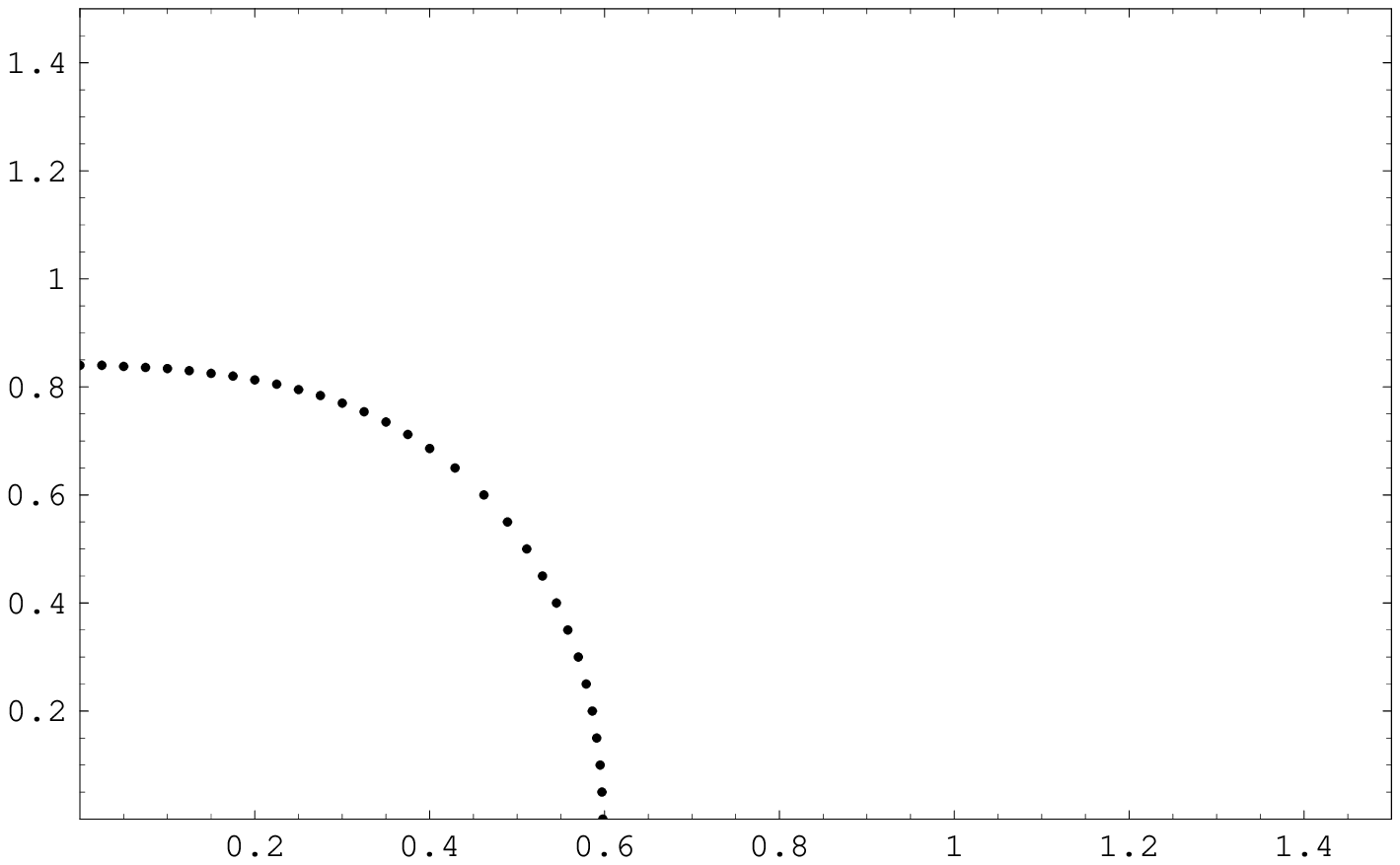}}\\
{\large $\lambda(M_t)/h_t(M_t)$}\\[2mm]
{\large\bfseries Fig.~2d}
\end{center}

\newpage
\vspace{2mm}
\hspace{-1cm}{\large $\lambda_{max}$}\\
\begin{center}
{\hspace*{-20mm}\includegraphics[height=100mm,keepaspectratio=true]{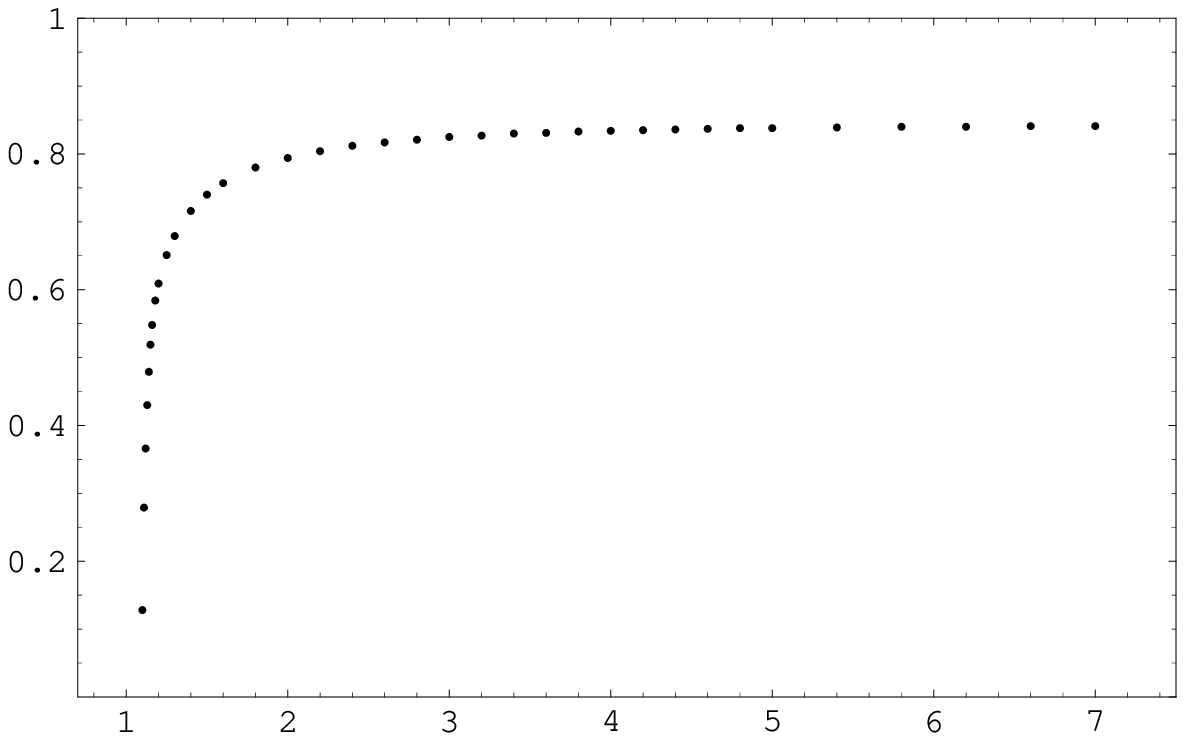}}\\
{\large $\tan\beta$}\\[2mm]
{\large\bfseries Fig.~3}
\end{center}

\newpage
\vspace{0mm}
\hspace{-1cm}{\large $m_{h_1}$}
\begin{center}
{\hspace*{-20mm}\includegraphics[height=90mm,keepaspectratio=true]{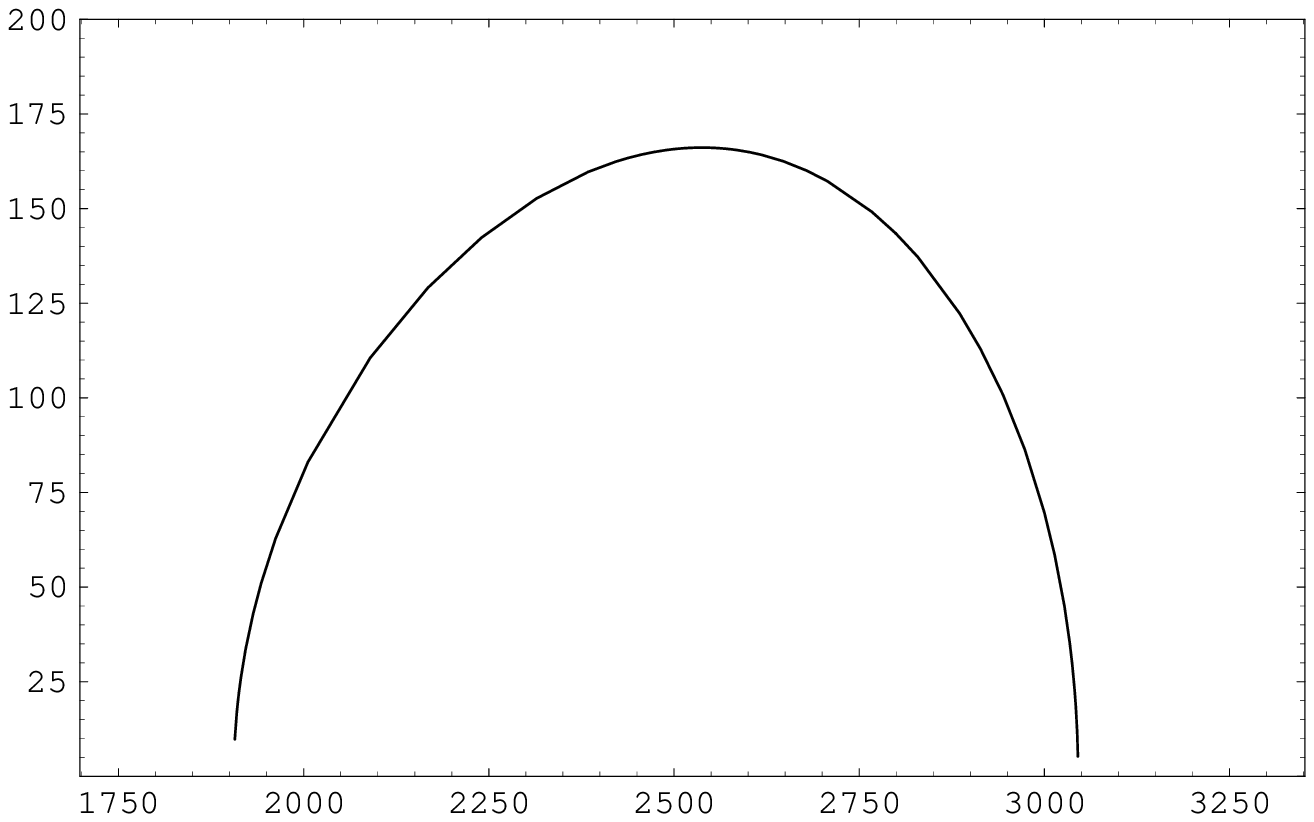}}\\
{\large $m_A$}\\[2mm]
{\large\bfseries Fig.~4a}
\end{center}
\hspace{-0.5cm}{\large $m_{h_i}$}
\begin{center}
{\hspace*{-20mm}\includegraphics[height=90mm,keepaspectratio=true]{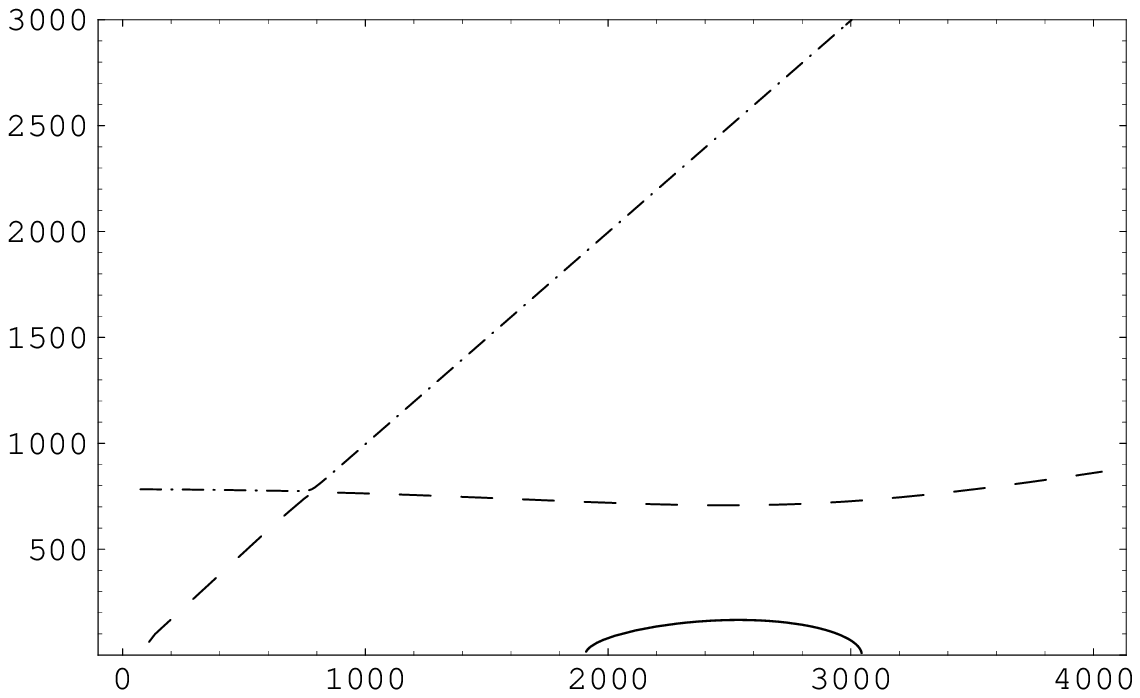}}\\
{\large $m_A$}\\[2mm]
{\large\bfseries Fig.~4b}\\ \vspace{0cm}
\end{center}

\newpage
\vspace{2mm}
\hspace{-1cm}{\large $m_{A,\,H^{\pm},\,h_3}$}\\
\begin{center}
{\hspace*{-20mm}\includegraphics[height=90mm,keepaspectratio=true]{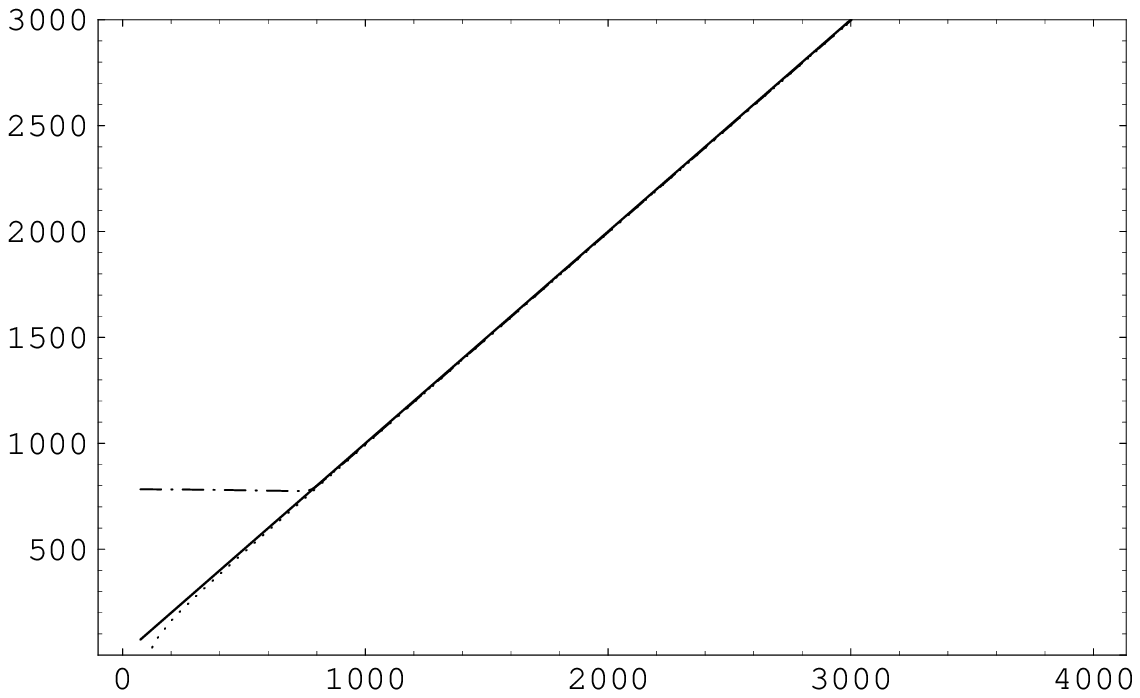}}\\
{\large $m_A$}\\[2mm]
{\large\bfseries Fig.~4c}
\end{center}

\newpage
\vspace{0mm}
\hspace{-1cm}{\large $|R_{ZZi}|$}
\begin{center}
{\hspace*{-20mm}\includegraphics[height=90mm,keepaspectratio=true]{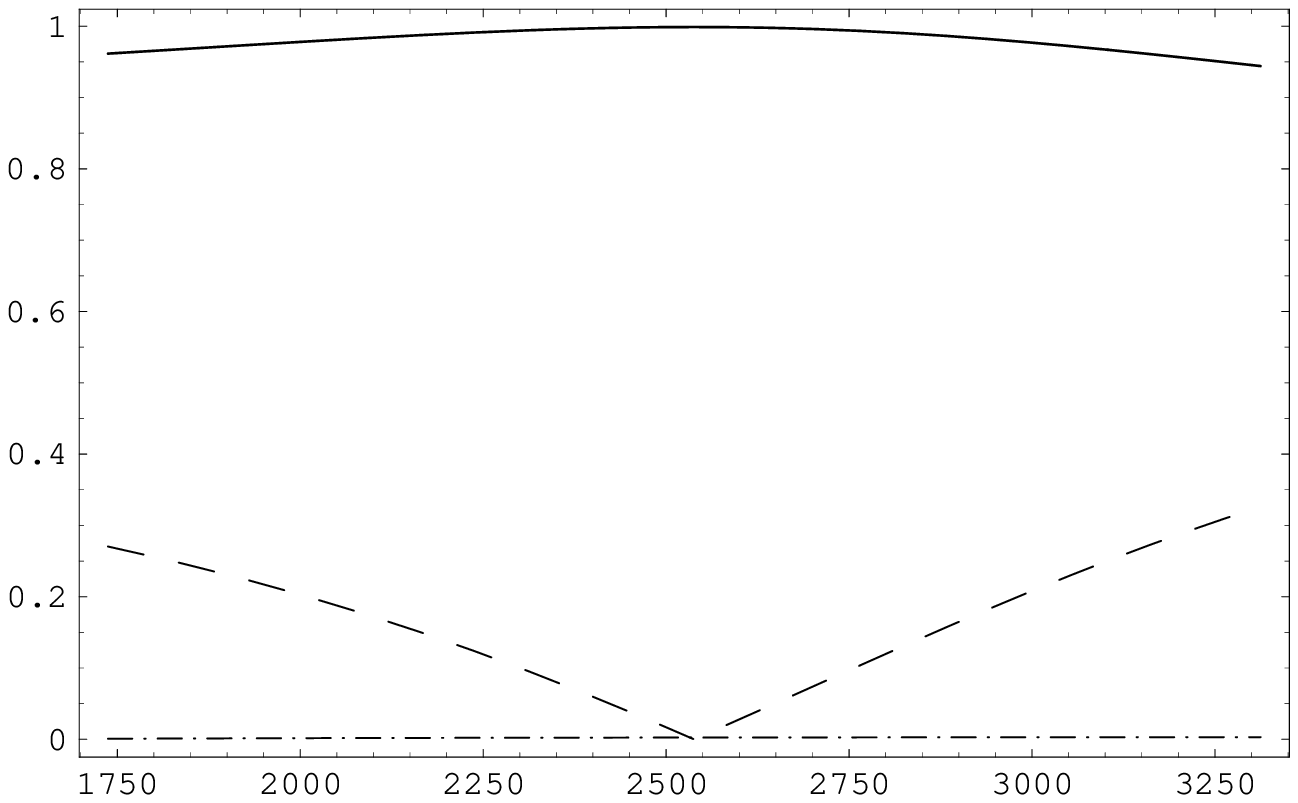}}\\
{\large $m_A$}\\[2mm]
{\large\bfseries Fig.~4d}\\ \vspace{0cm}
\end{center}
\hspace{-0.5cm}{\large $|R_{ZAi}|$}
\begin{center}
{\hspace*{-20mm}\includegraphics[height=90mm,keepaspectratio=true]{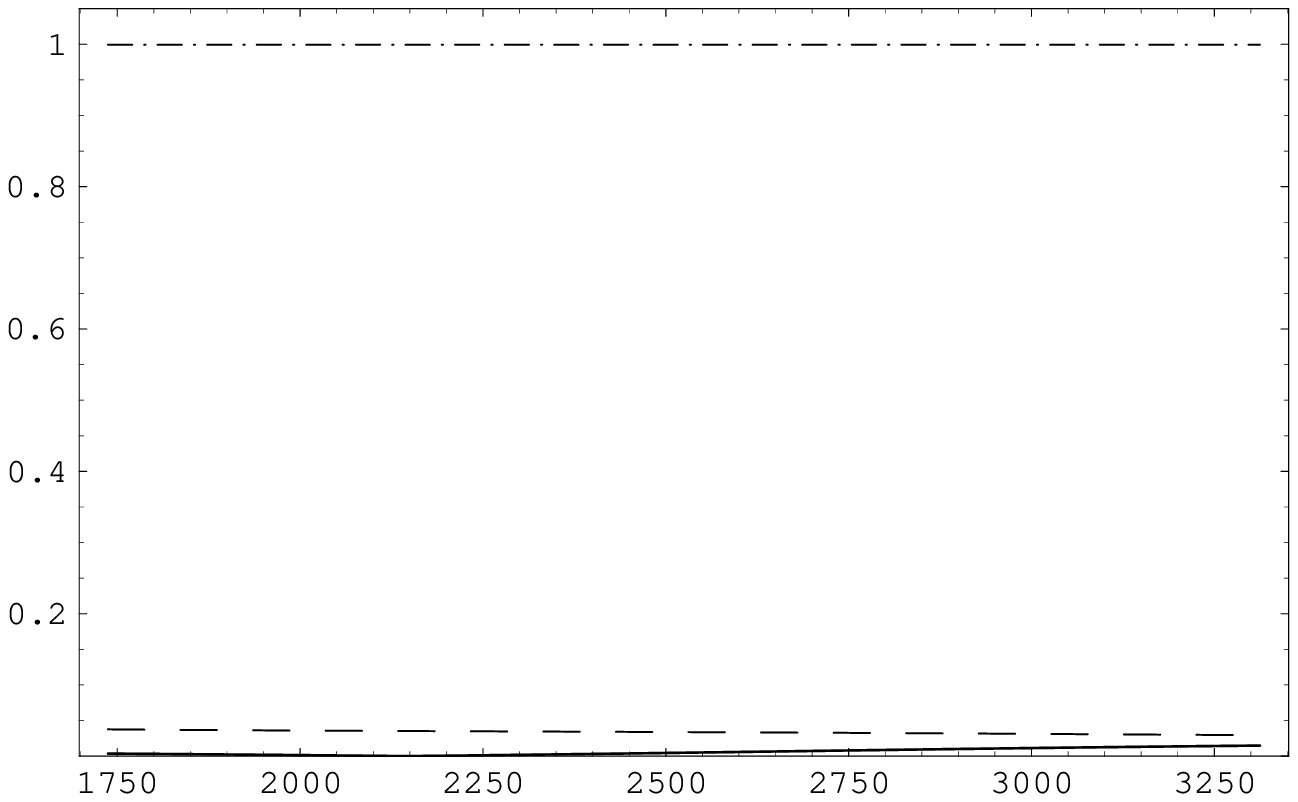}}\\
{\large $m_A$}\\[2mm]
{\large\bfseries Fig.~4e}
\end{center}

\newpage
\vspace{0mm}
\hspace{-1cm}{\large $m_{h_1}$}
\begin{center}
{\hspace*{-20mm}\includegraphics[height=90mm,keepaspectratio=true]{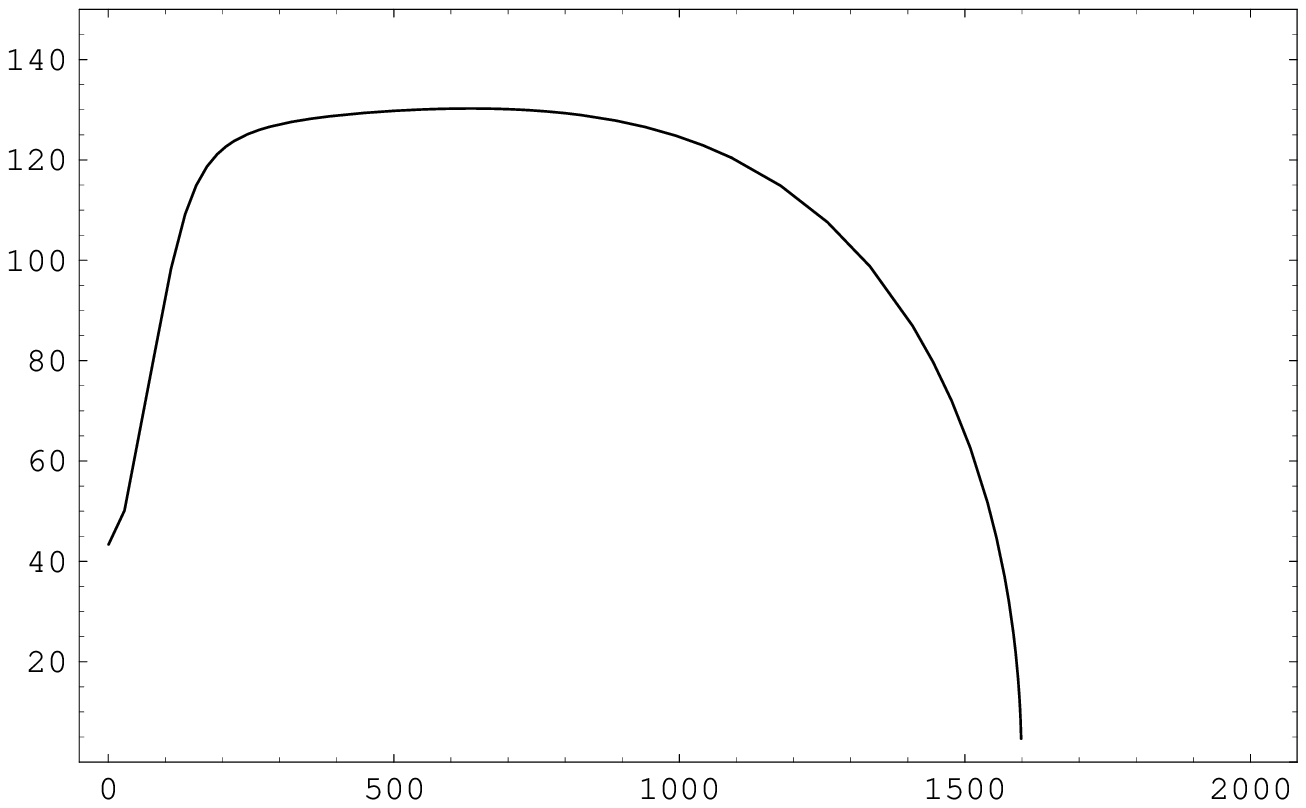}}\\
{\large $m_A$}\\[2mm]
{\large\bfseries Fig.~5b}
\end{center}
\hspace{-0.5cm}{\large $m_{h_i}$}
\begin{center}
{\hspace*{-20mm}\includegraphics[height=90mm,keepaspectratio=true]{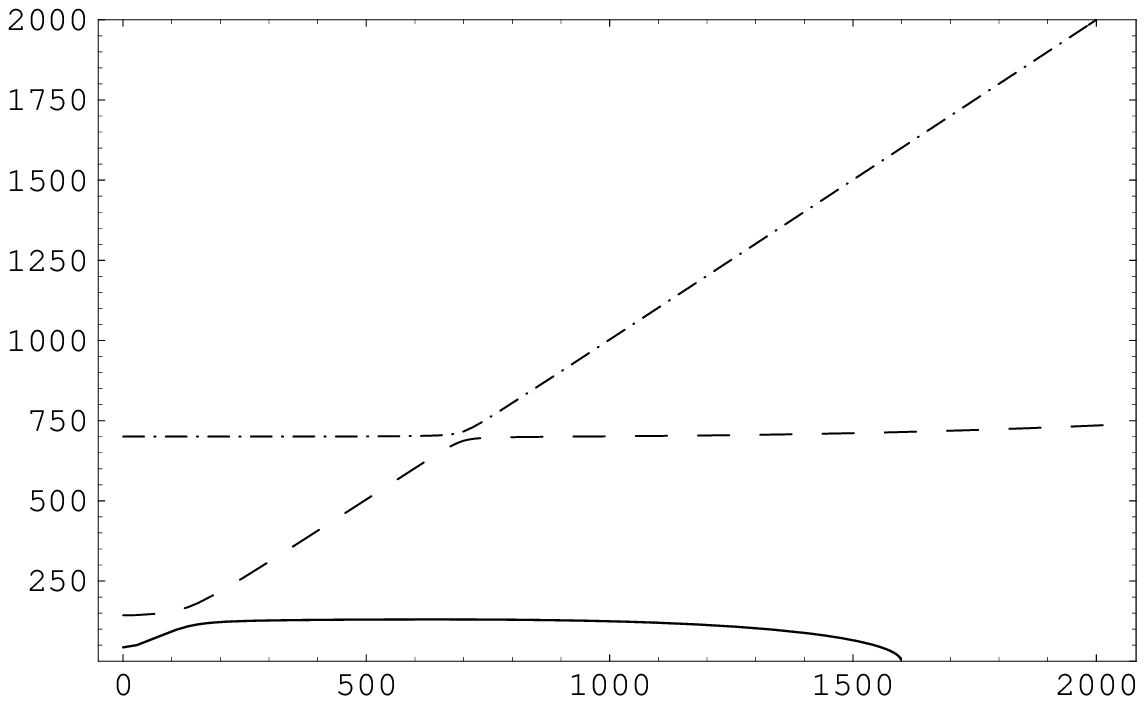}}\\
{\large $m_A$}\\[2mm]
{\large\bfseries Fig.~5a}\\ \vspace{0cm}
\end{center}

\newpage
\vspace{2mm}
\hspace{-1cm}{\large $m_{A,\,H^{\pm},\,h_3}$}\\
\begin{center}
{\hspace*{-20mm}\includegraphics[height=90mm,keepaspectratio=true]{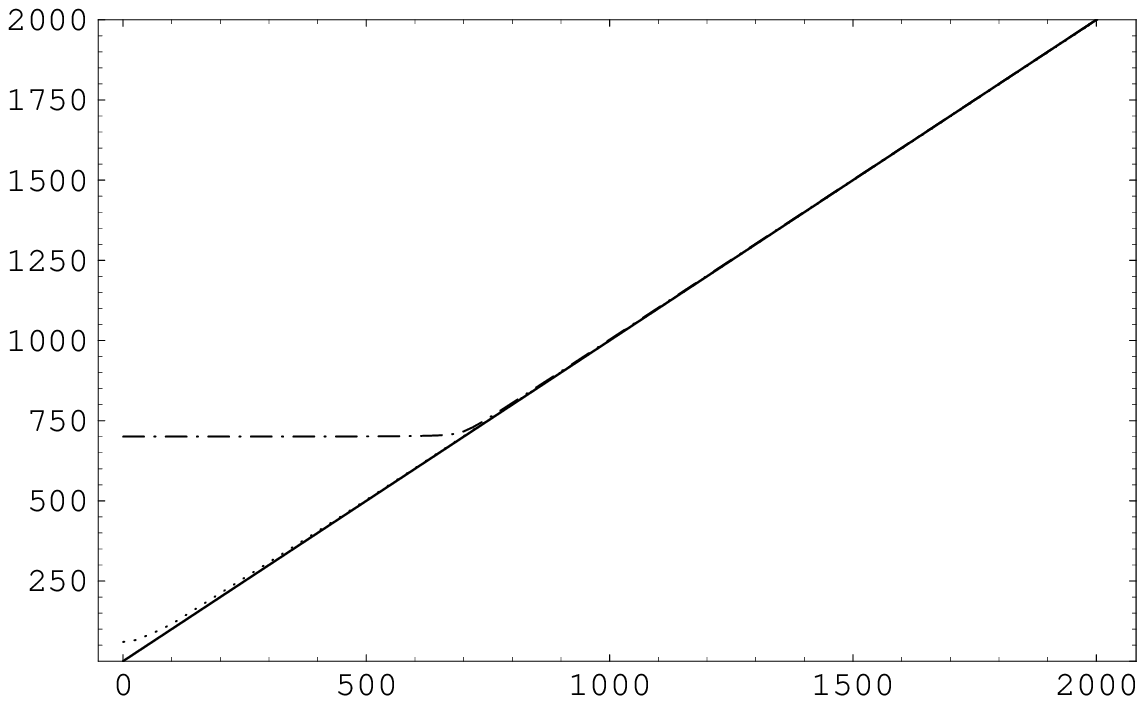}}\\
{\large $m_A$}\\[2mm]
{\large\bfseries Fig.~5c}
\end{center}

\newpage
\vspace{0mm}
\hspace{-1cm}{\large $|R_{ZZi}|$}
\begin{center}
{\hspace*{-20mm}\includegraphics[height=90mm,keepaspectratio=true]{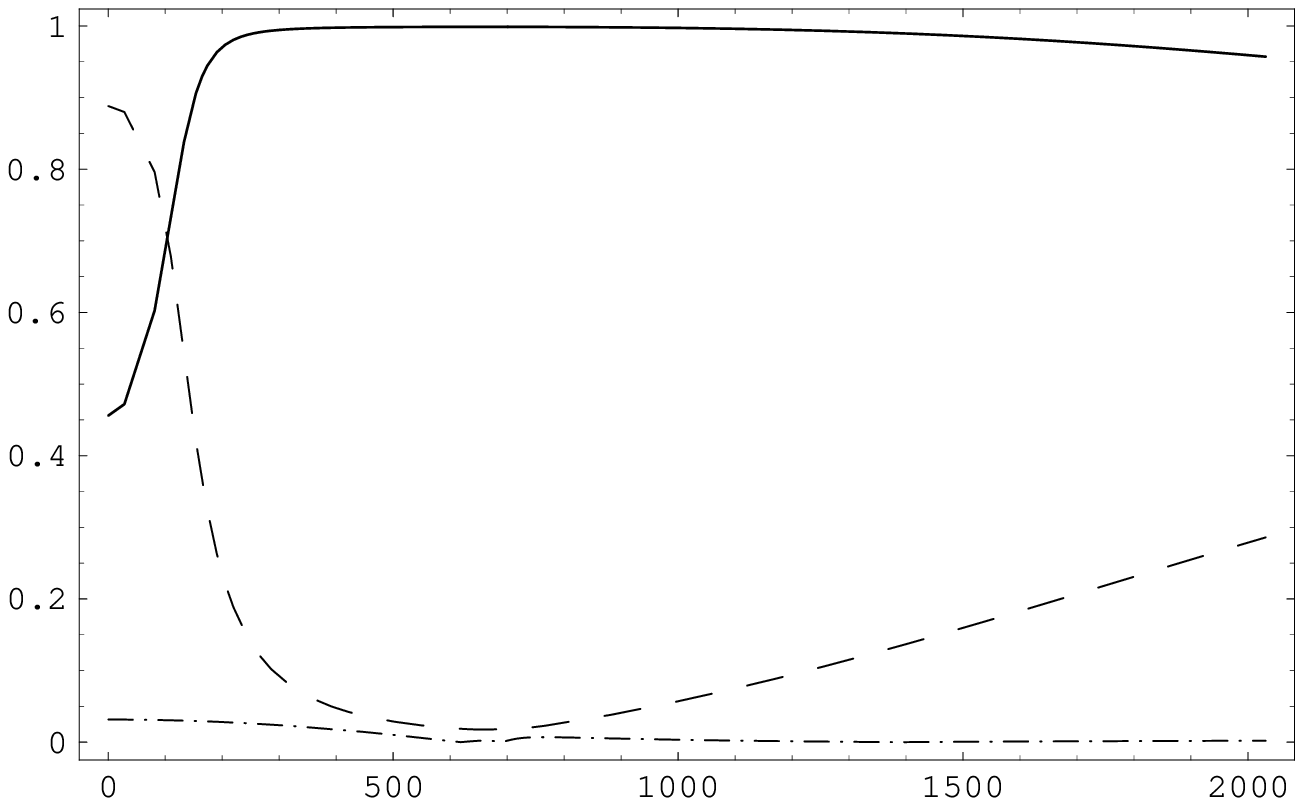}}\\
{\large $m_A$}\\[2mm]
{\large\bfseries Fig.~5d}\\ \vspace{0cm}
\end{center}
\hspace{-0.5cm}{\large $|R_{ZAi}|$}
\begin{center}
{\hspace*{-20mm}\includegraphics[height=90mm,keepaspectratio=true]{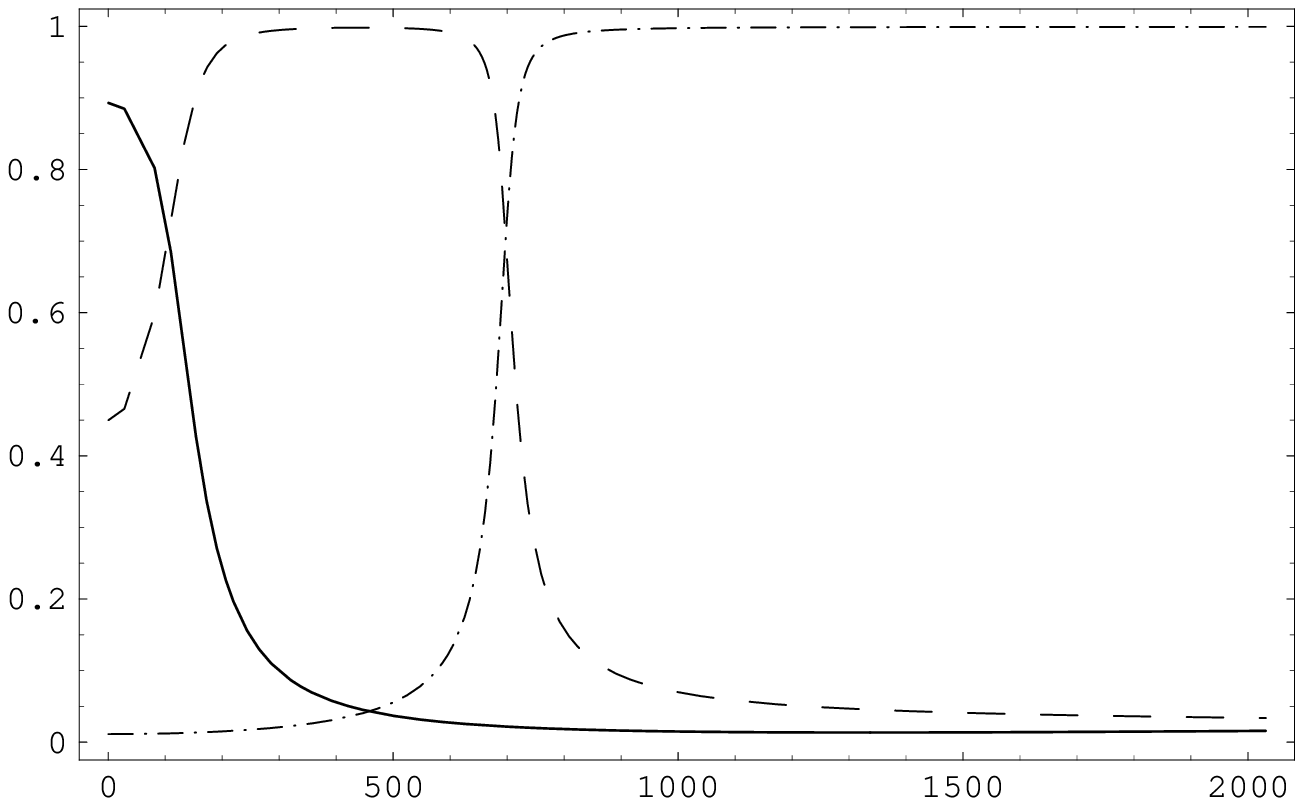}}\\
{\large $m_A$}\\[2mm]
{\large\bfseries Fig.~5e}
\end{center}

\newpage
\vspace{0mm}
\hspace{-1cm}{\large $m_{h_1}$}
\begin{center}
{\hspace*{-20mm}\includegraphics[height=90mm,keepaspectratio=true]{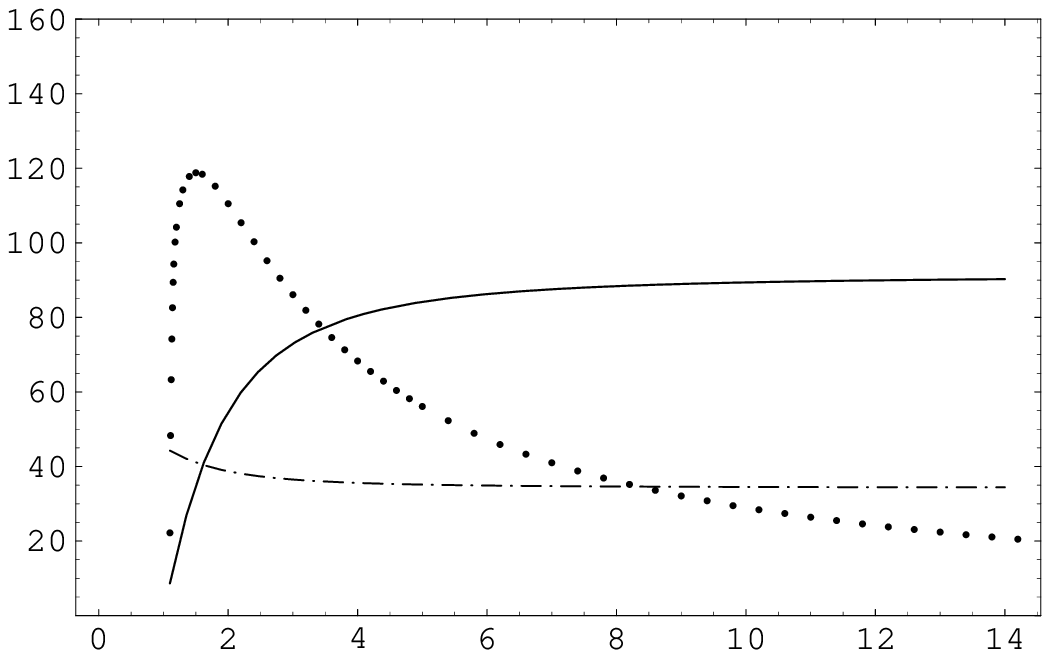}}\\
{\large $\tan\beta$}\\[2mm]
{\large\bfseries Fig.~6a}\\ \vspace{0cm}
\end{center}
\hspace{-0.5cm}{\large $m_{h_1}$}
\begin{center}
{\hspace*{-20mm}\includegraphics[height=90mm,keepaspectratio=true]{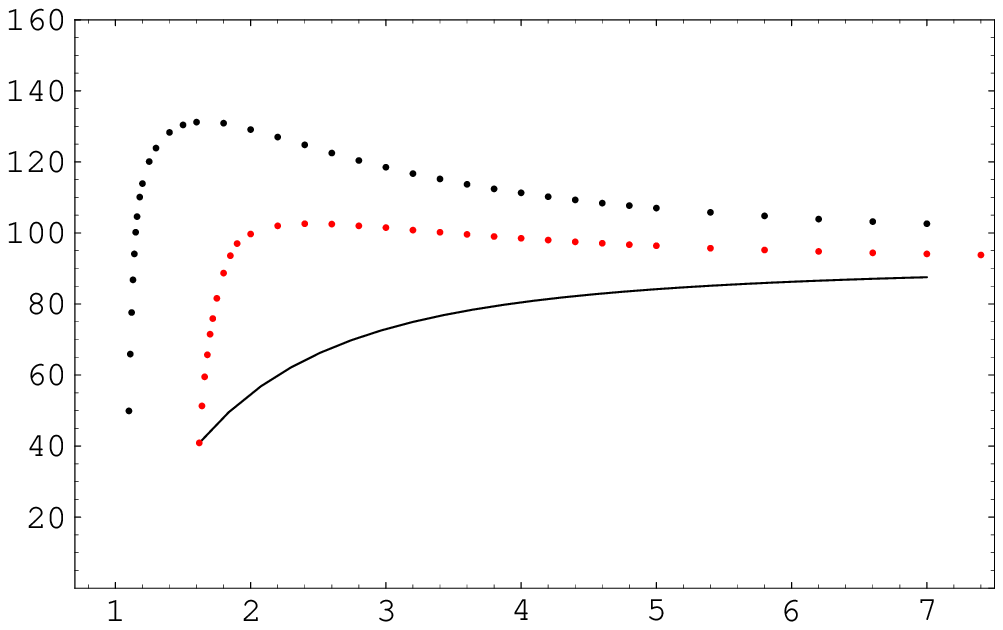}}\\
{\large $\tan\beta$}\\[2mm]
{\large\bfseries Fig.~6b}
\end{center}

\newpage
\vspace{0mm}
\hspace{-1cm}{\large $\ds\frac{\lambda(M_t)}{h_t(M_t)}$}
\begin{center}
{\hspace*{-20mm}\includegraphics[height=90mm,keepaspectratio=true]{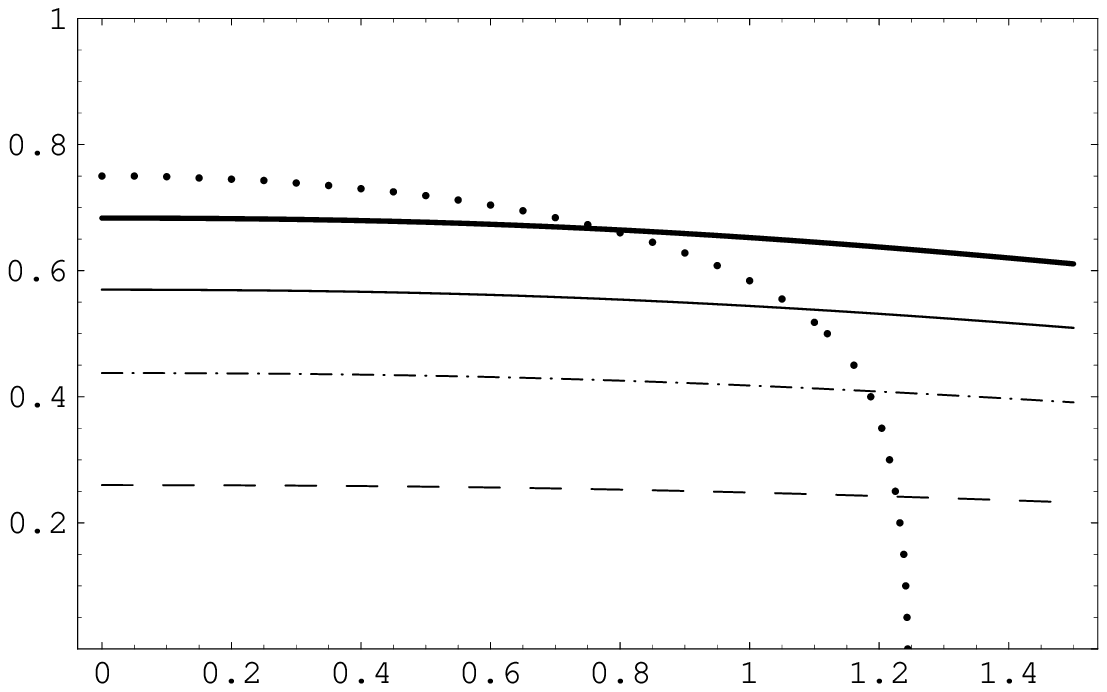}}\\
{\large $\kappa(M_t)/h_t(M_t)$}\\[2mm]
{\large\bfseries Fig.~7a}\\ \vspace{0cm}
\end{center}
\hspace{-0.5cm}{\large $\ds\frac{\lambda(M_t)}{h_t(M_t)}$}
\begin{center}
{\hspace*{-20mm}\includegraphics[height=90mm,keepaspectratio=true]{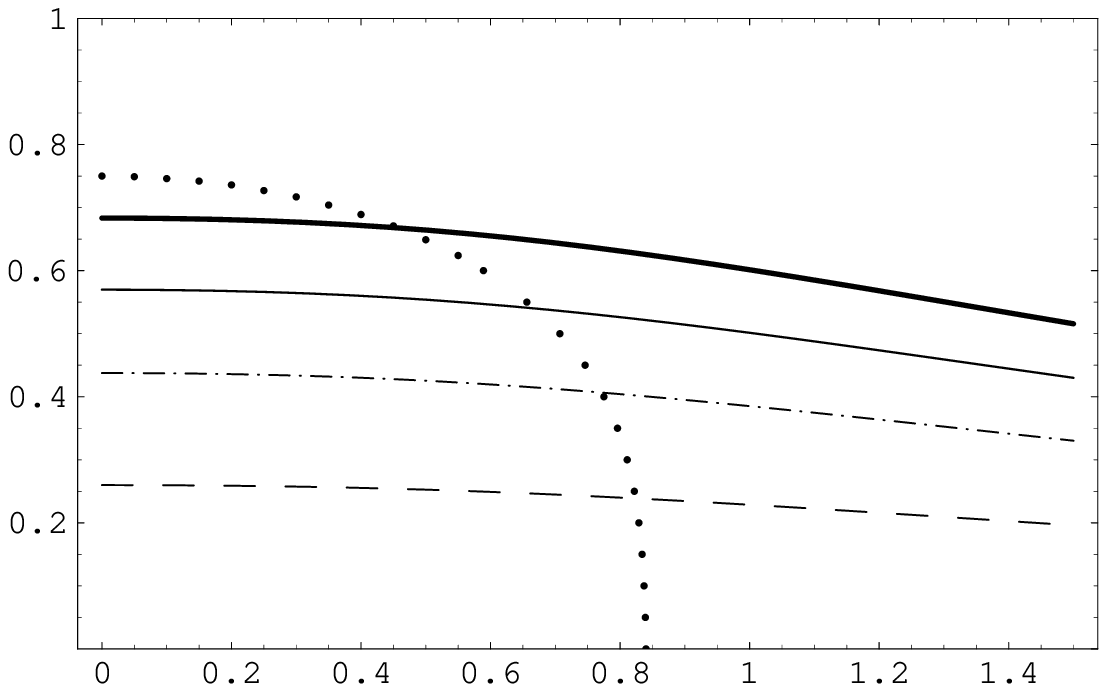}}\\
{\large $\kappa(M_t)/h_t(M_t)$}\\[2mm]
{\large\bfseries Fig.~7b}
\end{center}

\newpage
\vspace{0mm}
\hspace{-1cm}{\large $\ds\frac{\lambda(M_t)}{h_t(M_t)}$}
\begin{center}
{\hspace*{-20mm}\includegraphics[height=90mm,keepaspectratio=true]{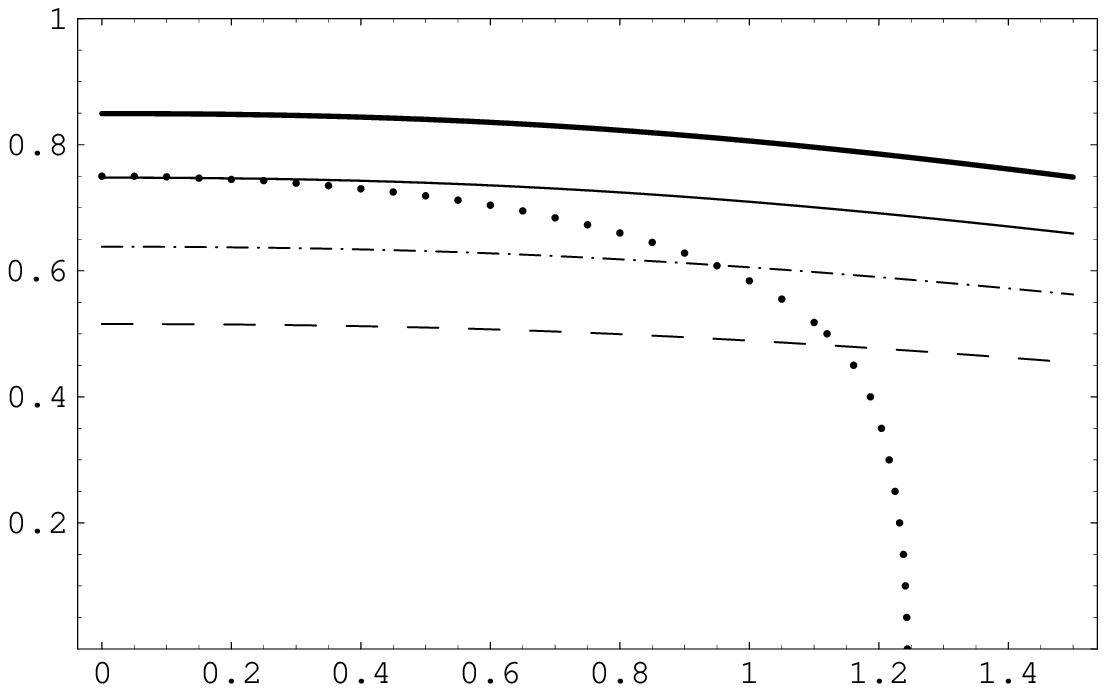}}\\
{\large $\kappa(M_t)/h_t(M_t)$}\\[2mm]
{\large\bfseries Fig.~7c}\\ \vspace{0cm}
\end{center}
\hspace{-0.5cm}{\large $\ds\frac{\lambda(M_t)}{h_t(M_t)}$}
\begin{center}
{\hspace*{-20mm}\includegraphics[height=90mm,keepaspectratio=true]{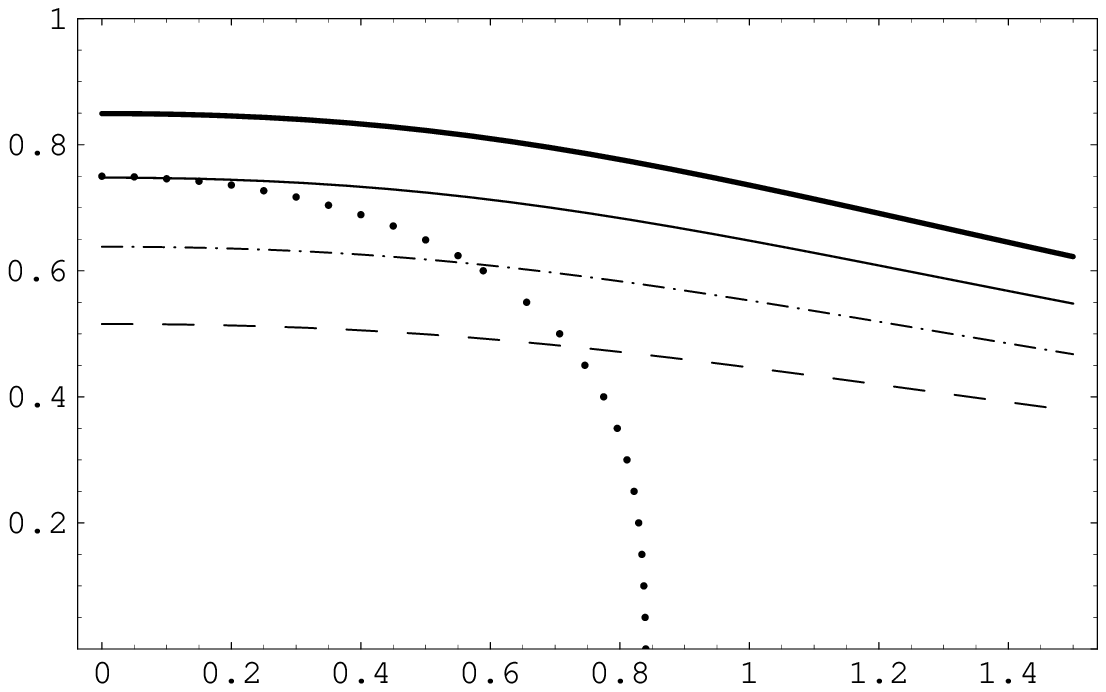}}\\
{\large $\kappa(M_t)/h_t(M_t)$}\\[2mm]
{\large\bfseries Fig.~7d}
\end{center}

\newpage
\vspace{0mm}
\hspace{-1cm}{\large $m_{h_1}$}
\begin{center}
{\hspace*{-20mm}\includegraphics[height=90mm,keepaspectratio=true]{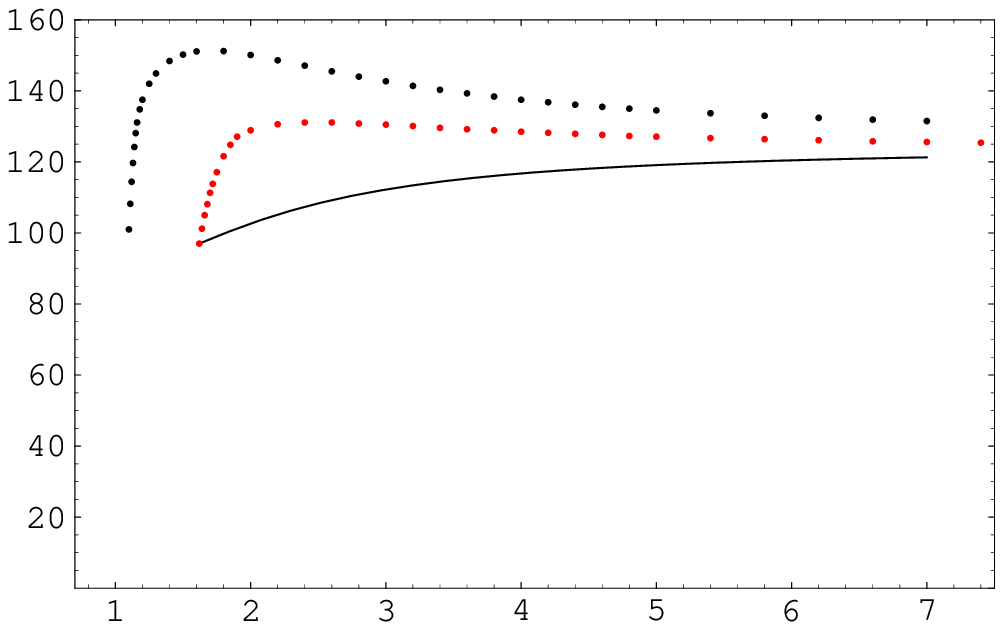}}\\
{\large $\tan\beta$}\\[2mm]
{\large\bfseries Fig.~8a}\\ \vspace{0cm}
\end{center}
\hspace{-0.5cm}{\large $m_{h_1}$}
\begin{center}
{\hspace*{-20mm}\includegraphics[height=90mm,keepaspectratio=true]{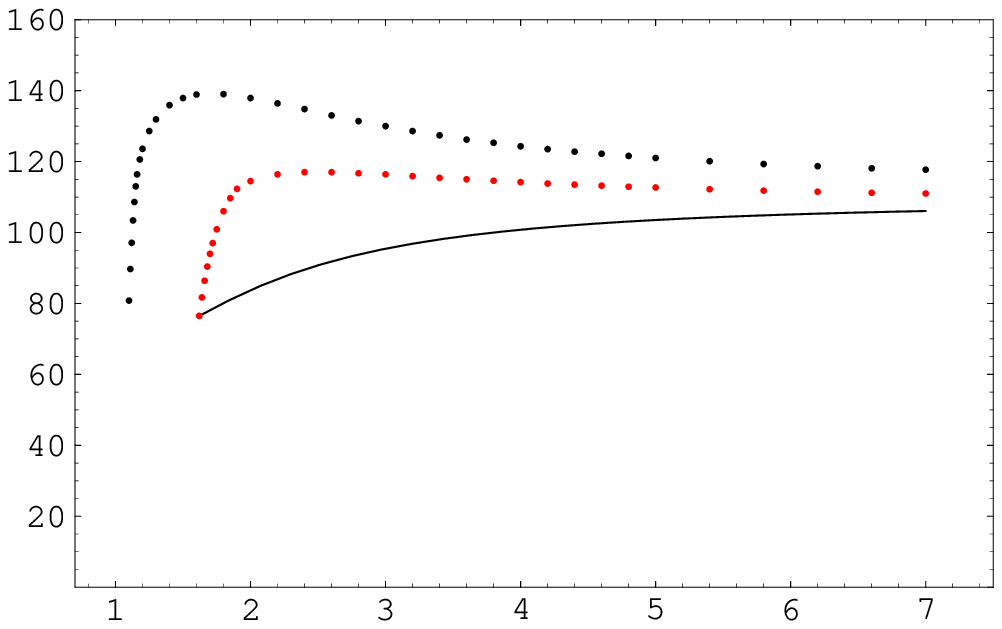}}\\
{\large $\tan\beta$}\\[2mm]
{\large\bfseries Fig.~8b}
\end{center}

\newpage
\vspace{0mm}
\hspace{-1cm}{\large $m_{\chi_i}$}
\begin{center}
{\hspace*{-20mm}\includegraphics[height=90mm,keepaspectratio=true]{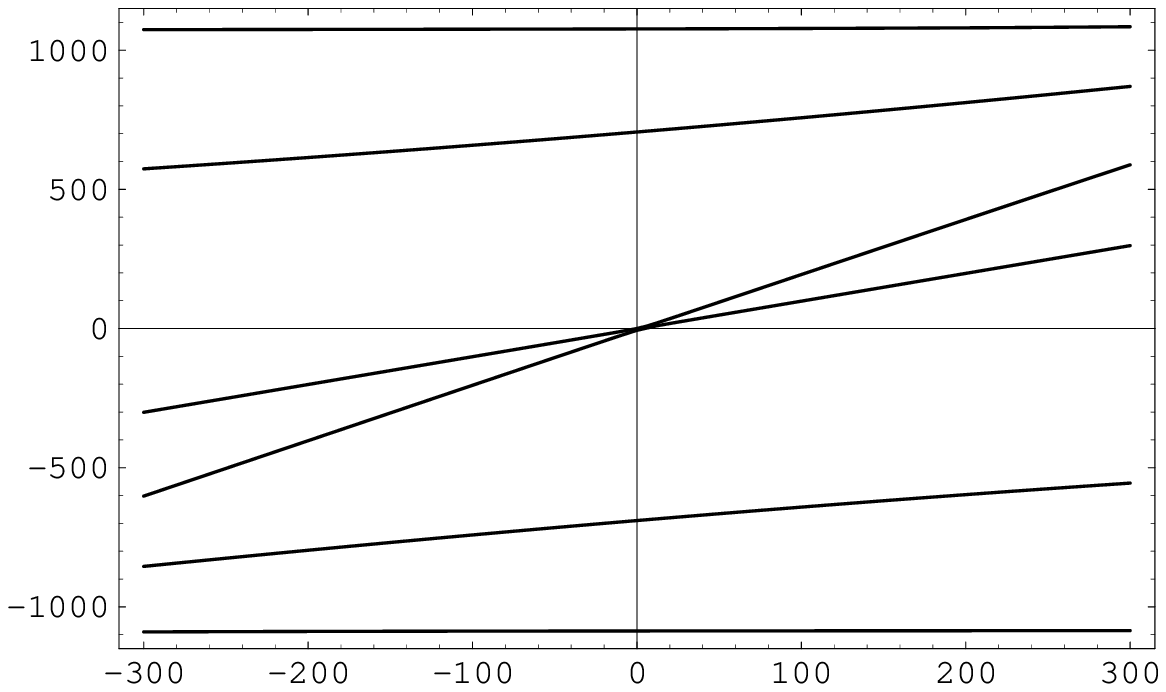}}\\
{\large $M_1$}\\[2mm]
{\large\bfseries Fig.~9a}\\ \vspace{0cm}
\end{center}
\hspace{-0.5cm}{\large $m_{\chi^{\pm}}$}
\begin{center}
{\hspace*{-20mm}\includegraphics[height=90mm,keepaspectratio=true]{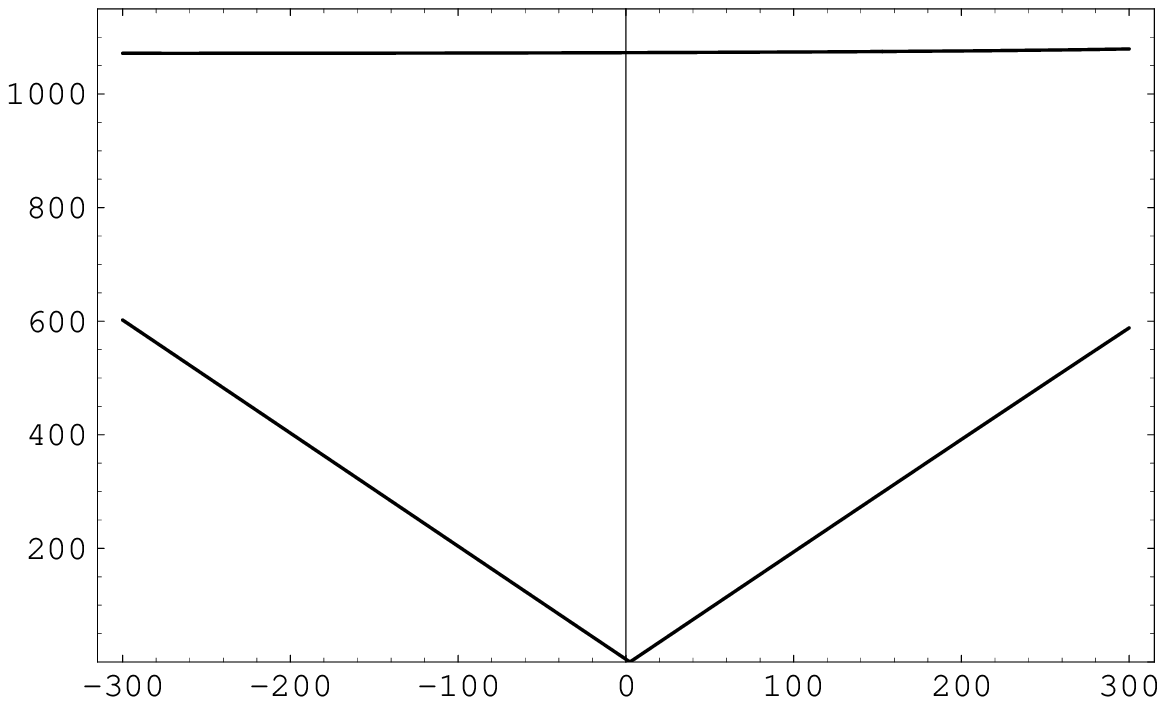}}\\
{\large $M_1$}\\[2mm]
{\large\bfseries Fig.~9b}
\end{center}

\newpage
\vspace{0mm}
\hspace{-1cm}{\large $m_{\chi_i}$}
\begin{center}
{\hspace*{-20mm}\includegraphics[height=90mm,keepaspectratio=true]{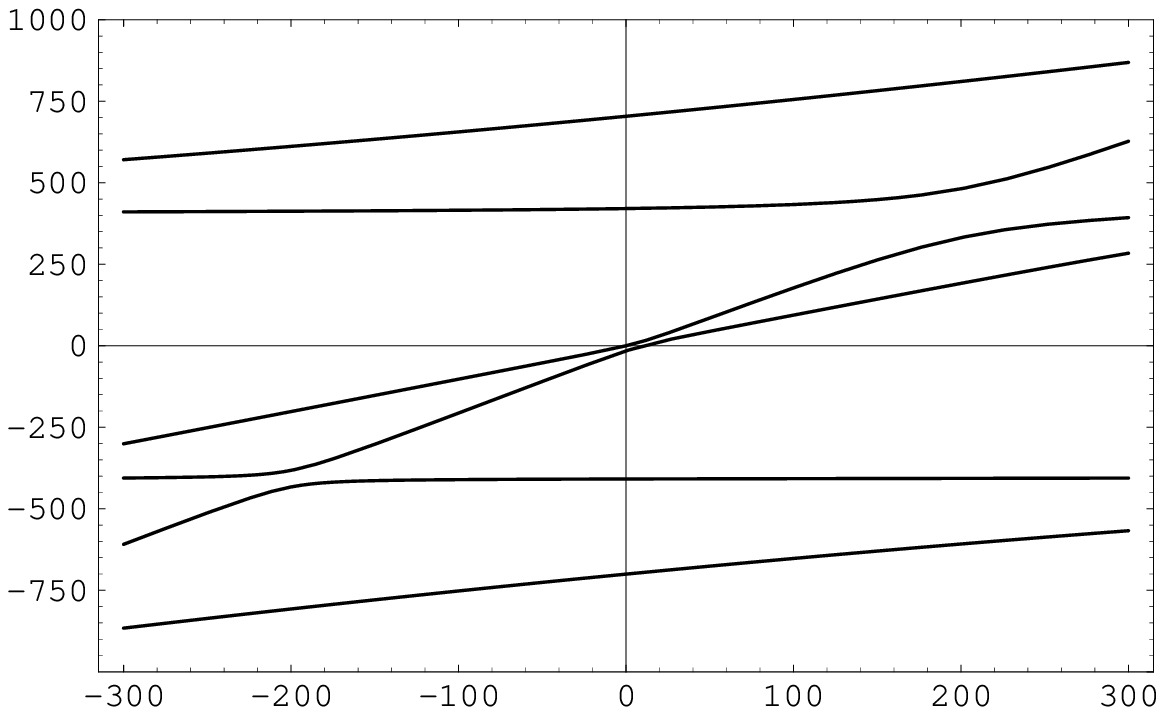}}\\
{\large $M_1$}\\[2mm]
{\large\bfseries Fig.~10a}\\ \vspace{0cm}
\end{center}
\hspace{-0.5cm}{\large $m_{\chi^{\pm}}$}
\begin{center}
{\hspace*{-20mm}\includegraphics[height=90mm,keepaspectratio=true]{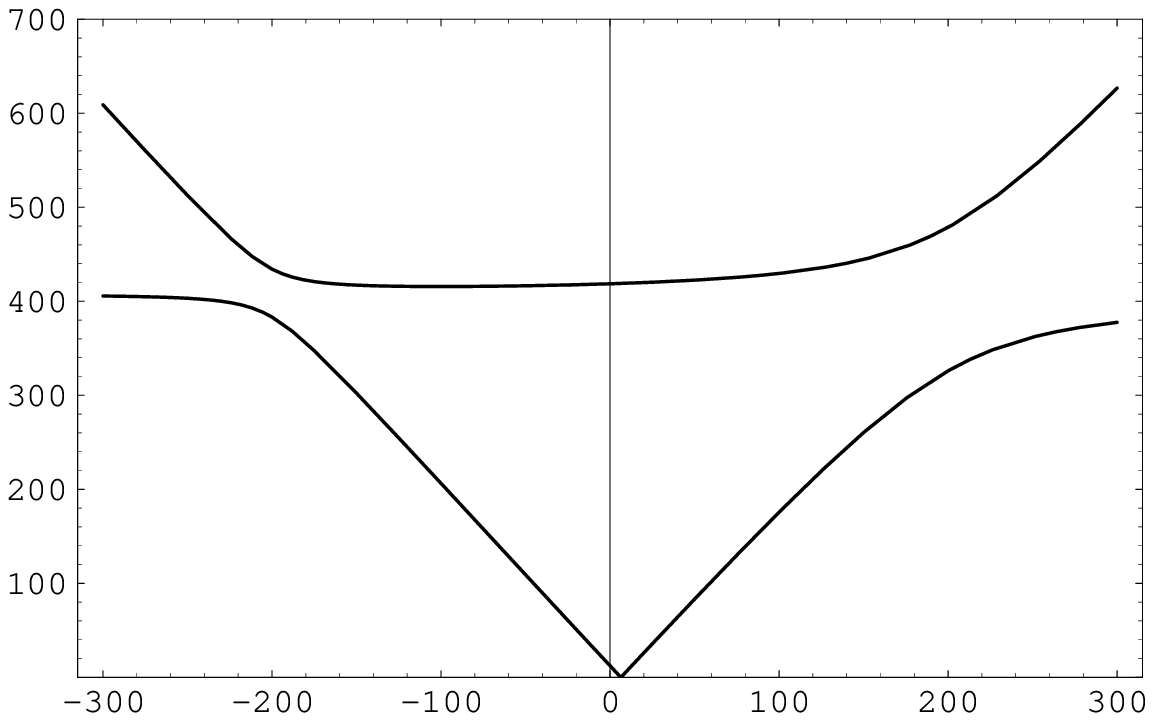}}\\
{\large $M_1$}\\[2mm]
{\large\bfseries Fig.~10b}
\end{center}

\clearpage

\begin{figure}[!t]
\begin{center}
  \epsfig{file=phenofig1.ps,height=16cm,angle=90}
\label{fig:DY}
\end{center}
\end{figure}
~\vspace{-2cm}
\begin{center}
{\large\bfseries
\qquad\qquad Fig.~11}
\end{center}

\clearpage

\begin{figure}[!t]
\begin{center}
  \epsfig{file=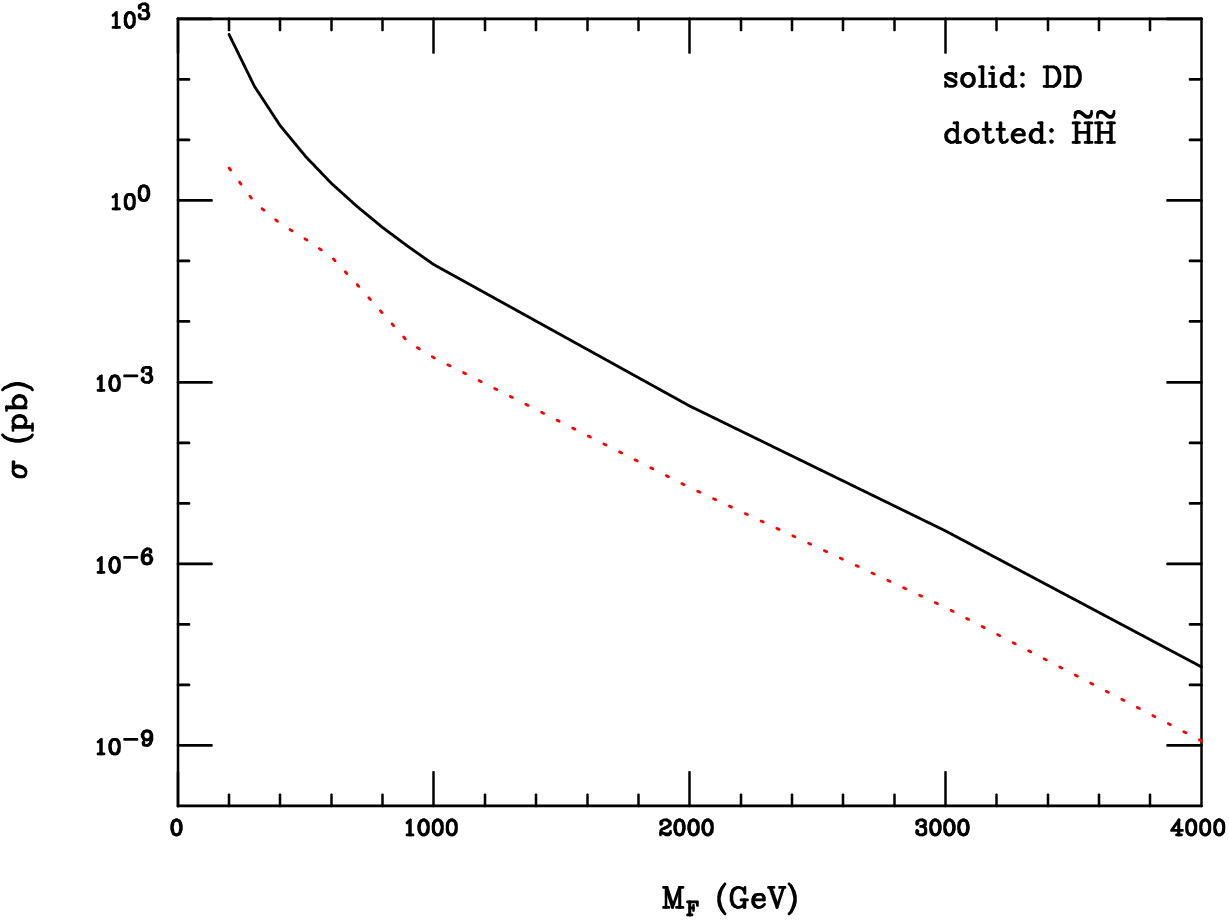,height=16cm,angle=90}
\end{center}
\label{fig:ExoticMass}
\end{figure}
~\vspace{-2cm}
\begin{center}
{\large\bfseries
\qquad\qquad Fig.~12}
\end{center}

\clearpage

\begin{figure}[!t]
\begin{center}
  \epsfig{file=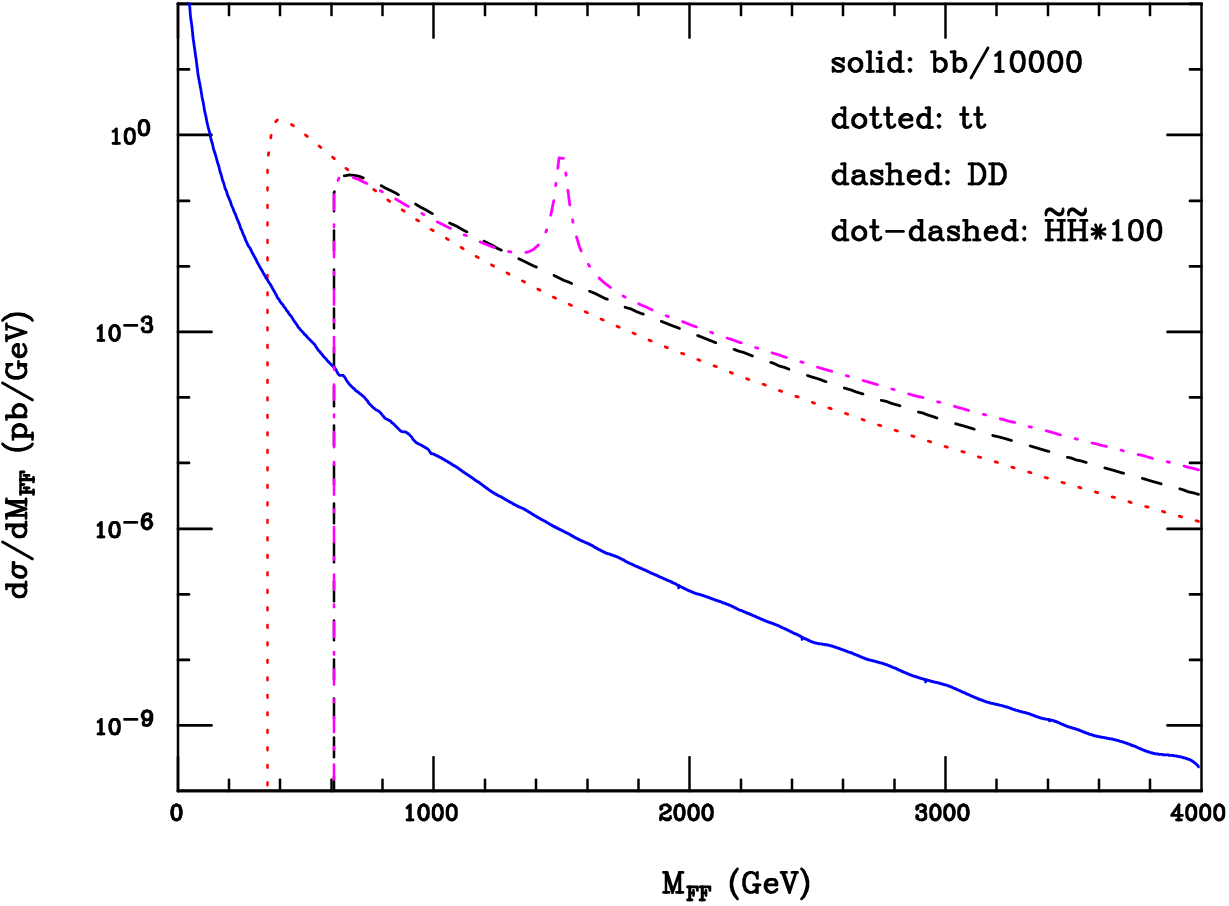,height=16cm,angle=90}
\end{center}
\label{fig:PairProdn}
\end{figure}
~\vspace{-2cm}
\begin{center}
{\large\bfseries
\qquad\qquad Fig.~13}
\end{center}

\clearpage

\begin{figure}[!t]
\begin{center}
  \epsfig{file=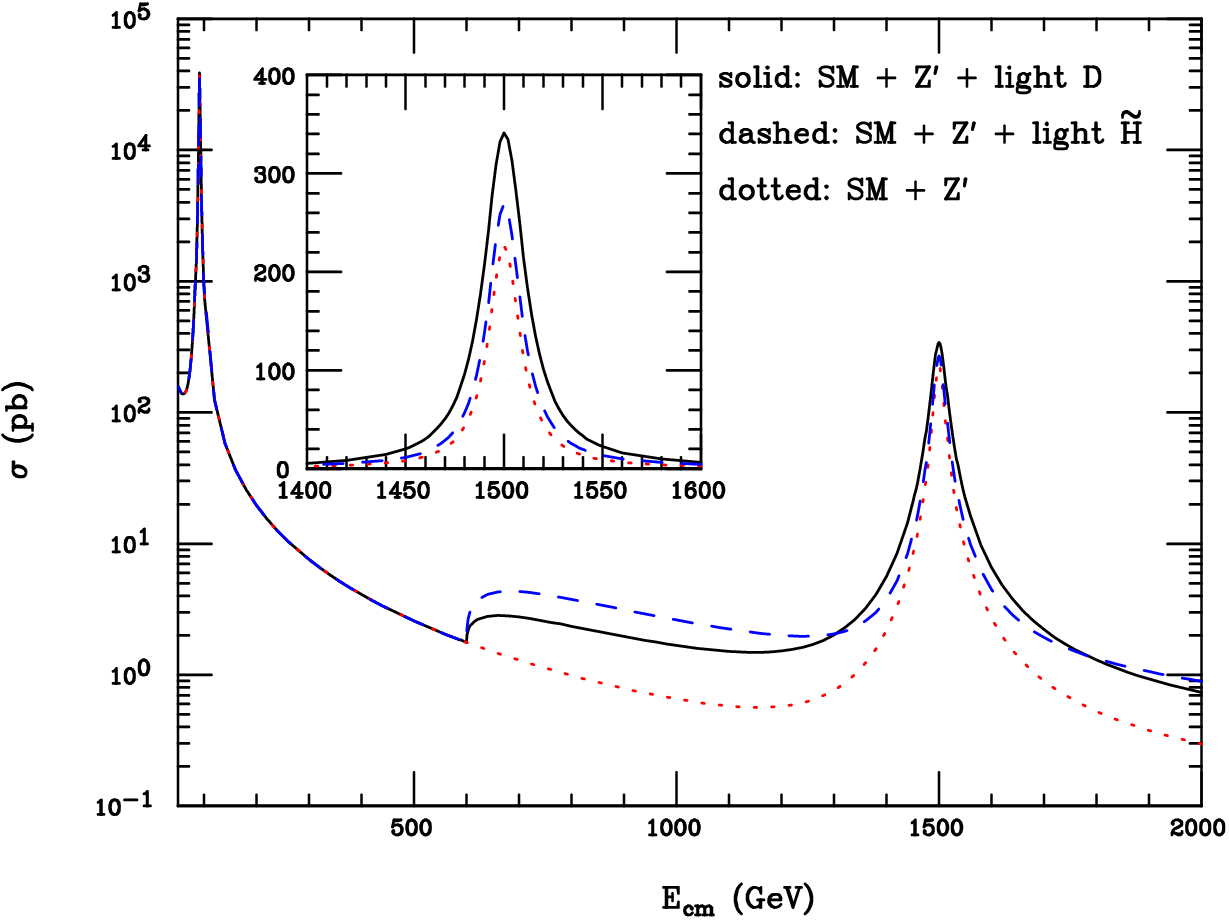,height=16cm,angle=90}
\end{center}
\label{fig:ILC}
\end{figure}
~\vspace{-2cm}
\begin{center}
{\large\bfseries
\qquad\qquad Fig.~14}
\end{center}

\clearpage

\begin{figure}[!t]
\begin{center}
  \epsfig{file=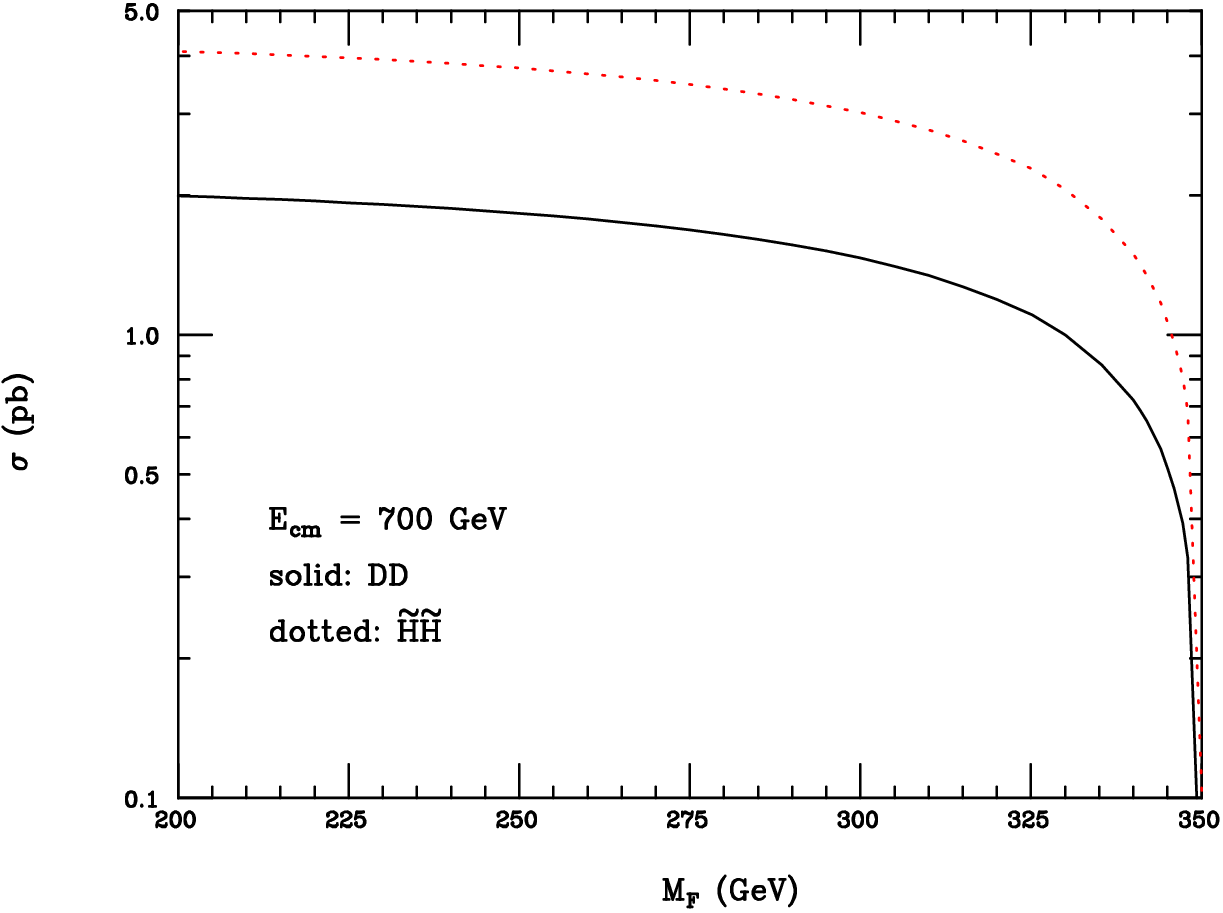,height=13.7cm,angle=90}
\end{center}
\label{fig:newILC1}
\end{figure}
{~\vspace{-14cm}
\begin{center}
{\large\bfseries
\qquad\qquad Fig.~15a}
\end{center}}
\begin{figure}[!t]
\begin{center}
{\vspace{1cm}\epsfig{file=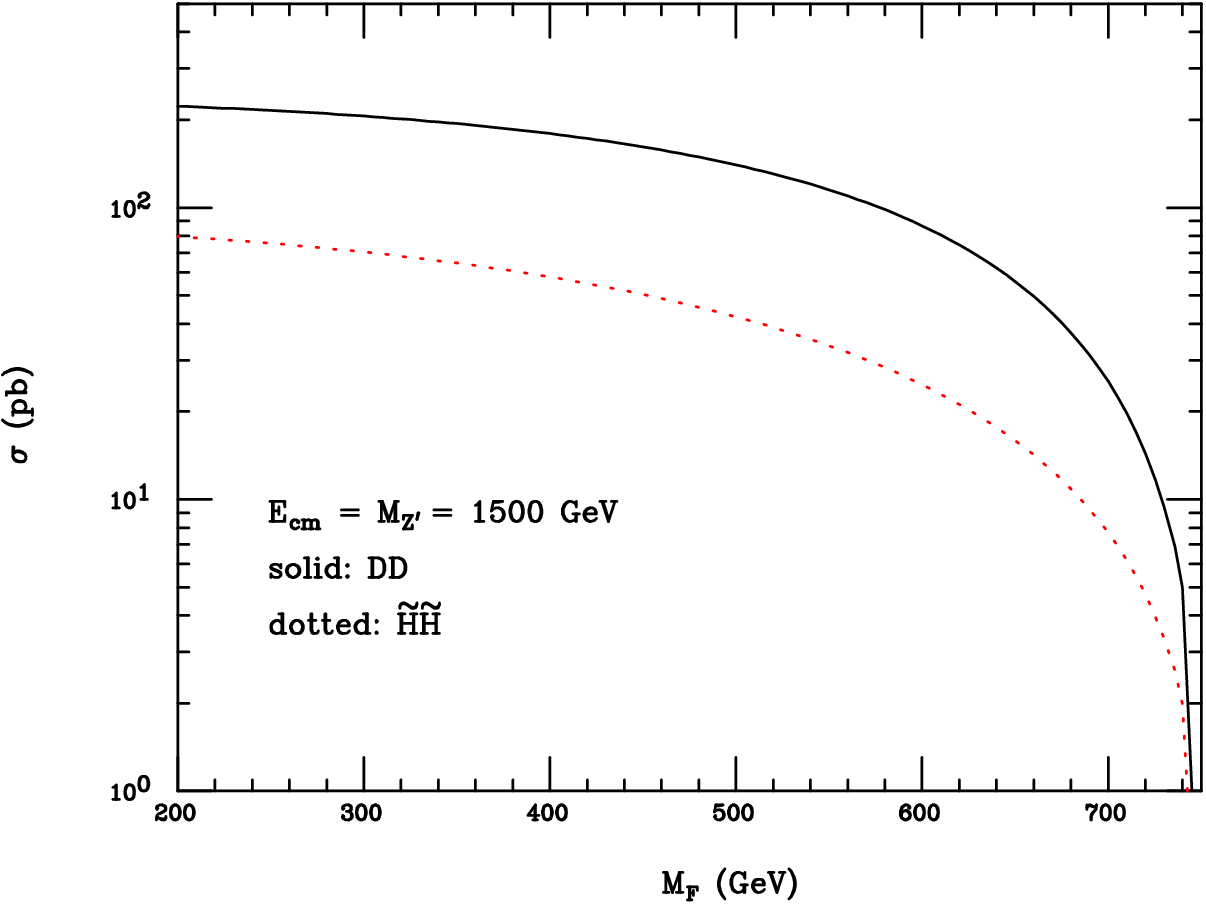,height=13.7cm,angle=90}}
\end{center}
\label{fig:newILC2}
\end{figure}
{\vspace{10.5cm}
\begin{center}
{\large\bfseries
\qquad\qquad Fig.~15b}
\end{center}}

\end{document}